# Enhanced Atomic Precision Fabrication by Adsorption of Phosphine into Engineered Dangling Bonds on H-Si Using STM and DFT

Jonathan Wyrick,[1*] Xiqiao Wang,[1,2] Pradeep Namboodiri,[1] Ranjit Vilas Kashid,[1] Fan Fei,[1,3] Joseph Fox,[1,3] Richard Silver[1]

[1]Atom Scale Device Group, National Institute of Standards and Technology, Gaithersburg, MD 20899, USA
[2]Joint Quantum Institute, University of Maryland, College Park, MD 20740, USA
[3]Department of Physics, University of Maryland, College Park, MD 20740, USA
[*]Corresponding Author: Jonathan.wyrick@nist.gov

**Abstract**

Doping of Si using the scanning probe hydrogen depassivation lithography technique has been shown to enable placing and positioning small numbers of P atoms with nanometer accuracy. Several groups have now used this capability to build devices that exhibit desired quantum behavior determined by their atomistic details. What remains elusive, however, is the ability to control the precise number of atoms placed at a chosen site with 100% yield, thereby limiting the complexity and degree of perfection achievable. As an important step towards precise control of dopant number, we explore the adsorption of the P precursor molecule, phosphine, into atomically perfect dangling bond patches of intentionally varied size consisting of 3 adjacent Si dimers along a dimer row, 2 adjacent dimers, and 1 single dimer. Using low temperature scanning tunneling microscopy, we identify the adsorption products by generating and comparing to a catalog of simulated images, explore atomic manipulation after adsorption in select cases, and follow up with incorporation of P into the substrate. For 1-dimer patches we demonstrate that manipulation of the adsorbed species leads to single P incorporation in 12 out of 12 attempts. Based on the observations made in this study, we propose this 1-dimer patch method as a robust approach that can be used to fabricate devices where it is ensured that each site of interest has exactly one P atom.

A quantum material or device is one whose functionality cannot be described by classical mechanics but rather fundamentally depends on the quantum properties that emerge from its precise geometrical and chemical makeup. In this regard, the ideal means to fabricate such a device would be to have absolute control over the placement of every constituent atom that makes a device, or at a minimum the key atoms which determine performance; for example, a transistor that operates based on the charge state of a single atom (Figure 1a) requires placement of exactly one P atom in a silicon matrix with nanometer precision positioning relative to nearby electrodes. While the necessary level of control is generally considered beyond the scope of standard fabrication techniques, it has been demonstrated for the class of electronic devices defined by precision doping of silicon. Both specialized ion implantation and scanning probe-based lithographic techniques have shown promise in the placement of phosphorus atoms at well-defined positions within a silicon crystal for use in quantum applications. In the case of ion implantation, a windowed mask can be used to achieve precision positioning control of the atoms while at the same time an electrical feedback mechanism can be used to monitor implantation events to ensure that precisely one atom gets imbedded;[1] however, the limitation of this method is that there is some positional uncertainty both in the z direction (the direction normal to the surface) as well as laterally, resulting in a positional uncertainty of ~30 – 50 nm. In contrast, the scanning probe-based hydrogen depassivation lithography (HDL) technique which uses a scanning tunneling microscope (STM) tip to controllably remove hydrogen atoms from the H-Si (100) surface (see Figure 1b) is able to position P atoms to within a nanometer (here the remaining H atoms act as a mask to subsequent $PH_3$ deposition). While HDL is in fact able to achieve atomically perfect lithography, it does not guarantee perfect dopant incorporation at a lithographic site so that true deterministic control over the number of P atoms placed has not been demonstrated.

It is important to point out that HDL has been used to fabricate devices that have resulted in single P atom precision; for example, there have been two single atom transistors made to date[2,3] where single

P atoms have been placed with nanometer precision relative to source/drain leads and gate electrodes patterned in the same atomic plane and embedded in Si (see Figure 1a). A number of donor-dot qubit style devices where 1 or a few P atoms are placed in proximity to an island of P atoms (the dot), allowing for spin selective loading and unloading on the donor, have also been fabricated;[4–7] for these devices, functionality is significantly impacted not only by precise control over tunnel coupling between the donor and the dot but also in requiring that the donor number be exactly 1 P atom (or some other controlled number of P atoms depending on the desired behavior), which is currently not possible as there is typically an uncertainty of one dopant atom per site. The reasons for uncertainty in the number of P atoms imbedded in these devices are twofold: the first is that in order to properly mask the surface the STM-based removal of hydrogen should be perfect to the last atom. This is a sufficiently challenging task that it was generally not attempted in the HDL devices mentioned above (as determined by inspection of post-lithography STM images from the respective studies). The second reason for potential failure in single atom placement is that even if the lithography is perfect, the chemistry that occurs when adsorbing and breaking down the precursor molecule ($PH_3$) is naturally subject to some degree of randomness and may not always produce the desired outcome.

Based on results from several STM and Density Functional Theory (DFT) studies of $PH_3$ adsorption on Si surfaces,[8–13] it has been inferred that if the STM is used to remove 6 H atoms from H-Si (100) as illustrated in Figure 1b, the likely outcome after depositing $PH_3$ and then heating to incorporate P into the Si lattice is that 1 P atom will be placed at the lithographic site. We refer to such lithographic patterns as 3-dimer patches because they are composed of 3 consecutive Si dimers along a dimer row, each dimer exposing 2 dangling bonds to which adsorbates can stick. One proposed mechanism[2,14] for why 3-dimer patches are the minimum needed for the standard process is that at most three $PH_3$ molecules can adsorb into the patch simultaneously, each giving up one H atom to a dangling bond within the patch; upon heating, one of the three $PH_2$ molecules then relinquishes its remaining two H atoms to its $PH_2$ neighbors, converting them to PH$_3$ which, with the additional energy from heating, desorb from the surface, leaving behind a lone P atom that readily incorporates into the Si lattice. Preliminary study of small lithographic structures[14] suggests that there is a 30% chance that 0 P atoms could incorporate at the chosen 3-dimer patch site while a more recent experimental study complemented with a model for incorporation kinetics[15] has refined this probability to 37% which would fundamentally undermine the functionality of a device intended to be comprised of single P atom components. Based on this known limitation, the single atom transistor of ref[3] was designed with a lithographic patch large enough to house more than one P atom; only after electrical transport measurements was it determined that the key transistor component in fact consisted of a single atom.

Within the field of research on HDL defined quantum devices this has thus far been the preferred way to deal with uncertainties in the number of atoms incorporated: specifically, to design and fabricate systems whose functionality is tolerant to such variation and construct lithographic structures larger than the ideal 3-dimer patch.[16] However, it will ultimately be necessary to move beyond the level of control currently available to the HDL method and develop ways to achieve greater precision. A case where this has become particularly apparent is in the use of donor arrays for analog quantum simulation (AQS) where we have been able to make significant progress with devices that include some disorder.[17] Such AQS systems can be qualitatively linked to an extended Hubbard model, but in order for them to be effective as a quantitative tool the chemical potential and on-site interactions at each array element will need to be precisely defined. This can only be done by fixing the number of P atoms exactly.

Achieving true single atom precision doping using HDL is a significant technological challenge requiring modification of the HDL fabrication process as it is currently implemented. In the present work we demonstrate the viability of using STM tip-based manipulation of adsorbed precursor molecules to ensure single atom incorporation, motivated by evidence for tip-induced H dissociation from PH$_x$ on clean Si (100).[18] Our key finding is that if manipulation is used, 1-dimer patches become the ideal lithographic structure for single-atom incorporation and a yield of 100% can be achieved.  In the absence of such modification techniques, the 3-dimer patch has historically been preferred.  It is worthwhile therefore to investigate adsorption, manipulation, and incorporation in this broader context, so we investigate both 3-dimer and 1-dimer patches as well as the bridging 2-dimer case.  There is no need to investigate adsorption into the only remaining smaller structure, that of a single dangling bond (half of a 1-dimer patch) as it is readily observed that $PH_3$ does not bind to lone dangling bonds.

In order to favorably alter the precursors it is necessary to be able to identify their atomic composition and placement on the surface reliably.  To address this, we focus first on the question of what adsorption configurations occur when $PH_3$ is deposited into patches of 3 or fewer dimers.  We use feedback-controlled lithography (FCL)[19]  to fabricate atomically perfect patterns as shown diagrammatically in Figure 1b and experimentally in Figure 1c for the 2 extremes (3-dimer patches and 1-dimer patches, respectively); this distinguishes the present work from previous efforts as each adsorption site is guaranteed to be defect free and have atomically precise, controlled geometry.  We utilize two distinct implementations of FCL, dictated by the behavior of the STM tip (two different tips were used).  For the first implementation we slowly ramp the sample bias while monitoring the feedback on tip height to determine when H atoms desorb as described previously in ref.[20]  For the second implementation, the STM tip reproducibly removed H atoms that were not directly underneath its apex so that it was not possible to detect any change in height during desorption events.  The precise reason for this behavior is not clear, though it is known that for some STM tips the tip-induced potential on a surface can have a maximum that is laterally offset from the tunneling apex.[21]  To account for this, we implemented a pulsed FCL method in which the sample bias is pulsed to 3V for 2s while the setpoint current is set to 2nA.  Immediately following a pulse the tip apex is automatically moved to the desorption site, followed by a second location where the relative height difference between the two sites is compared to a previous reading of the same two locations.  The process is repeated until this relative height difference changes, at which point the cycle is stopped (this FCL process is described in more detail in section S11 of the supporting information).

After the FCL step, we deposit PH$_3$ gas which selectively chemisorbs into the patches, undergoing a dissociation reaction in which at least one H atom is removed, typically attaching to a nearby dangling bond (notably we find that this might not always be the case). In order to identify what the resultant adsorption structures are we image them with STM (see methods for details). The interpretation of the acquired images is not straightforward and requires an in-depth analysis if the identity and placement of molecules is to be properly determined. We address this by performing DFT calculations to simulate possible adsorption configurations for PH$_x$ in 3-dimer (and fewer) patches and generating simulated STM images which can be compared to experimentally acquired images. If we can find a match between a simulated image and an experimental one, then we can conclude the atomic input geometry used to generate the DFT image represents the correct configuration. The supercell and basic atomic setup used for image simulation is shown in Figure 1b. After identification, we then use the STM tip to modify species in a subset of patches while leaving others unmodified as a control. We then heat the sample to incorporate P into the substrate and image the resultant surface. Based on these post incorporation images we determine for each case whether it yielded successful P incorporation.

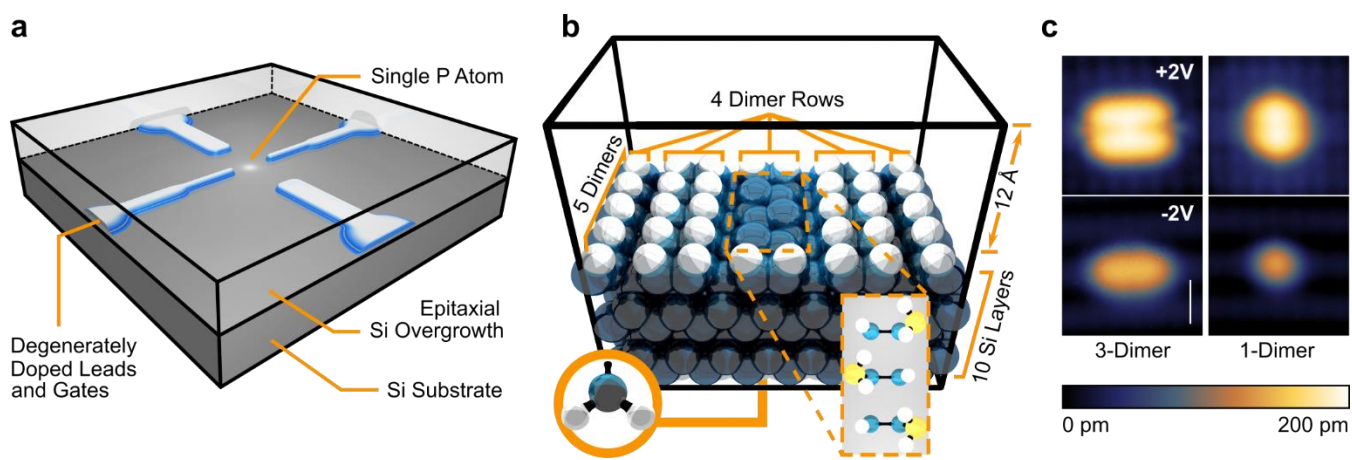

**Figure 1.** STM patterning for fabrication of single atom devices. (a) Schematic of an encapsulated single atom transistor device. (b) Atomic structure of the H-Si (100) surface where hydrogen (white atoms) has been removed

from 3 consecutive Si (blue atoms) dimers on a single dimer row, illustrating the geometric setup used for DFT calculation of simulated STM images in this study. Image simulation is performed on a 10 layer slab, with a reconstructed hydrogen terminated surface on top, and hydrogen termination on the bottom (bottom-left inset) that mimics coordination to bulk silicon atoms. The bottom-right inset shows a possible saturation dose adsorption configuration of 3 ($PH_2$ + H) molecules. (c) STM empty (top) and filled (bottom) states images acquired on a single 3-dimer (left) and a 1-dimer (right) lithographic patch; indicated voltages are the sample bias and the inset scalebar shows the separation between two dimer rows of 0.77 nm.

## Results and Discussion

In order to determine what adsorption configurations are possible, we first consider the likely molecular species that will result from attachment of the $PH_3$ precursor to the surface as well as to what sites a given molecule can bond. While $PH_3$ can bond to the surface directly, it dissociates quickly[9–11] into the products $PH_2$, PH, and P. We first use DFT calculations with the reduced supercell shown in Figure 2a to determine the energetic minimum geometries. To save computation time, we allow only the 6 Si atoms of a 3-dimer patch to relax in addition to the constituent atoms of the adsorbed molecule. By trying multiple possible adsites for each species (top, bridge, and hollow) and then allowing the system geometries to relax according to DFT calculated forces, we find that the preferred adsites are as shown in Figure 2b-f; these results are in good agreement with previous calculations.[9,10] We also confirm that Si dimers prefer to buckle as shown in the side view of Figure 2g, and that they do so in an alternating manner along the dimer row consistent with the standard c(4x2) reconstruction of the Si (100) surface.[22] Once a Si atom is coordinated to at least one additional atom, however, its associated dimer flattens as shown in the side view of Figure 2h.

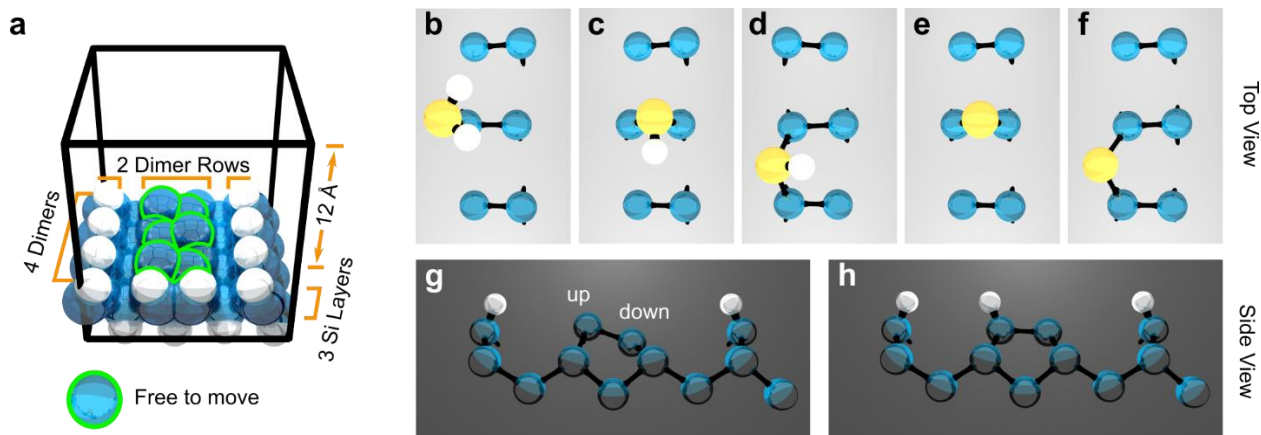

**Figure 2.** DFT relaxation of atomic coordinates on a reduced supercell. (a) Reduced setup details: the slab consists of 3 Si layers, with all atoms fixed in position except for the 6 Si atoms for which hydrogen termination has been removed (outlined in green), as well as any adsorbates placed into the 3-dimer patch. (b – f) Final energetic minimum geometries for adsorbed $PH_x$ species demonstrating that (b) $PH_2$ prefers top-sites, (c,d) PH prefers bridge sites, and (e,f) P also prefers bridge sites. (g) Side view of a relaxed dimer showing that the energetically preferred configuration is in a buckled state. (h) Side view of a relaxed dimer with 1 hydrogen atom adsorbed demonstrating that the dimer becomes un-buckled once at least 1 Si atom has coordinated to an additional species.

Based on this information, we generate a comprehensive list of all possible adsorption configurations that might occur upon dosing $PH_3$, with the following additional assumptions: 1) a surface Si atom can be coordinated to at most 1 additional species (the Si atoms are already bonded to 3 other Si atoms), and 2) Si-Si bonds, including the dimer bonds, will not break (unless DFT atomic coordinate relaxation predicts it) in order to accommodate bonding to $PH_x$. The first assumption is reasonable in so much as this is the most likely coordination for Si dictated by its chemistry. Since the sample is maintained at room temperature and no external energy is input beyond the kinetic energy of the dosed molecules, it is unlikely (though not impossible) that the activation energy for breaking bonds will be surmounted in ways not accounted for by DFT relaxation. These are simplifying assumptions and exceptions to them are physically possible, however we do note that within these constraints we have been able to find matching configurations for all of the experimental images obtained to date. This set of rules results in 754 total adsorption configurations (see section S1 of the supporting information for more details), which are then each geometrically relaxed using the DFT setup of Figure 2a. Once we have the set of relaxed coordinates for a configuration, we transfer the coordinates for the 6 Si atoms of the lithographic patch as well as those of the adsorbed molecules onto the larger slab of Figure 1b. For the larger system, we require all atomic positions to remain fixed, calculate the electronic ground state, and then use the integrated local density of states from the Kohn-Sham orbitals to simulate STM images for both filled and empty states. The result is a catalog of simulated STM images that can be compared to experiment as a means to identify observed adsorption configurations; the full catalog is presented in section S13 of the supporting information. Importantly, because we allow additional H atoms to be adsorbed into the 3-dimer patch, the generated catalog covers all sizes and geometries of patches consisting of 3 dimers and fewer (*i.e.* structures in the smaller patches of this study are also identifiable).

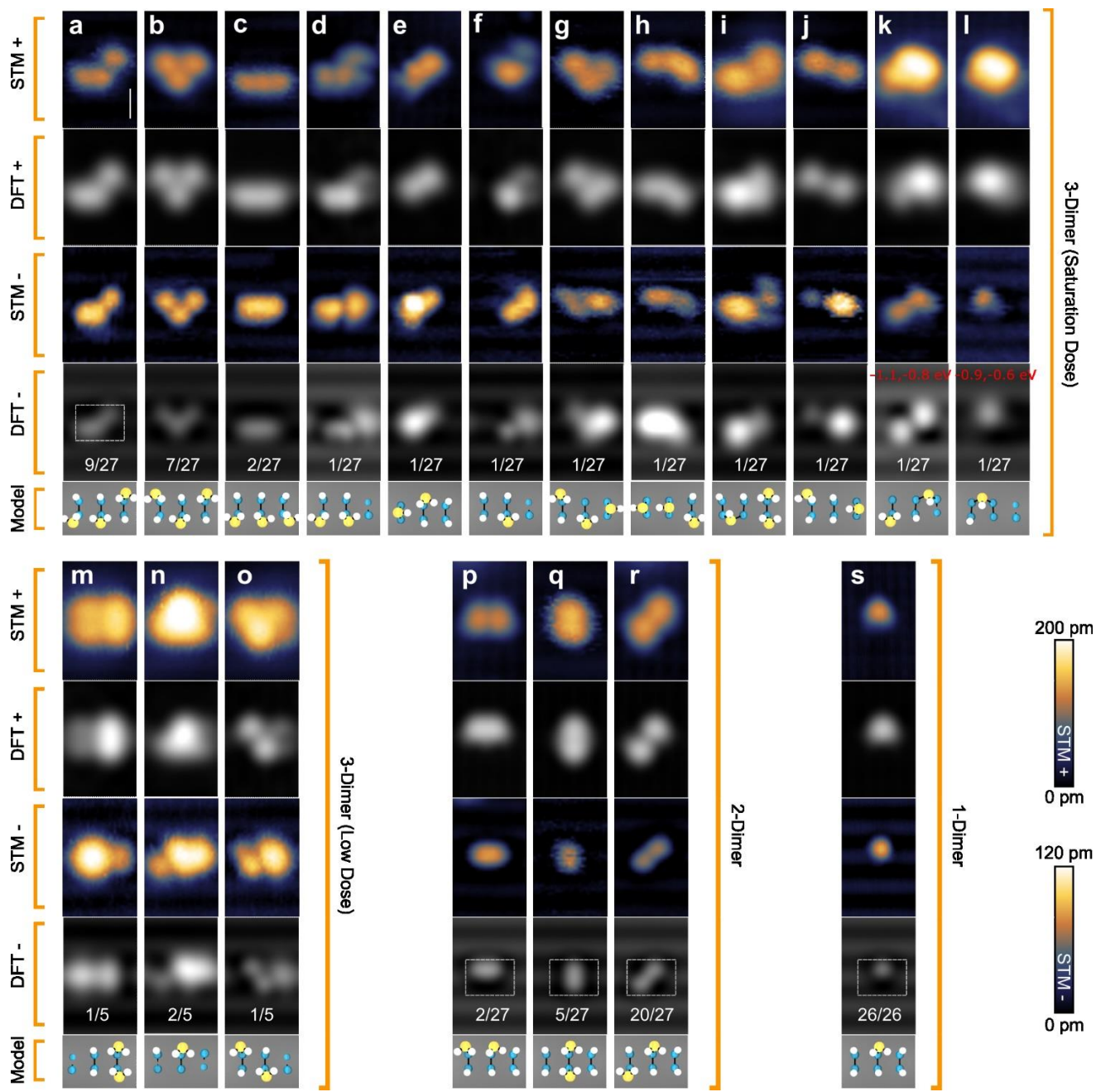

**Figure 3.** Experimental and simulated STM images for saturation (a-l) and low dose (m-o) adsorption of PH$_x$ into 3-dimer patches, 2-dimer patches (p-r), and 1-dimer patches (s). Image type is organized by row: experimental and simulated positive sample bias STM image, experimental and simulated negative sample bias STM image, and ball-and-stick model of the identified adsorption configuration, respectively. All STM images in the STM+ rows have a z (tip height) scale according to the top color bar, while those of the STM- rows are scaled according to the bottom color bar. The scalebar in (a) is 0.77 nm in length, spanning exactly 1 dimer row (this can be seen more clearly in the supporting information, Figure S7, where the image color scale has been adjusted to highlight dimer rows). The dashed box in the DFT- image of (a) shows the size of the ball and stick model as overlayed on the simulated image. Equivalent dashed boxes are shown for (p-s). The frequency of appearance for each configuration is overlaid on the DFT- images and is expressed as a ratio of the number of times a configuration was seen divided by the total number of corresponding patches of a given size and dose that were patterned (*i.e.* there were 27 saturation dose 3-dimer patches, 5 low dose 3-dimer patches, 27 saturation dose 2-dimer patches, and 26 saturation dose 1-dimer patches).

Red labels for (k,l) indicate modified limits of integration with the lower limit on the left and the upper on the right (see methods).

Figure 3 shows comparrisons of the experimentally observed STM images with the DFT simulated images that have been identified as being the most likely matches. Three experimental runs were performed, the first of which involved a low dose of $PH_3$ into a total of five 3-dimer patches (Figure 3m-o). The second and third runs were saturation dosed and in combination consisted of a set of 27 3-dimer patches (Figure 3a-l), 27 2-dimer patches (Figure 3p-r), and 26 1-dimer patches (Figure 3s). One patch had no adsorption for the low dose, while all patches showed adsorption for the saturation doses.

With the ability to identify the adsorption configurations we now have insight into the chemistry of how $PH_3$ reacts with the optimal lithographic structures for single P atom incorporation (3-dimer patches if no manipulation is performed, as predicted by previous DFT and un-passivated surface adsorption studies,[9–12] and 1-dimer patches if manipulation is performed). The expected adsorption behavior was that dissociation products should form as a result of 1 H atom from a $PH_3$ molecule attaching to an empty site in a 3-dimer patch while $PH_2$ attaches to another empty site. Since there are 6 top sites in a 3-dimer patch, with a saturation dose there should be 3 $PH_2$ molecules and 3 H atoms in each patch; the proposed dominant configuration[2,12] being that shown in the inset of Figure 1b and seen experimentally in Figure 3b, where the $PH_2$ molecules bond to separate dimers and alternate which side of the row they are on. We find, however, that the configuration of Figure 3a with two $PH_2$ molecules on one side of the dimer row and the third on the other is equally common.

We also observe that in several cases the number of H atoms is not conserved (Figure 3f-I,k,l). For example, in Figure 3l there should be a total of 4 H atoms at top sites (2 from the PH, and 1 from each of the $PH_2$ molecules); in a 3-dimer patch, there are not enough top sites available for this to occur yet we still observe such configurations, implying that by some mechanism the remaining 2 H atoms must have left the vicinity of the 3-dimer patch. Given that this behavior is not seen in the low dose sample, we conclude that the additional H atoms are being driven off or reacting with other $PH_3$ molecules trying to bond to the patch, though the precise mechanism by which this happens is unclear.

A number of important observations can be made with respect to the data of Figure 3 which we discus here. Identification of adsorbed $PH_x$ structures in STM images can be quite difficult, particularly since there are so many possible configurations. While it is not perfect (see section S4 of the supporting information for an example), comparison to the catalog of simulated images significantly narrows down the set of possible matches. Inconsistencies between the simulated STM image and the experimental image can arise for a number of reasons, the primary ones being: uncertainty as to the actual tip height above the sample (an arbitrary estimate must be chosen for image simulation), uncertainty in the mapping from DFT Kohn-Sham energies to sample bias, and tip-induced charging effects. It is possible to improve the tip height estimate by comparing to simulations of known features (*e.g.* dangling bonds, common defects, *etc.*) and adjusting the simulated height to get the best fit, however tip-induced charging and Kohn-Sham energy mapping can be confounded with this; more importantly, the purpose of the catalog is to be used not only with the present study, but with future experiments in which the details of tip height, *etc.* may be different. The fact that not all atoms were relaxed (to keep computation time practical) and, that they were only allowed to relax on the reduced supercell of Figure 2a, may also play an important role in differences between simulated and calculated STM images in some cases, though notably relaxation on a larger supercell did not result in a significant change to final atomic coordinates for a test case (see supporting information section S2). Additionally, details of the tip can affect details of the resultant image based on what tip orbitals are overlapping with sample states (*e.g.* s, p, d, or some mixture of them). A reasonable future improvement to this catalog could be made by expanding it to include atom resolved AFM studies and corresponding simulations; in a number of cases this method has proven to resolve detailed atomic structure better than STM,[23] and has been used to identify common defects on H-Si (100) surfaces.[24]

For the low dose configurations shown in Figure 3 as well as saturation dosed configurations d,f, and l, where each has a Si dimer that does not participate in bonding, buckling behavior has an effect on the resultant STM image because imaging with an STM tip can potentially induce the buckled state to change;[20,25,26] in other cases the *up/down* atom (*up* and *down* in the context of buckling are illustrated in Figure 2g) can alternate more rapidly than the data acquisition rate, resulting in an image that is a thermal average of the 2 configurations, and in other cases the presence of a nearby defect can pin the dimer atoms in a fixed buckled state. For each such configuration of Figure 3 we find that at positive sample bias the resultant image is best represented by averaging the 2 *up/down* buckled configurations. Averaging also gives the best results for the negative bias images of Figure 3m,o, but not for Figure 3n. For the latter, it appears that the dimer is pinned in the configuration shown: the *up* atom is closest to the $PH_2$ (see supporting information section S5 for a comparison that illustrates this).

For a subset of patches on the saturation dosed samples, we performed feedback-controlled manipulation (FCM) followed by heating the sample to 367 °C for 2 minutes in order to achieve incorporation (*i.e.* at least 1 P atom substitutionally replaces an underlying Si atom which is ejected to the surface). FCM uses the same algorithmic control of STM current and voltage as our two described methods of FCL, however we distinguish the procedure as manipulation, rather than lithography, because in this case we manipulate the adsorbed species instead of creating dangling bond patches for subsequent precursor molecule adsorption. For three of the 3-dimer patches we performed FCM with the tip positioned directly above a $PH_2$ molecule, while for three other 3-dimer patches unintended modification occurred as a result of an unstable tip-sample junction during imaging; because all of these cases resulted in H desorption from the adsorbed species (as well as some of the surrounding surface H) we treat them as manipulation events, but it should be noted that these were performed in an uncontrolled manner.

Using a new STM tip, we performed controlled manipulation studies on 12 1-dimer patches, leaving 14 other 1-dimer patches unmodified for comparison. This set of tests is the ultimate focus of the present work; they are a demonstration of a repeatable method for 100% yield single P incorporation. Figure 4a shows the details of the 1-dimer single atom incorporation method. In comparison to 3- and 2-dimer patches there is an immediate advantage to deposition into 1-dimer patches as evidenced by the result of Figure 3s, that the adsorbed species is always the same: a single $PH_2$ molecule occupies one of the two Si top sites while a single H atom occupies the other. Previous DFT results suggest that the barrier to incorporation will be smallest if we can remove the H atoms from the $PH_2$ leaving an isolated P atom.[12,13] Based on the preferred adsites illustrated in Figure 2, it will be necessary to make bridge sites available for the P to occupy first. We remove the two H atoms from the neighboring dimer by placing the tip at site 1 of Figure 4a and applying pulsed FCM. Following this, we move the tip to site 2 and once more repeat the pulsed FCM algorithm, effectively manipulating the underlying $PH_2$ and because the tip is placed at the center of the dimer row it is likely that the H covering the other Si atom will also desorb (opening up an additional possible bridge site). In the ideal case, the resultant structure is that of Figure 4b, a single P atom surrounded by the H-terminated surface. It is typical at this point that at least one of the H atoms (and often both) from the original $PH_2$ takes the place of one the previously removed H atoms while still leaving a bridge site for the P. This exact behavior however is not guaranteed and does not appear to be a requirement in so much as pulsing the tip often creates spurious dangling bonds and the final resultant structure may appear more complicated than the ideal case of Figure 4b (some examples are given in supporting information, section S8). So long as the final configuration includes a single P atom, which can be identified by its distinctive appearance in positive and negative bias STM images (Figure 4b), we find that upon heating, the P atom will incorporate into the lattice.

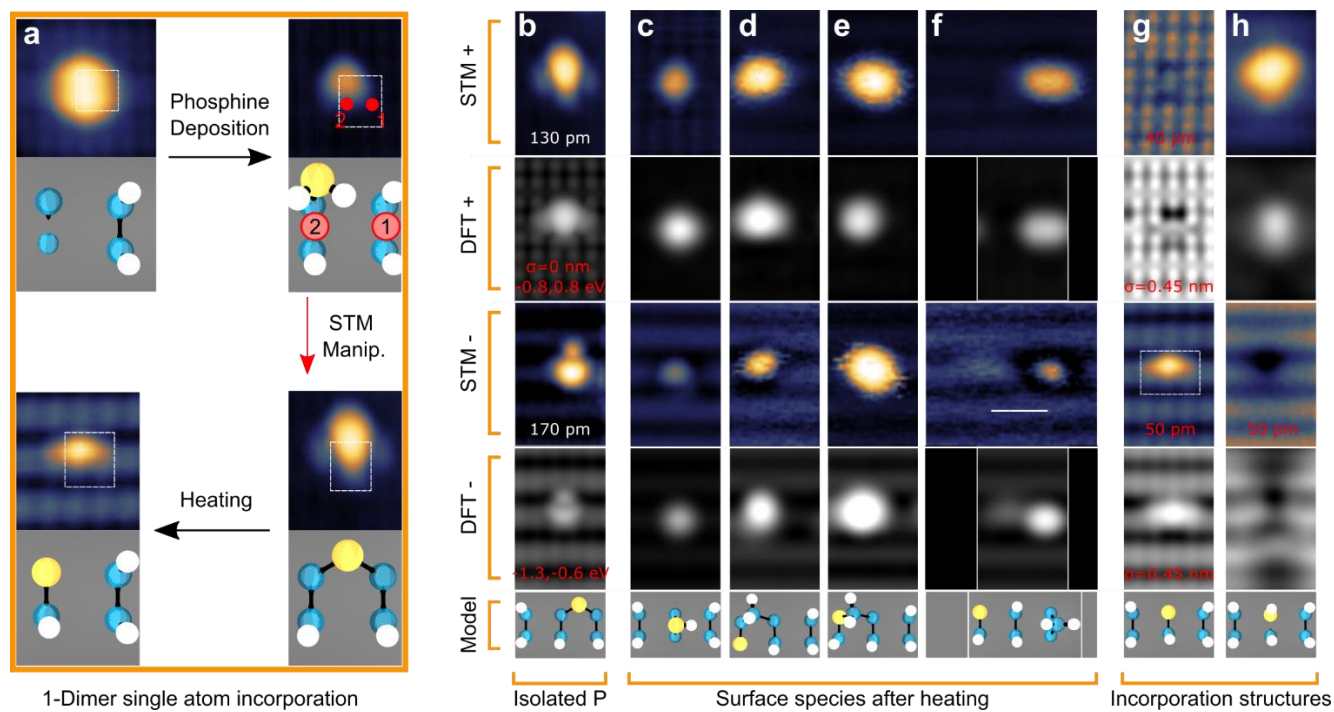

**Figure 4.** Single atom incorporation and identification of resultant surface structures. (a) Procedure for incorporation of a single P atom using a 1-dimer patch: phosphine is deposited and adsorbs as PH$_2$, FCL is used at an adjacent dimer (site 1) to remove two H atoms, FCM is used at site 2 to strip H atoms from the PH$_2$, resulting in P at a bridge site which is then heated to incorporate into the lattice. (b) Isolated P at a bridge site off the center line of the dimer row as seen after FCM. (c-f) Surface species seen after heating the substrate, including indirect evidence of incorporation in the form of ejected Si (d-f). Separation between the bridging SiH$_2$ in (f) and the subsurface P is 1.155 nm along the dimer row direction as indicated by the inset scale bar, corresponding to a separation of 3 dimers; the corresponding simulated image has a 2 dimer separation, but is qualitatively similar. (g-h) Direct evidence for P incorporation, without (g) or with (h) an H atom atop the incorporated P. The image types of (b-h) are arranged into rows with the same organization as Figure 3. The color scale for the heights of experimental images is the same as Figure 3 with the maximum positive bias image height of 130 pm and negative bias image height of 170 pm as indicated by the inset labels of (b). Cases where different image processing parameters were used (for clarity) are noted by red labels with experimental images having variation in maximum height, and simulated STM images having variation in the amount of Gaussian blur used, and, in the case of (b), variation in the limits of integration over Kohn-Sham orbitals (see methods).

Determination of successful incorporation requires imaging with STM after the heating step and analysis of the resultant structures at each patterned site. Figure 4c-h shows the types of structures that we observed for which the STM image after incorporation appears different from that before incorporation. Among these, there are cases of indirect evidence for incorporation in which the top layer Si atom that was replaced by P is ejected to sit on the surface (Figure 4 d-f). Notably, it appears that during the 2 minutes that the sample is heated, while the surface H does not desorb, it is sufficiently mobile so that in each case that we've observed it is able to enter the lithographic patch and decorate the ejected Si atom to create SiH$_2$.

We also find that the ejected Si species are mobile enough that in some cases they diffuse away from the lithographic site, leaving behind direct evidence for incorporation as illustrated in in Figure 4g,h. In the case of 1-dimer patches where FCM was used to convert $PH_2$ into P, all ejected Si atoms left the lithographic site.

Table 1 summarizes our post-incorporation findings for each type of lithographic patch studied, further divided into cases where manipulation was used or not. The first two columns represent cases where the P was successfully incorporated with "Ejected Si" corresponding to outcomes matching Figure 4d-f and "Direct" corresponding to those matching Figure 4g,h, while the remaining three columns are for cases where P failed to incorporate: the adsorption structure changed but no P incorporated, the adsorption structure remained unchanged, or all species desorbed, respectively. In the cases of structures that changed but did not incorporate, we generally found products where some $PH_x$ had desorbed while some remained in the patch but underwent dissociation reactions. Figure 4c shows such an example, where the final product is PH; this assignment is further supported by a follow-up annealing step (367 °C for 2 minutes) that was performed in which we saw the PH desorbed rather than incorporate. Importantly, Table 1 shows that a variety of non-ideal outcomes (the last 3 columns) occur when a tip-based manipulation step is not included, and by comparison shows favorable results when manipulation is used. Two exceptions to this come from the manipulated 3-dimer patches: the desorption event depicted in the table occurred when attempting FCM on a 3-dimer patch structure where the tip picked up all of the adsorbed species, and in a separate case manipulation resulted in two ejected Si atoms upon incorporation, suggesting that more P atoms were incorporated at the site than intended. In this respect, manipulation of 3-dimer patches is not a viable method for single-atom incorporation as there is a risk of causing more than 1 P atom to incorporate. Since the goal of the present study is to improve single atom incorporation, we have investigated only a small set of 3-dimer manipulation cases and no such cases for 2-dimer patches which would suffer from the same problem.

**Table 1.** Incorporation outcomes. "Ejected Si" refers to cases as seen in Figure 4d-f. "Direct" Refers to P-Si heterodimers as seen in Figure 4g,h. "Changed" refers to a changed adsorption structure but no incorporation such as PH in Figure 4c. "Unchanged" refers to cases where the adsorption structure did not change upon incorporation. "Desorbed" refers to cases where only clean H-Si (100) surface is found after incorporation.

| Preparation | Ejected Si | Direct | Changed | Unchanged | Desorbed | Totals |
|---|---|---|---|---|---|---|
| 3-Dimer Unmanipulated | 7 | 3 | 9 | 2 | 0 | 21 |
| 3-Dimer Manipulated | 5 | 0 | 0 | 0 | 1 | 6 |
| 2-Dimer | 0 | 0 | 15 | 8 | 4 | 27 |
| 1-Dimer Unmanipulated | 1 | 0 | 2 | 11 | 0 | 14 |
| 1-Dimer Manipulated | 0 | 12 | 0 | 0 | 0 | 12 |

When compared to previous 3-dimer patch studies,[14,15] the unmanipulated 3-dimer patch row shows reasonably consistent results, particularly when the prevalence of PH molecules that make up the "changed" column is considered. We found that individual PH molecules (Figure 4c) appear as single protrusions of similar size at both positive and negative bias, with similar apparent height to that of ejected Si. Of the "changed" species, four were PH. Following the convention adopted by Ivie *et al*.[15] in which "unchanged" species are not included in the count, we report a yield of 53% (out of a sample of 19) for unmanipulated 3-dimer patches. However, if PH were misidentified as ejected Si, the reported yield would instead be 74%. Given sample sizes, both of these interpretations are consistent with previous reported yields. Importantly, these numbers as well as the significantly lower yields associated with unmanipulated 2-dimer and 1-dimer patches corroborate the notion that these are not practical for robust single donor incorporation.

Based on the results of Table 1, the preferred method for single P atom incorporation is the 1-dimer method depicted in Figure 4a. The contrast between unmanipulated 1-dimer adsorption structures and those that were manipulated is immediately evident. Out of 14 unmanipulated structures, only 1 showed evidence of incorporation, while all 12 manipulated 1-dimer structures incorporated. Notably, it appears that incorporation of a lone P atom is relatively quick on the timescale of the 2 minute anneal as the ejected Si atoms diffused away from the 1-dimer patch in all 12 cases, leaving behind direct evidence for incorporation in the form of the P-Si heterodimers of Figure 4g,h. Furthermore, there is no risk of unintentionally incorporating multiple P atoms at a site since there can be at most one P present in a 1-dimer patch due to $PH_3$ adsorption behavior (*i.e.* one H must dissociate from the $PH_3$ and the two product species, $PH_2$ and H, will sit at the two available Si top sites).

While not seen in any of the 1-dimer patch adsorption structures, because of the unpredictability of STM tip-sample junctions there is always the increased possibility of unintentional P desorption during manipulation attempts as did occur in one case for a 3-dimer structure. If such an event occurs, it is still possible to re-passivate the site if necessary and follow up with FCL fabrication of a 1-dimer patch and redeposit $PH_3$, at which point manipulation can be attempted once more. An iterative procedure of this nature (where steps are repeated until all sites are incorporated as intended) would be very costly in terms of time, potentially adding many hours per iteration, but may be considered worthwhile for fabrication of an atomically perfect device. Additionally, the presence of direct evidence for subsurface P in all 12 manipulated 1-dimer patches suggests a preferred standard for incorporation evidence: since the P atoms appear to remain at their original incorporation site while the ejected Si diffuses away, future post-incorporation testing may benefit from longer anneal times as necessary (*i.e.* 2 minutes is already sufficient for 1-dimer patches but longer times may be needed for 2- and 3-dimer patches), targeted at removing the obfuscating ejected Si atoms.

**Conclusion**

Summarizing the key results, we have studied adsorption of $PH_3$ into 3-dimer patches and smaller, following up with FCM on selected adsorption structures and subsequent P incorporation. We find that the most common adsorption configurations for 3-dimer patches are those of Figure 3a,b, while for 2-dimer patches the most common configuration is that of Figure 3r, and for 1-dimer patches there is only one possible configuration, that of Figure 3s. We find that tip-based manipulations using FCM significantly improve incorporation probabilities upon annealing, even enabling incorporation of 2 P atoms at one 3-dimer patch site. Based on these results, we find that the 1-dimer combined with FCM protocol (Figure 4a) is a robust and reliable method for the fabrication of 100% yield single P atom-per-site structures using the HDL technique. This finding is of particular significance because it advances the HDL technique from being unreliable with regards to single atom placement where previous results showed a 30-37% failure rate at each site.[14,15] The catalog of simulated STM images (see supporting information section S13) that was used in this study was instrumental in determining both adsorption and post incorporation configurations, enabling a more in-depth look at the chemistry of $PH_x$ in designer lithographic patches. Additionally, the ability to determine the species at each site ensures STM operators can make educated decisions as to how to respond to adsorption configurations seen while fabricating using HDL. As HDL techniques develop toward greater device complexity, it will become increasingly important to also control the number of P atoms at each site to be greater than one (*e.g.* ensuring exactly 2 P atoms incorporate); knowing where to place the tip and what manipulations to perform will require the precision identification enabled by the simulated image catalog. Some examples of devices that would benefit are singlet-triplet qubits[27] consisting of a single P atom at one site within tunneling range of a second site with 2 P atoms, and AQS arrays[17] where selected sites have a predetermined number of P atoms intended to induce a different chemical potential than sites with precisely 1 P atom. In this study we have presented a pathway to use HDL fabrication in development of devices and materials where the details of the Hamiltonian are engineered with absolute precision based on the number and placement of atoms in quantum structures.

## Methods

**Sample and Tip Preparation.** H-Si (100) surfaces (the low dose and saturation dose studies were on separate samples) were prepared from 2.5 mm × 10 mm × 0.25 mm chips that were lightly *p* type boron doped at a density of $10^{15}$ to $10^{16}$ cm$^{-3}$. The samples underwent standard chemical cleaning consisting of Base Piranha, RCA-1, and RCA-2, followed by introduction to ultrahigh-vacuum (UHV) at a base pressure of $6 \times 10^{-11}$ mbar ($6 \times 10^{-9}$ Pa). Samples were degassed at 600 °C for 12 hours via direct current heating, then flashed to 1200 °C for 45 s; flashing was aborted to 600 °C whenever pressures went above $1 \times 10^{-8}$ mbar ($1 \times 10^{-6}$ Pa) and then resumed once pressures went below $2 \times 10^{-10}$ mbar ($2 \times 10^{-8}$ Pa). Hydrogen passivation was performed at 367 °C (see supporting information section S10 for details on temperature calibration) with H$_2$ gas backfilled into the vacuum chamber at a pressure of $2.8 \times 10^{-6}$ mbar ($2.8 \times 10^{-4}$ Pa) for 20 minutes; at the same time a W filament (7 W, 0.7 A) in line of sight to the sample was used to crack the H$_2$ molecules into their constituent atoms. The sample was then transferred to the STM UHV chamber with a base pressure below $4 \times 10^{-11}$ mbar ($4 \times 10^{-9}$ Pa).

The tips were prepared by electrochemically etching a polycrystalline W wire in KOH solution. They were then cleaned in vacuum by annealing for several hours before use. Final tip preparation was done by *in situ* modifications such as voltage pulses (up to 10 V) and current setpoint pulses (up to 300 nA).

**STM Setup.** All lithography and scans were performed using a Scienta Omicron LT STM at a temperature of 77 K. (Certain commercial equipment, instruments, or materials are identified in this paper to foster understanding. Such identification does not imply recommendation or endorsement by the national institute of standards and technology, nor does it imply that the materials or equipment identified are necessarily the best available for the purpose.) All biases quoted in the main text are specified as sample bias by convention, though in our setup the physically applied bias is to the tip while the sample is grounded. All images were acquired in constant current mode with the current setpoint specified at 50 pA. All positive bias images (empty states) were acquired at 2 V sample bias, and all negative bias images (filled states) were acquired at -2 V.

**$PH_3$ Dosing Details.** Phosphine was dosed directly into the STM chamber via a leak valve. The line preceding the leak valve (the dosing line) was filled with $PH_3$ gas to a pressure of $4 \times 10^{-3}$ mbar ($4 \times 10^{-1}$ Pa). The leak valve was then opened to backfill the STM chamber to a maximum of $4 \times 10^{-8}$ mbar ($4 \times 10^{-6}$ Pa). At cryogenic temperatures, it is not possible to maintain this pressure without refilling the dosing line, whereas if the STM chamber is at room temperature a refill is not necessary, implying that maintaining a fixed pressure for a set amount of time is not an accurate method for determining dose at low temperature in our setup. As a means to estimate the dose, rather than refill the line we instead tracked the pressure drop. We applied a pressure drop of $1 \times 10^{-3}$ mbar in the dosing line which corresponds to a dose of 5.4 Langmuir ($4 \times 10^{-6}$ Pa for 3 minutes when the chamber is at room temperature). To achieve a saturation dose, we placed the sample on a wobble-stick in line of site to the leak valve. For the low dose, we left the sample in the STM, reducing exposure sufficiently that the same dosing line pressure drop resulted in a low dose; notably therefore, the saturation doses occurred on room temperature samples (on the wobble stick) while the low dose was performed on a sample at 77 K. After dosing, the saturation dosed samples required relocation of the adsorption sites relative to fiducial marks; this was done by using large area scans to first relocate STM fabricated relocation features that were 400 nm × 1000 nm and had a known position relative to the lithographic patch sites (see section S9 of the supporting information for more details).

**FCL and FCM Conditions.** The two STM tips used in this study required different implementations of FCL/FCM. For the first implementation, the tip apex was positioned at the site where lithography (center of dimer) or manipulation (above a $PH_2$ molecule for 3-dimer patches) was to be performed and the setpoint current was immediately set to 2 nA. The bias was then ramped at a rate of 0.05 V/s from 2 V to 3 V, with feedback on. When a change in tip height greater than 0.02 nm in magnitude was observed, the current and bias were immediately set back to imaging conditions. For the second implementation, we positioned the STM tip at a predefined offset from the desired lithographic site (center of dimer for both lithography and 1-dimer patch $PH_2$ manipulation), set the setpoint current to 2 nA and pulsed the sample bias to 3 V for 2 s. The STM was then set to imaging conditions and the tip apex was moved to sample two sites in succession: the location of the intended lithography/manipulation, and a second position for comparison. These sites were measured once before any bias pulses were performed in order to determine an initial height difference between the two. By comparing the relative heights of the two sites after pulsing, it was possible to detect whether any change had occurred. If the height difference changed by more than 0.02 nm, then the process was stopped, otherwise it was repeated. A visualization of this process is given in the supporting information, section S11.

**DFT Calculations.** As detailed in the main text, Figures 1b and 2a show the atomic setups used for simulated STM images and adsorption geometries, respectively. We used the Perdew-Burke-Ernzerhof generalized gradient approximation[28,29] for the exchange-correlation functional and a projector augmented wave basis with energy cutoff of 800 eV as implemented in GPAW.[30–32] Initial atomic coordinates for Si and H atoms in the slabs of both setups were taken from previously relaxed slabs of optimum size for the relaxation problem as described in detail in reference.[20] For subsequent atomic coordinate relaxation of adsorbates on the slab of Figure 2a, a force cutoff of 0.05 eV/Å was used. Both setups used a k-point mesh consisting of the Γ point only. Simulated STM images were generated by integrating the local density of states (LDOS) as calculated by DFT, following the Tersoff-Hamann approach for an s-wave tip.[33] It is also possible to consider alternative tip wavefunctions (*i.e.* linear combinations of s,p,d,f orbitals), however this was unnecessary as the s-wave simulated images already enabled identification of the adsorption configurations of this study from the simulated images. We assumed a tip height of 0.4 nm above the H-terminated surface at the lower-left corner of an STM scan and extracted the integrated LDOS at that position. Images were then determined as surfaces of constant integrated

LDOS using this value. Upon generating an image we additionally applied a Gaussian blur of $\sigma$ = 0.77 nm (cutoff radius = 2 $\sigma$) to mimic the blurring seen in the experimental images. It has been previously noted[34] that Kohn-Sham eigen-energies can differ (often linearly with a scale factor and offset) from those inferred from experimental bias voltages. With the exceptions of lone dangling bonds (see supporting information S3), and species specifically noted in Figures 3 and 4, we found good agreement using integration limits of 0 to 1.1 eV for positive bias, and –0.9 to 0 eV for negative bias. For the species of Figures 3 and 4 whose integration limits deviate, values are included as labels on the respective simulated images. As discussed in the supporting information, section S3, such shifts can be at least partially considered to be due to tip-induced charging (band banding).

**Acknowledgement**

This research was funded in part by the Department of Energy Advanced Manufacturing Office Award Number DE-EE0008311 and by a National Institute of Standards and Technology (NIST) Innovations in Measurement Science award, "Atom-Based Devices: Single Atom Transistors to Solid State Quantum Computing." This work was performed in part at the Center for Nanoscale Science and Technology NanoFab at NIST. We also acknowledge useful discussions with James Owen and John Randall of Zyvex Labs regarding strategies for tip enhanced incorporation.

**Supporting Information:** Additional computational details on generating adsorption configurations; the effect of supercell size on relaxation calculations; validating STM image simulation and P incorporation; examples of catalog use in practice; pinning versus averaging of buckled dimers; rescaled color scale for Figure 3 to improve dimer row visibility; the effectiveness of using negative bias height to determine adsorbed species; spurious dangling bonds from 1-dimer patch manipulation; relocating dimer patches; temperature calibration using onset of H desorption from H-Si (100); pulsed FCL/FCM where the lithographic site is displaced from the tip apex; full catalog of simulated STM images.

# Supporting Information:

# Enhanced Atomic Precision Fabrication by Adsorption of Phosphine into Engineered Dangling Bonds on H-Si Using STM and DFT

Jonathan Wyrick,[1*] Xiqiao Wang,[1,2] Pradeep Namboodiri,[1] Ranjit Vilas Kashid,[1] Fan Fei,[1,3] Joseph Fox,[1,3] Richard Silver[1]

[1]Atom Scale Device Group, National Institute of Standards and Technology, Gaithersburg, MD 20899, USA
[2]Joint Quantum Institute, University of Maryland, College Park, MD 20740, USA
[3]Department of Physics, University of Maryland, College Park, MD 20740, USA
[*]Corresponding Author

# Contents

# S1. Generating configurations.

We determine the set of all possible 3-dimer patch adsorption configurations by first considering the 3 possible species, the number of Si atoms they effectively occupy (e.g. bridge sites occupy 2 Si), and the maximum number of H atoms that as a result can occupy the remaining Si atoms in the 3-dimer patch. The maximum number of H atoms is determined by how many must have dissociated from a $PH_3$ molecule in order to create the adsorbed species. Table S1 shows the possibilities considered.

**Table S1.** Site occupation and H dissociation characteristics for each $PH_3$ derivative on the Si (100) surface. The "Si atoms occupied" column specifies how many Si atoms participate in bonding to the molecule, while the "Max. H adsorption" column specifies how many additional H atoms may appear on the surface as a result of dissociating from the original $PH_3$ molecule.

| Species | Si atoms occupied | Max. H adsorption |
|---|---|---|
| $PH_2$ | 1 | 1 |
| $PH$ | 2 | 2 |
| $P$ | 2 | 3 |

To further determine the set of possible configurations, we construct the set of all physically possible combinations of numbers of $PH_2$, $PH$, $P$, and H adsorbed into a patch together. The key requirement is that the total number of Si atoms occupied is not allowed to exceed 6, or equivalently the number of sites available for H adsorption after the number of $PH_2$, $PH$, and $P$ have been allotted must not be less than 0. This delineation of possibilities is shown in Table S2.

**Table S2.** Adsorption of multiple molecules into a 3-dimer patch. The first 3 columns are the number of each species adsorbed into the 3-dimer patch. "Top sites avail." is the number of Si atoms still available to be bonded to after adsorption of the phosphine derivative species. "Max. H adsorbed" is the number of H atoms that must be removed from phosphine upon adsorption (first number) and the number that can remain in the 3-dimer patch based on site availability (second number). The final column are the numbers of H atoms in the 3-dimer patch used to generate configurations.

| $PH_2$ | PH | P | Top sites avail. | Max. H adsorbed | H |
|---|---|---|---|---|---|
| 6 | 0 | 0 | 0 | 6 → 0 | 0 |
| 5 | 0 | 0 | 1 | 5 → 1 | 1,0 |
| 4 | 1 | 0 | 0 | 6 → 0 | 0 |
| 4 | 0 | 1 | 0 | 7 → 0 | 0 |
| 4 | 0 | 0 | 2 | 4 → 2 | 2,1,0 |
| 3 | 1 | 0 | 1 | 5 → 1 | 1,0 |
| 3 | 0 | 1 | 1 | 6 → 1 | 1,0 |
| 3 | 0 | 0 | 3 | 3 | 3,2,1 |
| 2 | 2 | 0 | 0 | 6 → 0 | 0 |
| 2 | 1 | 1 | 0 | 7 → 0 | 0 |
| 2 | 1 | 0 | 2 | 4 → 2 | 2,1,0 |
| 2 | 0 | 2 | 0 | 8 → 0 | 0 |
| 2 | 0 | 1 | 2 | 5 → 2 | 2,1,0 |
| 2 | 0 | 0 | 4 | 2 → 4 | 4,3,2,1,0 |
| 1 | 2 | 0 | 1 | 5 → 1 | 1,0 |
| 1 | 1 | 1 | 1 | 6 → 1 | 1,0 |
| 1 | 1 | 0 | 3 | 3 | 3,2,1,0 |
| 1 | 0 | 2 | 1 | 7 → 1 | 1,0 |
| 1 | 0 | 1 | 3 | 4 → 3 | 3,2,1,0 |
| 1 | 0 | 0 | 5 | 1 → 5 | 5,4,3,2,1,0 |
| 0 | 3 | 0 | 0 | 6 → 0 | 0 |
| 0 | 2 | 1 | 0 | 7 → 0 | 0 |
| 0 | 2 | 0 | 2 | 4 → 2 | 2,1,0 |
| 0 | 1 | 2 | 0 | 8 → 0 | 0 |
| 0 | 1 | 1 | 2 | 5 → 2 | 2,1,0 |
| 0 | 1 | 0 | 4 | 2 → 4 | 4,3,2,1,0 |
| 0 | 0 | 3 | 0 | 9 → 0 | 0 |
| 0 | 0 | 2 | 2 | 5 → 2 | 2,1,0 |
| 0 | 0 | 1 | 4 | 3 → 4 | 4,3,2,1,0 |

Using the numbers of $PH_2$, PH, P, and H specified in Table S2, we construct all possible configurations. As a simple example, in the case of ($PH_2$, PH, P, H) = (1,0,0,0), there is only one molecule to be placed in the 3-dimer patch, a single $PH_2$. We generate a list of possible configurations where the $PH_2$ is placed at each of the 6 top sites. The remaining 5 top sites in each case are labeled with "up" or

"down" to indicate dimer buckling, except for the Si atom across from the PH$_2$ which is labeled with an empty string, "", to indicate that dimer has flattened (see Figure 2h of the main text).  We also check the resultant list of configurations against itself and remove any configurations that are equivalent upon reflection across either the dimer row center axis, the perpendicular axis that runs through the two central Si atoms of the patch, or a combination of these two; notably the presence of up/down designations becomes important for low dose configurations where two cases that would otherwise be equivalent can be different from one another because of the buckled state of an empty dimer.  When placing more than one species into the 3-dimer patch, we additionally apply the occupation rules illustrated in Figure S1.  These rules ensure that each Si atom can participate in only 1 additional bond beyond its default surface configuration.  The full catalog is shown in section S13.

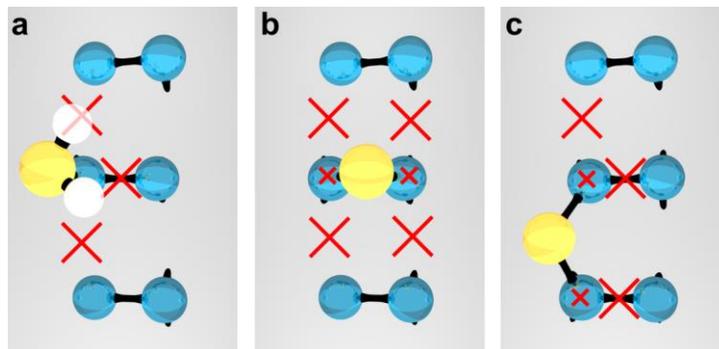

**Figure S1.**  Disallowed neighboring sites for subsequent adsorbates.  Given an adsorbate placed at (a) a top site, (b) a bridge site on the center axis of the dimer row, or (c) a bridge site on the side of a dimer row, the large red X's represent disallowed bridge sites while the small red x's represent disallowed top sites.

## S2. Effect of supercell size on relaxation calculations.

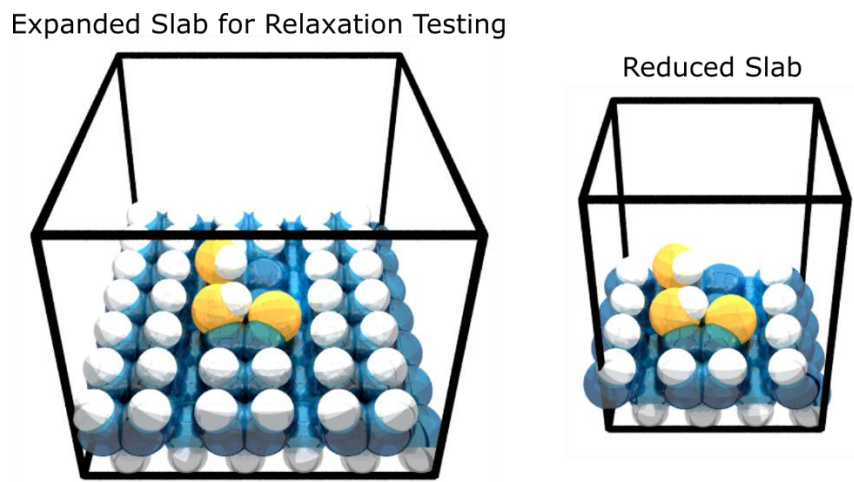

**Figure S2.** Expanded and reduced slabs for atomic coordinate relaxation testing.

As noted in the main text (see Figure 2a), a smaller slab and supercell were used for DFT calculations in which the atomic coordinates of adsorbates and surface atoms were relaxed. In doing so, there is a risk that adsorbates could potentially interact with their periodic repeats (due to periodic boundary conditions) in ways that result in incorrect minimum energy structures. Since surface atoms and adsorbates would then be transferred to the larger supercell for more accurate electronic structure determination and ultimately STM image simulation, the primary concern for such errors is that the atomic coordinates determined from the smaller supercell calculation would be sufficiently wrong to generate qualitatively different simulated STM images. To check for this possibility, we chose a representative configuration and relaxed the relevant subset of atoms on a larger slab/supercell combination (left of Figure S2). When compared to relaxation results from the analogous reduced slab (right of Figure S2), we found that the coordinates of each relaxed atom agreed between the two to within less than 0.1 Å. Based on this finding, we infer that the simulated STM images are not significantly altered by the choice of a smaller slab and supercell for relaxation.

# S3. Validating STM image simulation and P incorporation.

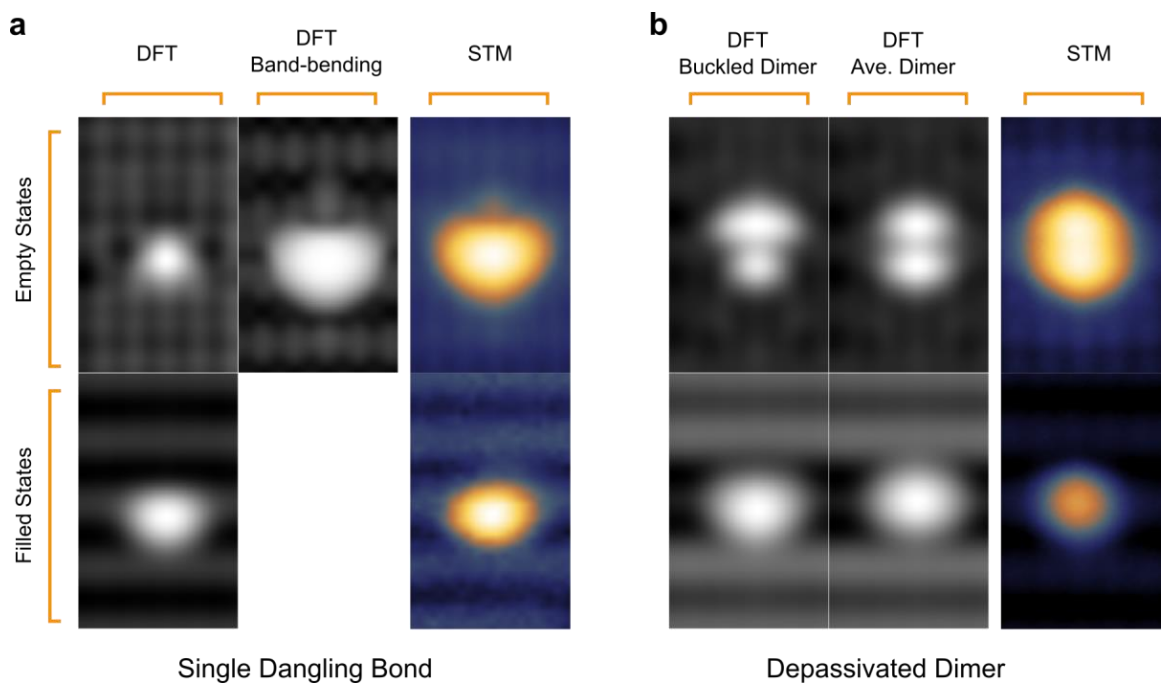

**Figure S3.** (a) Simulation of a single dangling bond, imaged at +2 V (empty states) and -2 V (filled states), including variation in appearance depending on the absence (left) or presence (right) of tip-induced band bending. (b) Simulation of a single depassivated dimer, buckled (left) and as the average of the two buckled states (right).

One way to ensure that the DFT calculations are generating meaningful simulated STM images is to compare to known defects on the H-Si (100) surface. Since they are the building blocks of the HDL technique, dangling bonds and depassivated dimers are of particular interest. Figure S3a shows the example of a single dangling bond where we find it is necessary to include some filled states in the positive bias image in order to get agreement with experimentally acquired +2 V STM images. At lower bias (+1.3 V) a single dangling bond appears more like the leftmost simulated image,[1] supporting the idea that this apparent difference between calculation and experiment is related to tip-induced band bending.

Figure S3b shows the calculated STM images for a fully depassivated dimer compared to experiment. As noted in the main text (see Figure 2g,h), the two Si atoms of a depassivated dimer will experience buckling. When imaging buckled dimers, if they are not pinned by a nearby defect (see section S5 for an example), they will either switch back and forth faster than image acquisition producing an image that is the average of the two buckled configurations, or they will take on a configuration of minimum current flow between tip and sample producing an image that is the minimum of the two buckled configurations.[2] In the case of Figure S3b, the former occurs.

Figure S4a shows the results of additional STM image simulations of common defects found on H-Si (100). The chosen defects and their respective models are based on recent work by Croshaw et al.[1] including both STM and qPlus AFM images of these common defects. More types of defects are presented in that work, however the subset of defects selected here are those that were amenable to the existing calculation setup used to generate the full $PH_x$ adsorption catalog. In Figure S4 we adopt the same naming conventions as used in ref.[1] For each image we find good agreement between DFT simulation and experiment (several positive bias examples are highlighted in Figure S4c).

Figure S4b,c shows evidence for ejected Si island formation after incorporation which serves as validation that the incorporation anneal was successful. Neglecting height variation from step edges, there are 3 primary heights apparent in Figure S4a corresponding to H-Si (100), the depassivation region (i.e. Si (100)), and ejected Si islands, respectively. This picture can be simplified by performing an additional re-passivation step as was done for Figure S4b. In this case it is clear that the direction of island growth is perpendicular to the dimer rows, consistent with what is observed with direct epitaxial Si overgrowth.

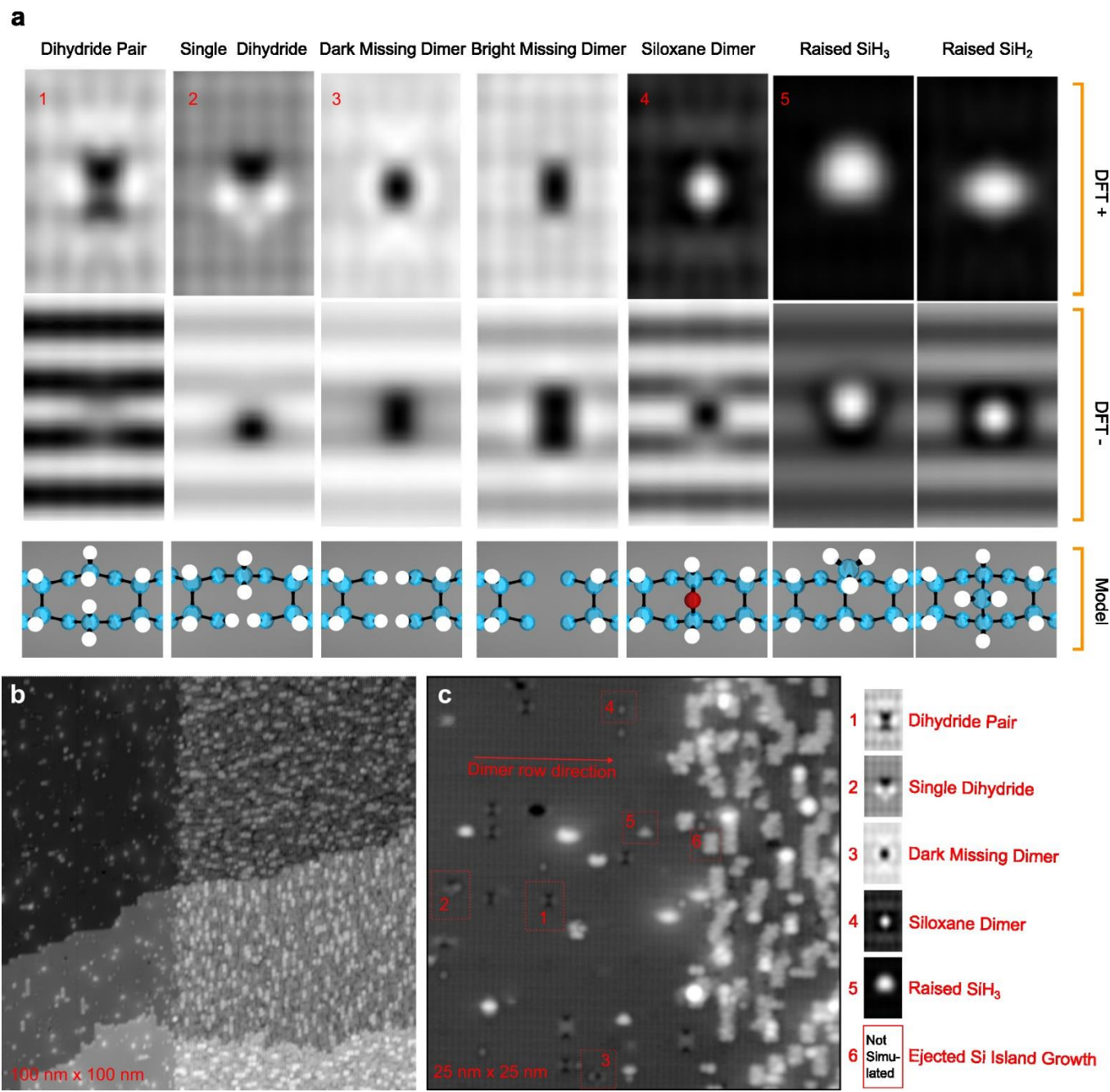

**Figure S4.** (a) Simulated STM images of common defects found on H-Si (100). (b) Formation of ejected Si islands (right side of image) as seen after incorporation. (c) Ejected Si islands (right) and common defects after incorporation and re-passivation. The key at right includes simulated images for the identified defect structures. No simulated image is given for island growth since the Si islands shown have greater extent than the supercell used in DFT calculations.

# S4. Catalog use in practice.

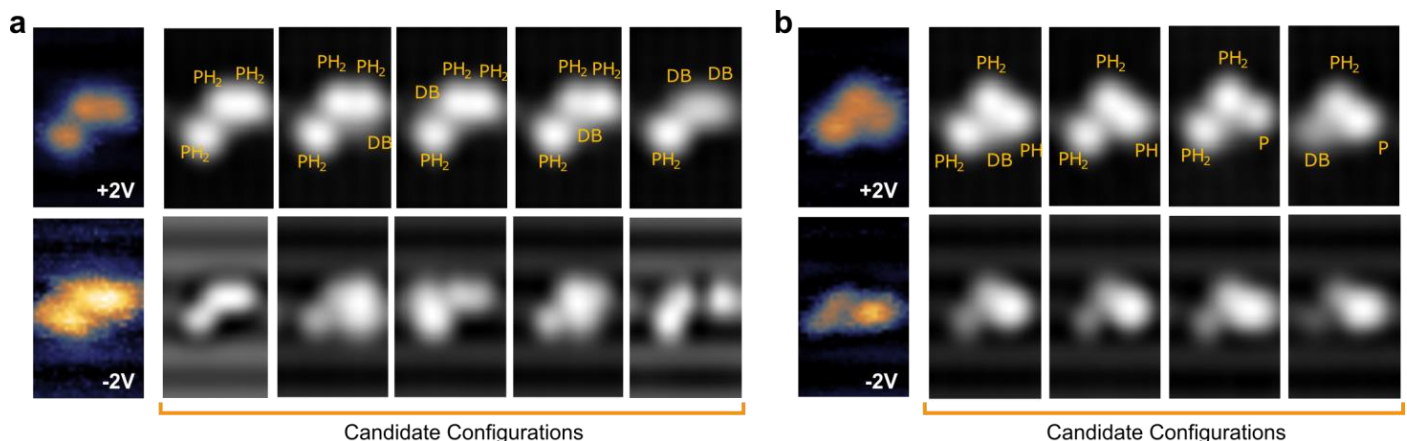

**Figure S5.** Identifying adsorption configurations using the catalog. Candidate configurations are those for which the positive bias simulated image appears to match experiment. Further narrowing down of possible matches can be done by comparing the negative bias images of the candidates.

There are many simulated images in the full catalog (see section S13) that appear similar to one another, complicating the process of making a precise identification. As currently implemented, comparisons are done manually by the operator, where similarity between simulated and experimental images is judged based on the number of lobes observed, their geometric arrangement, and their relative brightness when compared to one another. A valuable future research direction would be in the development of automated algorithms that can eliminate the risk of human subjectivity in making identifications.

In cases like the example shown in Figure S5a, there may be multiple matches at positive bias but only one negative bias case agrees with the geometry seen experimentally: here, the first candidate configuration appears to be the most reasonable choice. This identification is complicated by the fact that in a number of cases the tunnel junction was unstable while imaging phosphine derivatives at negative bias, resulting in significant smearing along the fast scan direction. Despite this difficulty, none of the other candidate configurations of S5a present geometries that could reasonably be linked to what was seen experimentally.

For the case of Figure S5b, identification is not so straightforward as both positive and negative bias show multiple matches. We attribute the first candidate configuration as being the correct one, though this identification is not certain. Notably, even with the reduced certainty of S5b, the set of possible candidates has been reduced down to 5 (the key uncertainty is whether the 3$^{rd}$ species is PH or P). Given the increased brightness of the right-most lobe at positive bias in S5b which is not seen in the experimental image (in fact, the leftmost lobe appears brightest in STM), we can likely also rule out the 4$^{th}$ configuration. All configurations shown are stable adsorption structures, meaning that each configuration is a physically possible outcome from dosing. We cannot however compare relative binding energies as a means to determine which candidate is correct because the difference between each case of Figure S5b is related to the total number of H atoms in each structure; the patch/molecule system is not a closed system during dosing because H atoms may be able to leave from or arrive into the patch from elsewhere. Several of the observed adsorption structures in Figure 3 of the main text are only possible because the number of H atoms is not conserved (i.e. the number of H atoms in the patch after adsorption is different from the number of H atoms that desorbed from the original PH$_3$, in some cases more and in others less).

## S5. Pinning vs. averaging of buckled dimers.

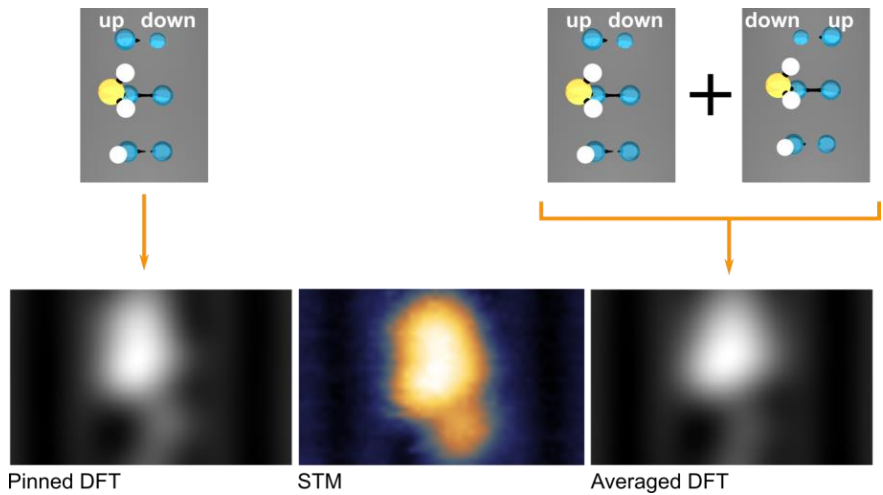

**Figure S6.** Comparison of simulated STM images for pinned and thermally averaged dimer buckling to the experimental STM image. Configurations used to generate each simulated image are shown above with "up" and "down" labels that denote the dimer buckling state corresponding to the same designations in Figure 2g of the main text. All images are at positive bias.

In the main text it was noted that the configuration of Figure 3n likely exhibits a pinned dimer (whereas the other low dose configurations do not). This determination was made by comparing the "pinned DFT" image to the "averaged DFT" image in Figure S6. We note that the upper lobe of the pinned image is more elongated in the vertical direction, matching more closely with what has been seen experimentally (the center image).

## S6. Rescaled color scale for Figure 3 to improve dimer row visibility.

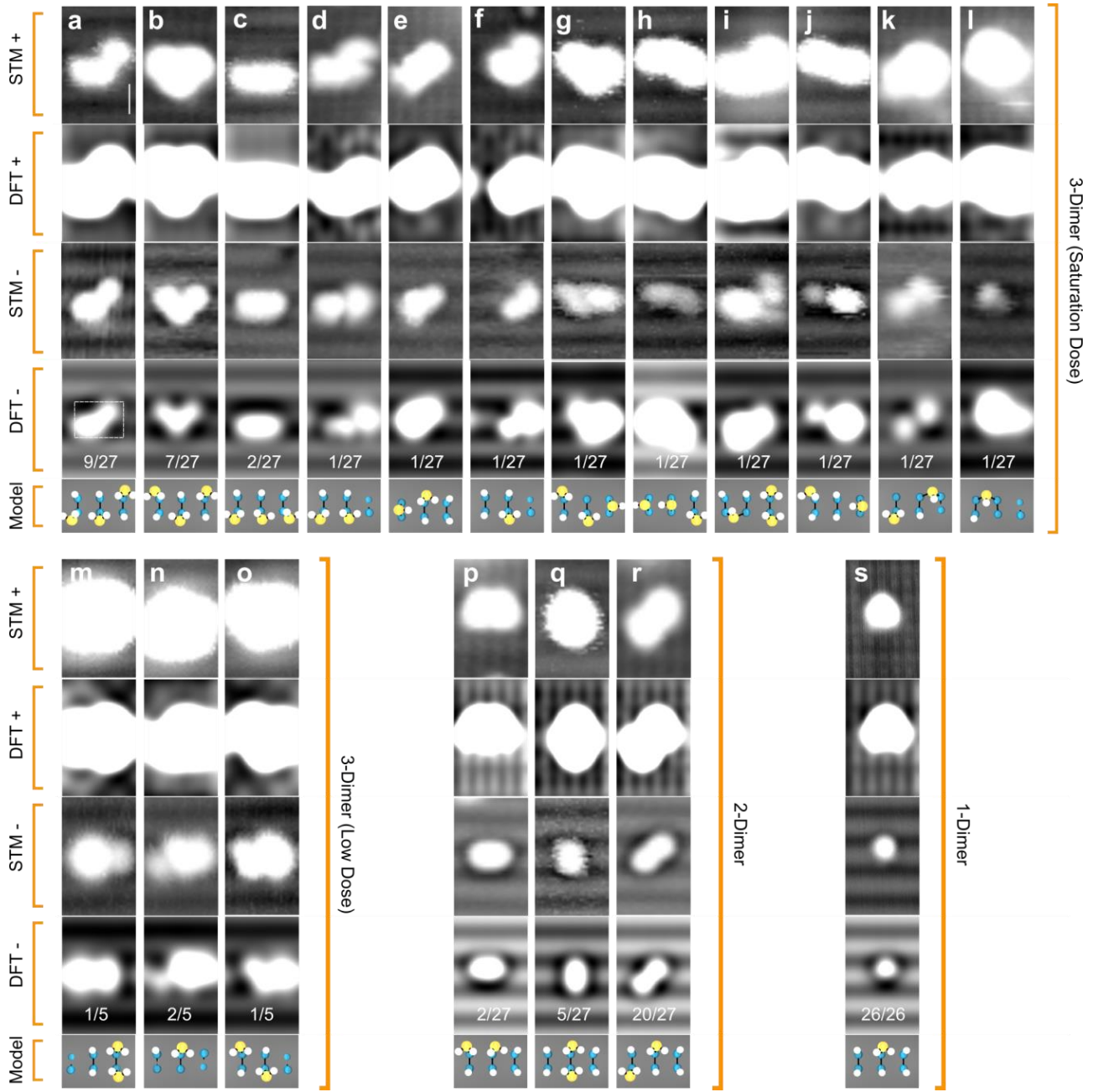

**Figure S7.** Rescaled Figure 3 to make dimer rows more visible.

Figure S7 shows the same experimental and simulated images as Figure 3 of the main text, but with a rescaled color scale so that dimer rows are visible. With this version of the figure, any alterations in the appearance of the surrounding Si-H due to attachment of the $PH_x$ species is apparent. These variations may also be useful in identifying the adsorbed species.

## S7. Using negative bias height to determine species.

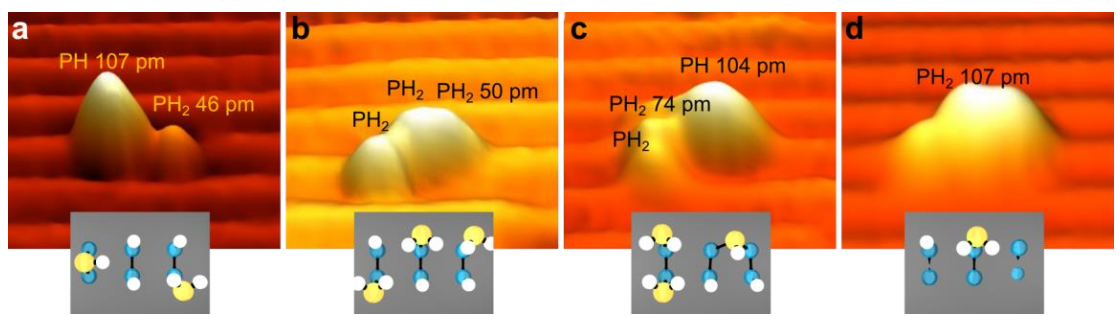

**Figure S8.** Comparison of apparent heights from selected configurations: (a) PH and $PH_2$, (b) 3 $PH_2$, (c) 2 $PH_2$ and 1 PH, and (d) 1 $PH_2$ and 4 dangling bonds. Labels are placed close to their respective species in the angled topographic view of the STM images, specifying the molecule type and its apparent height. For images with multiple $PH_2$'s of the same height, only one height is specified. Insets below each image show the corresponding adsorption configuration.

As a general rule of thumb, the apparent height of $PH_2$ should be less than that of PH which should be less than that of P. These expected height variations are based on the observation that H atoms on the molecules tend to cause a reduction of apparent height when imaged by STM (e.g. a dangling bond looks taller/brighter than a H terminated Si atom). When attempting identifications where a molecule is fully surrounded by H atoms (i.e. there are no neighboring molecules or dangling bonds), the apparent height should be sufficient to distinguish the adsorbed species. However, as shown in Figure S8, once there are neighboring species (including dangling bonds), there is overlap in the range of apparent heights measured for the different species. In the set of images shown, $PH_2$ ranges from 46 pm in height to 107 pm; this overlaps with the range seen for PH which varies from 104 pm to 107 pm, making it impossible to distinguish the two species by height alone.

## S8. Spurious dangling bonds from 1-dimer patch manipulation.

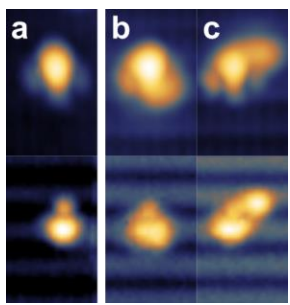

**Figure S9.** Appearance of lone P atom without (a) and with (b,c) nearby spurious dangling bonds created during tip-base manipulation. Top images are taken at positive sample bias while those on the bottom are the corresponding negative bias image.

When using FCM to induce H dissociation from $PH_2$, it is often the case that additional dangling bonds can result close to the final lone P, as can be seen in the examples of Figure S9b,c. The isolated P atom has several distinctive features that make it relatively easy to identify (Figure S9a): at positive bias there are 3 lobes, with the central one being the brightest and pointing in the "downward" direction, while at negative bias there are 2 lobes consisting of a smaller one on top and a larger one on the bottom. Even when there are dangling bonds nearby as in Figure S9b,c, the lobe structure of the isolated P atom is clear.

## S9. Relocating dimer patches.

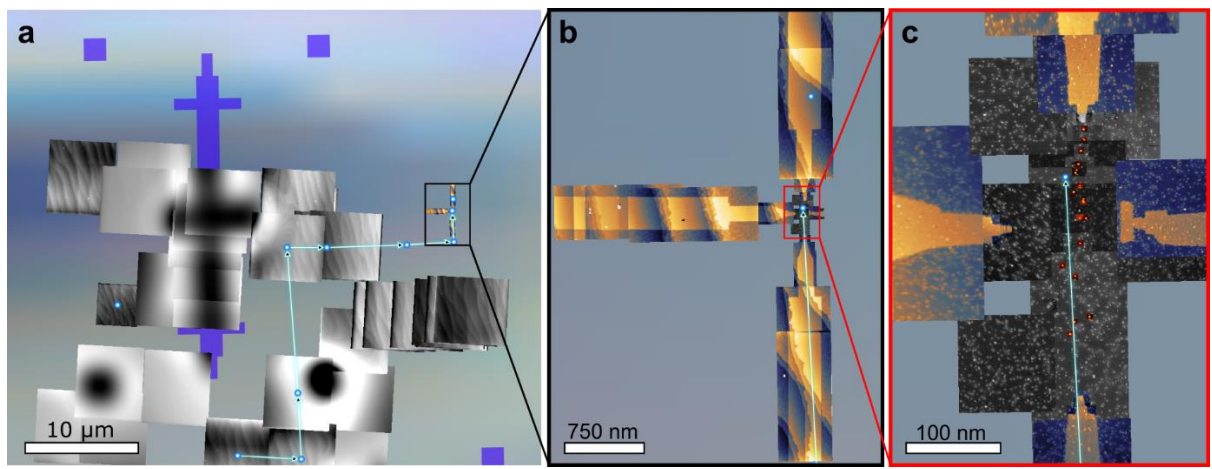

**Figure S10.** Relocating 3-dimer patches. (a) Micron scale overlay of large range STM scans (grayscale) on fiducials (the file used to write the fiducials is displayed as purple marks). One example coarse walk path taken is shown in blue. (b) Zoom-in showing large scale pads written using HDL to aid in relocating the patches. (c) Ends of large scale pads pointing towards the sites where patches (orange) were written. Patches in this example include test structures used for FCL calibration, 3-dimer patches, and 2-dimer patches.

In our STM setup, procedures such as dosing and annealing (for incorporation) require removal of the sample from the STM stage. This means that after each of these steps, once the sample is remounted on the STM stage, the STM tip will no longer be positioned as it was previously: the location of the 3-dimer patches is generally shifted by 10s to 100s of microns relative to the STM tip. Our Si chips have etched fiducial marks which can be seen optically while approaching the tip and are also safe for scanning as can be seen in Figure S10a.

As a first step before fabricating any patches, we land on a fiducial mark and determine the scan region's location relative to the fiducial. We coarse walk the scanner away from the fiducial while recording the number of steps traveled. Because we have calibrated our coarse motion, the recorded number of steps is a good estimate for the distance away from the fiducial that the FCL patches will be fabricated. We then write and image a series of patches, followed by the use of HDL to write large pad features approximately 400 nm × 1000 nm (Figure S10b). These large scale features point towards the location of the patches as shown in Figure S10c.

When relocating the patches after dosing or annealing, we land on a known fiducial, then coarse walk to the recorded location of the patches. Using large area scans, we can see the pads that were written previously. To aid in this process we typically acquire both a constant current topography image as well as a dI/dV map. In some cases (particularly after incorporation), it may be difficult to see the pad in topography while dI/dV still shows good contrast. Once a large-scale pad has been relocated, we switch to small image sizes (e.g. 100 nm × 100 nm or less).

# S10. Temperature calibration using onset of H desorption from H-Si (100).

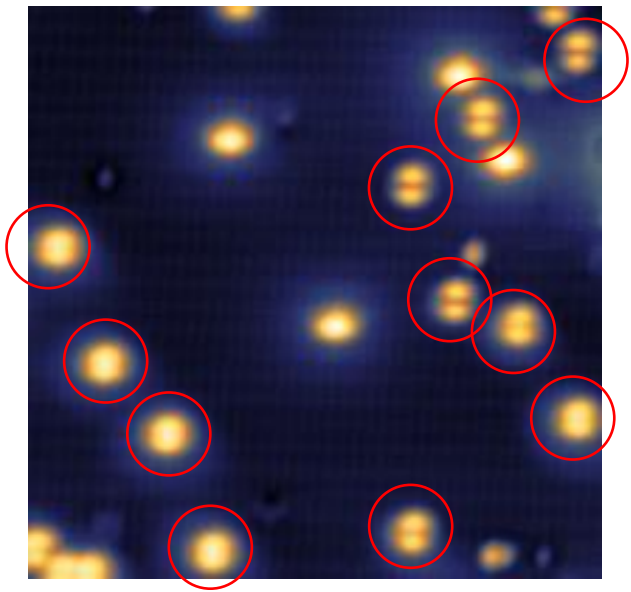

**Figure S11.** Positive bias image of H-Si (100) surface after a 2 minute anneal at a nominal pyrometer reading of 285°C. Red circles highlight fully depassivated dimers as evidence for the onset of H desorption. Samples processed at lower temperatures do not show significant quantities of fully depassivated dimers.

To ensure the work described here is reproducible, it is important to accurately report relevant process temperatures. In this case, those are the temperatures for H passivation and for P incorporation. Our reported passivation and incorporation temperatures are corrected pyrometer readings based on the known temperature for onset of H desorption. Temperature programmed desorption experiments on H-Si (100)[3] have found that H desorption occurs at temperatures ≳ 377 °C. On our system the onset of H desorption (see Figure S11) is at a nominal pyrometer reading of 285°C, while our H termination and incorporation processes are carried out at a nominal reading of 275°C. Given the proximity in temperature between the desorption onset and our process temperatures (10 degrees difference), we apply a simple shift to the known desorption temperature to arrive at our corrected process temperature of 367 °C. We also suggest that the preferred method for reproduction of these processes in a different system should therefore be to first determine experimentally the onset of H desorption and then set process temperatures relative to it.

# S11. Pulsed FCL/FCM where the lithographic site is displaced from the tip apex.

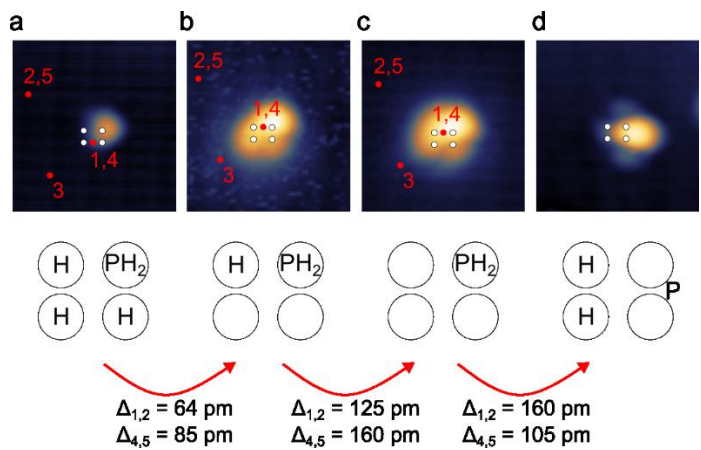

**Figure S12.** Illustration of FCL and FCM to change a PH$_2$ molecule into a lone P atom overlaid with the tip positioning and measurement/pulse sequence used for a case where the site at which lithography/manipulation occurs is displaced from the tip's tunneling apex. (a) A PH$_2$ molecule is surrounded by hydrogen atoms (the 4 white circles overlaid on the STM image indicate the 4 most relevant top sites on the vertically running dimer row). (b) A PH$_2$ molecule adjacent to a fully depassivated dimer. (c) A PH$_2$ molecule where H has been removed both from the adjacent dimer as well as from its neighboring Si atom on the same dimer. (d) A lone P atom after the 2 H atoms from the PH$_2$ have been manipulated to neighboring unoccupied Si atoms. The 4 overlaid white circles of each STM image correspond to the 4 circles below each image where occupying species are labeled. The red dots on each STM image and corresponding numbers indicate the sequence of tip positions used during lithography and manipulation.

Figure S12 shows a series of positive bias STM images that comprise the STM manipulation step described in Figure 4a of the main text. Also shown in this series of images is the sequence of tip positions (red dots and numbers) used for FCL/FCM for a tip that consistently removes H atoms at a fixed displacement away from its tunneling apex. This displacement is given by the separation between site 3 (the tip apex position) and sites 1,4 (where lithography/manipulation occurs); this separation is approximately 1.3 nm in the present work. The sequence corresponding to the red numbers is as follows:

1) Measure tip height at site 1

2) Measure tip height at site 2

3) Pulse bias at site 3

4) Measure tip height at site 4 (same as 1)

5) Measure tip height at site 5 (same as 2)

The difference between heights measured in steps 1 and 2 is given by $\Delta_{1,2}$ and the difference between steps 4 and 5 is given by $\Delta_{4,5}$ in Figure S12. If the difference between $\Delta_{4,5}$ and $\Delta_{1,2}$ is less than 20 pm, then the bias pulse is repeated until there is a change, otherwise the area is imaged to verify that the intended lithography or manipulation has occurred.

## S12. References.

## S13. Full catalog of simulated images.

See figure caption on the final page for a description of the layout of the following simulated images:

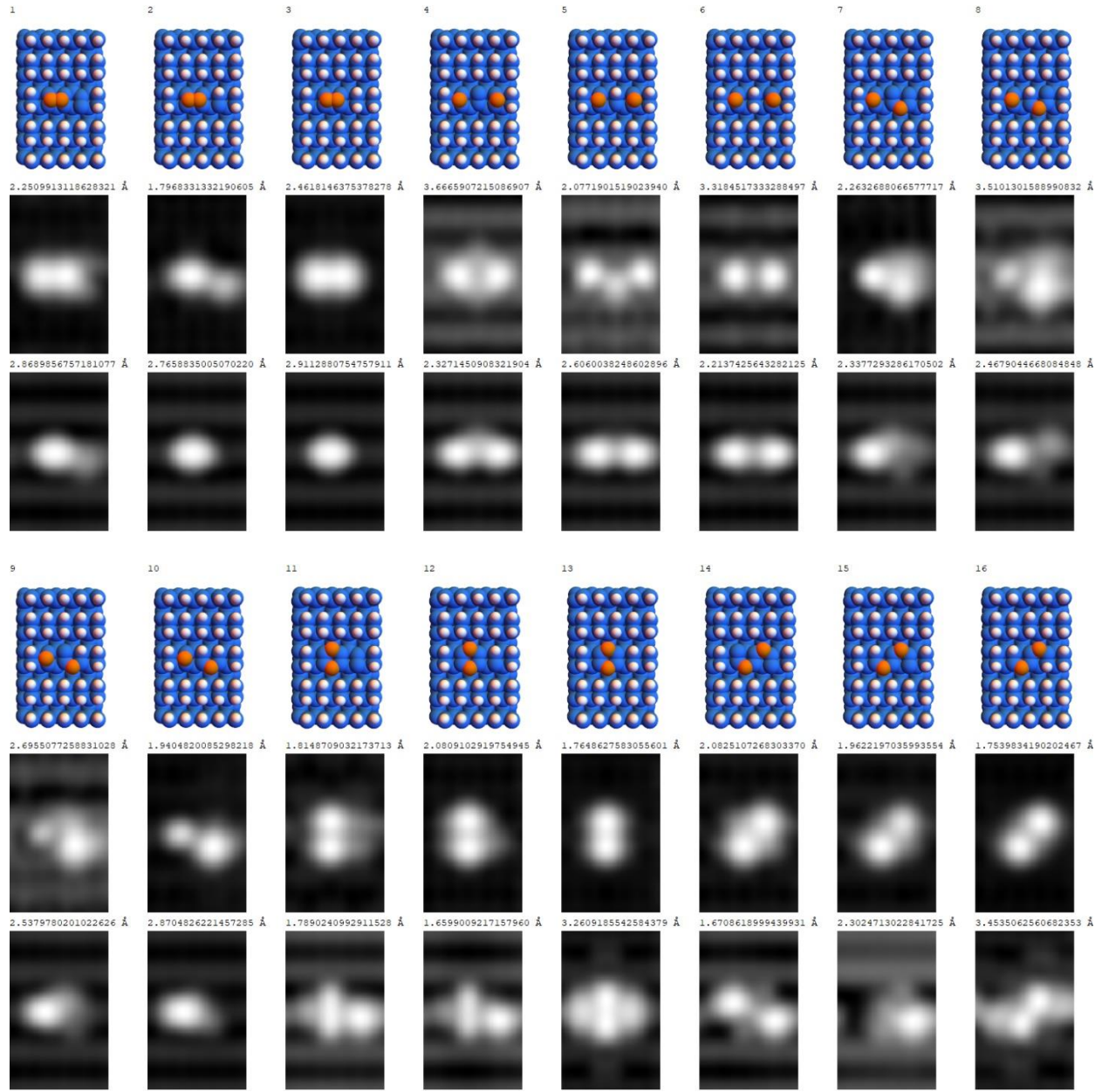

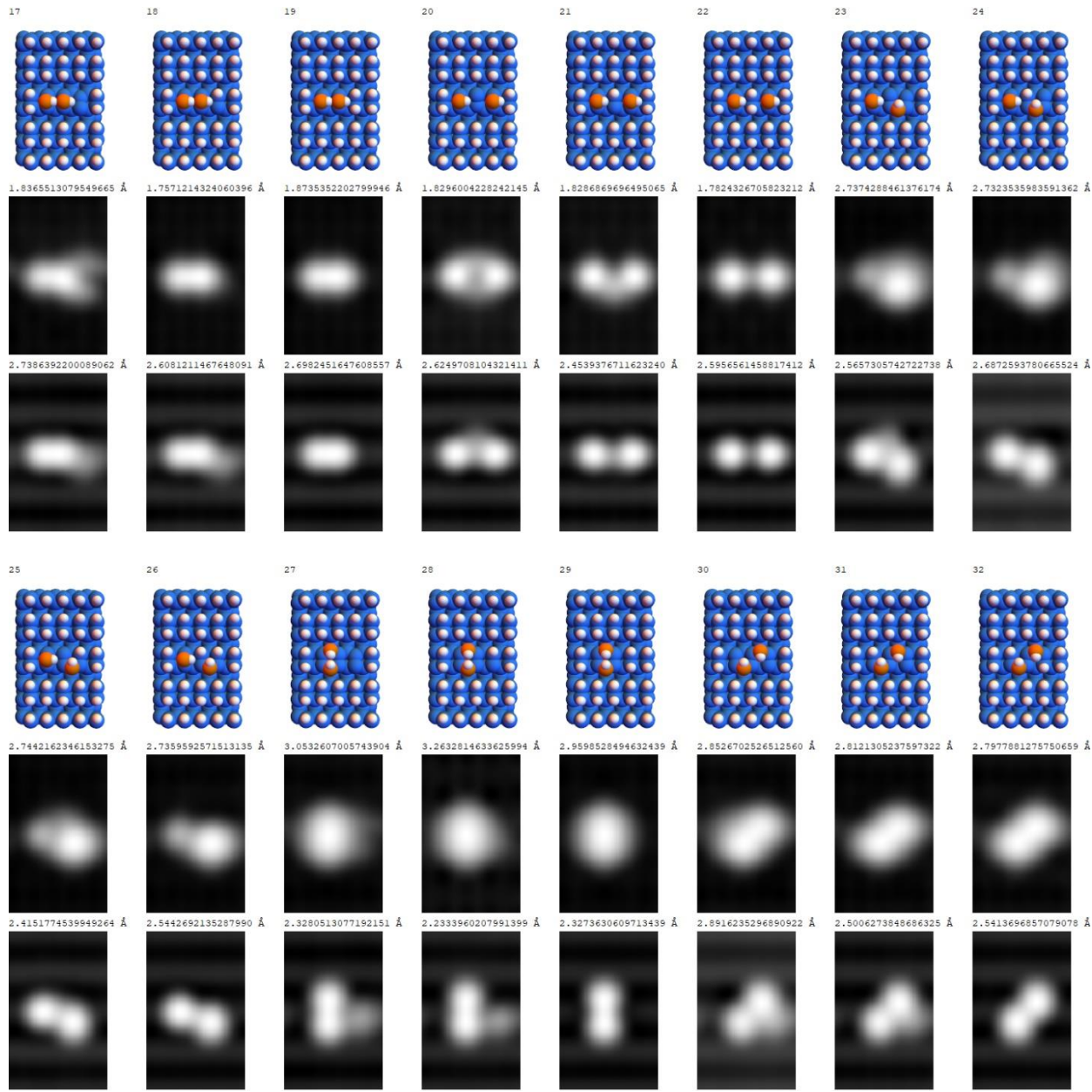

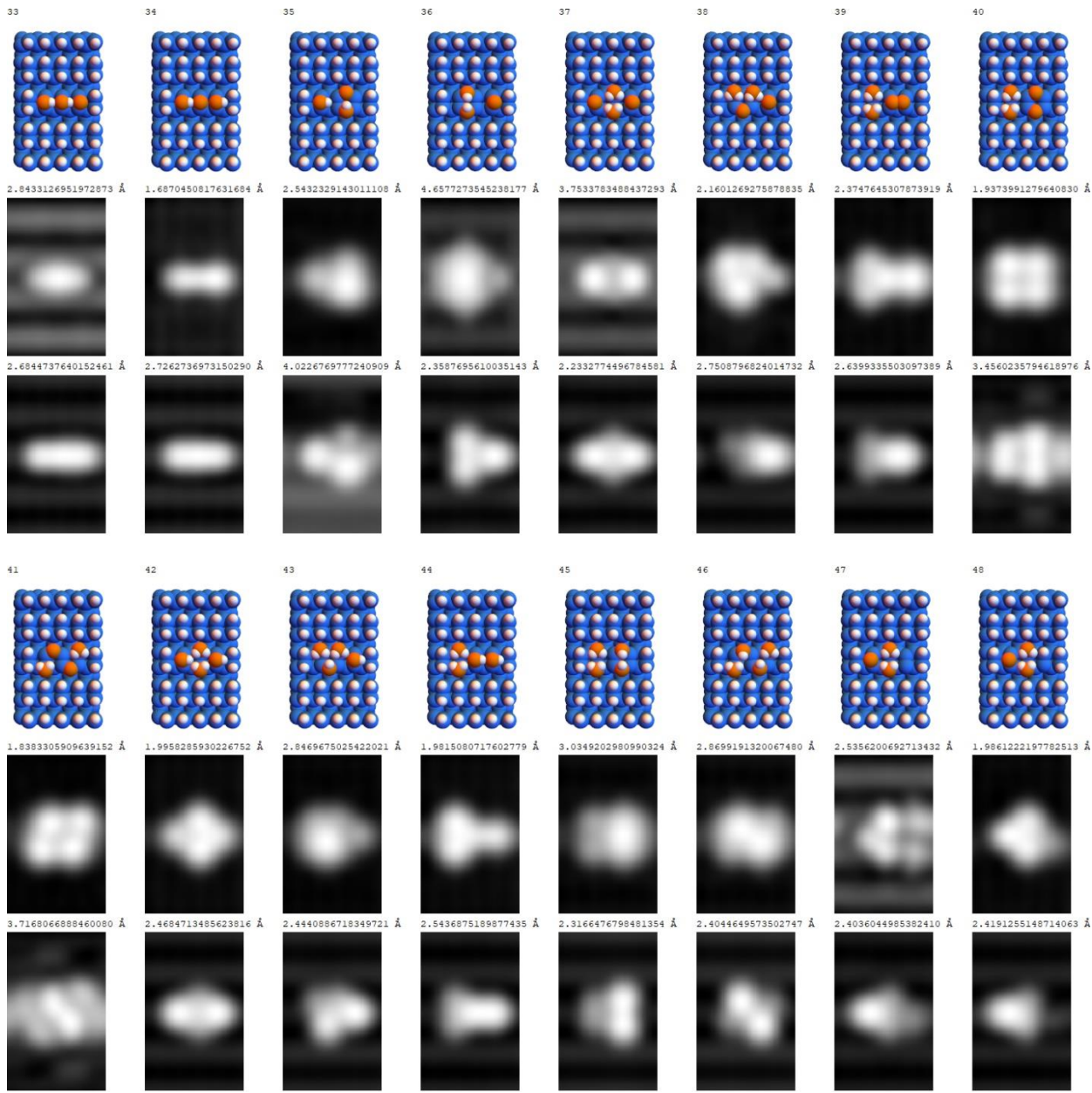

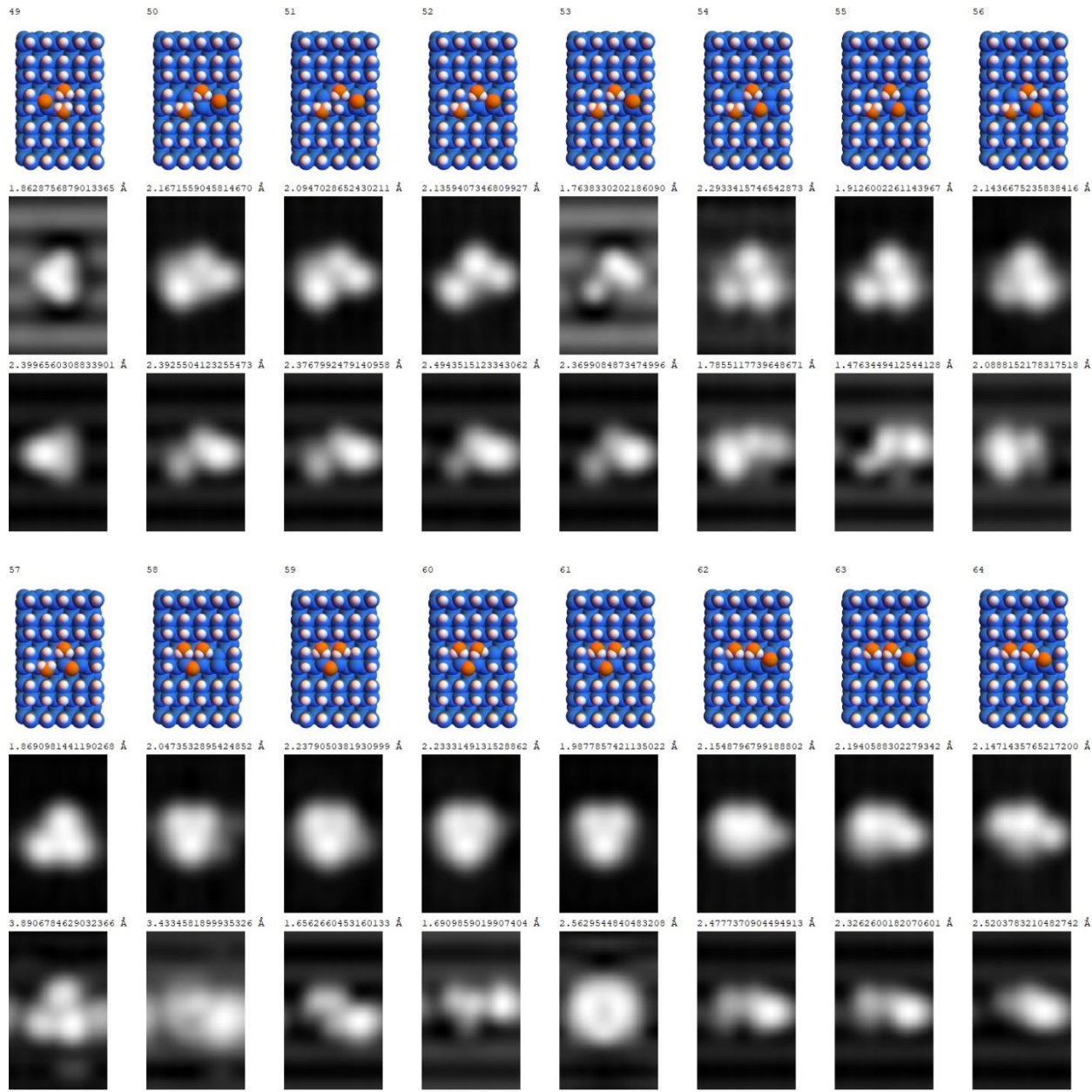

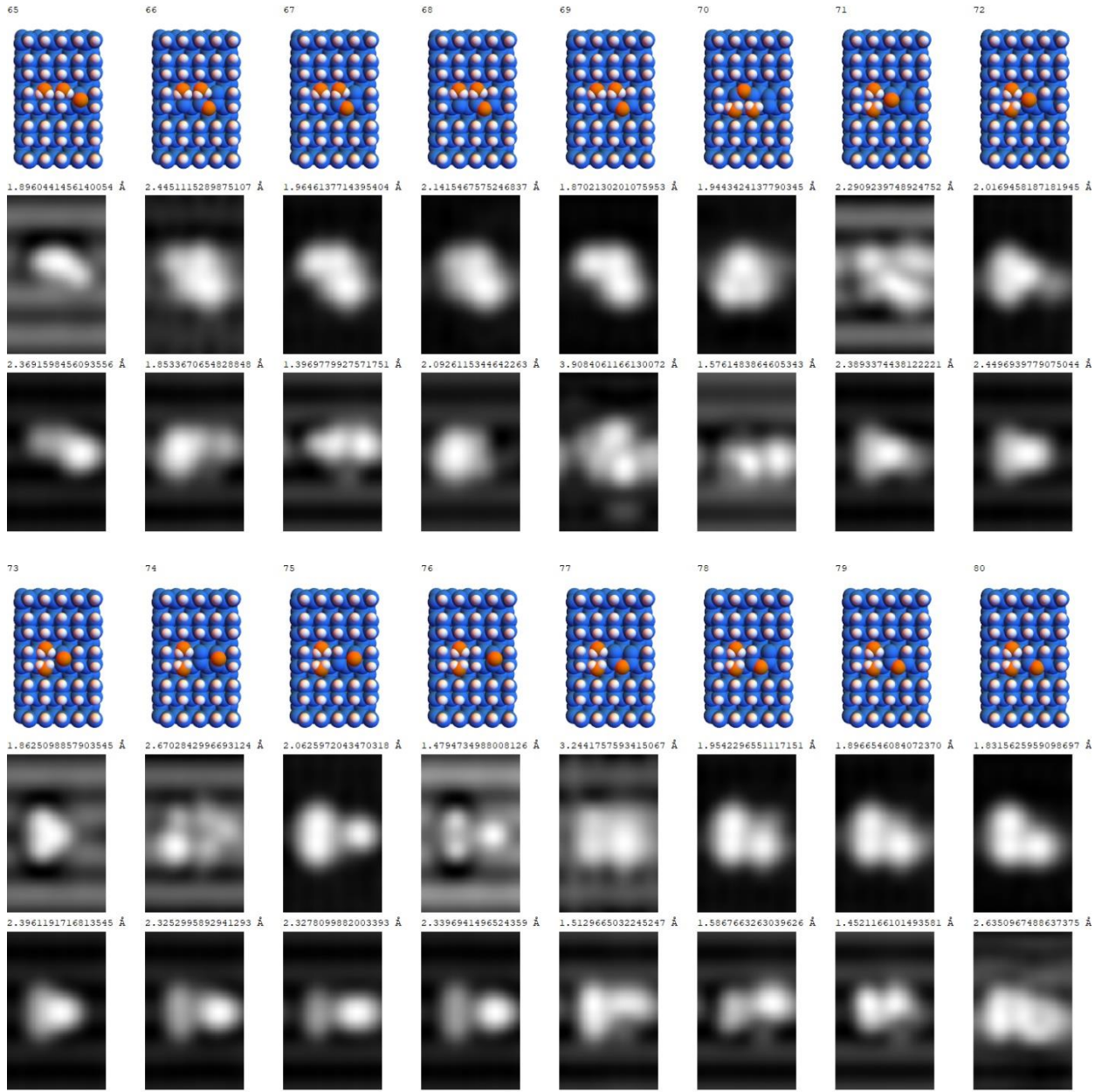

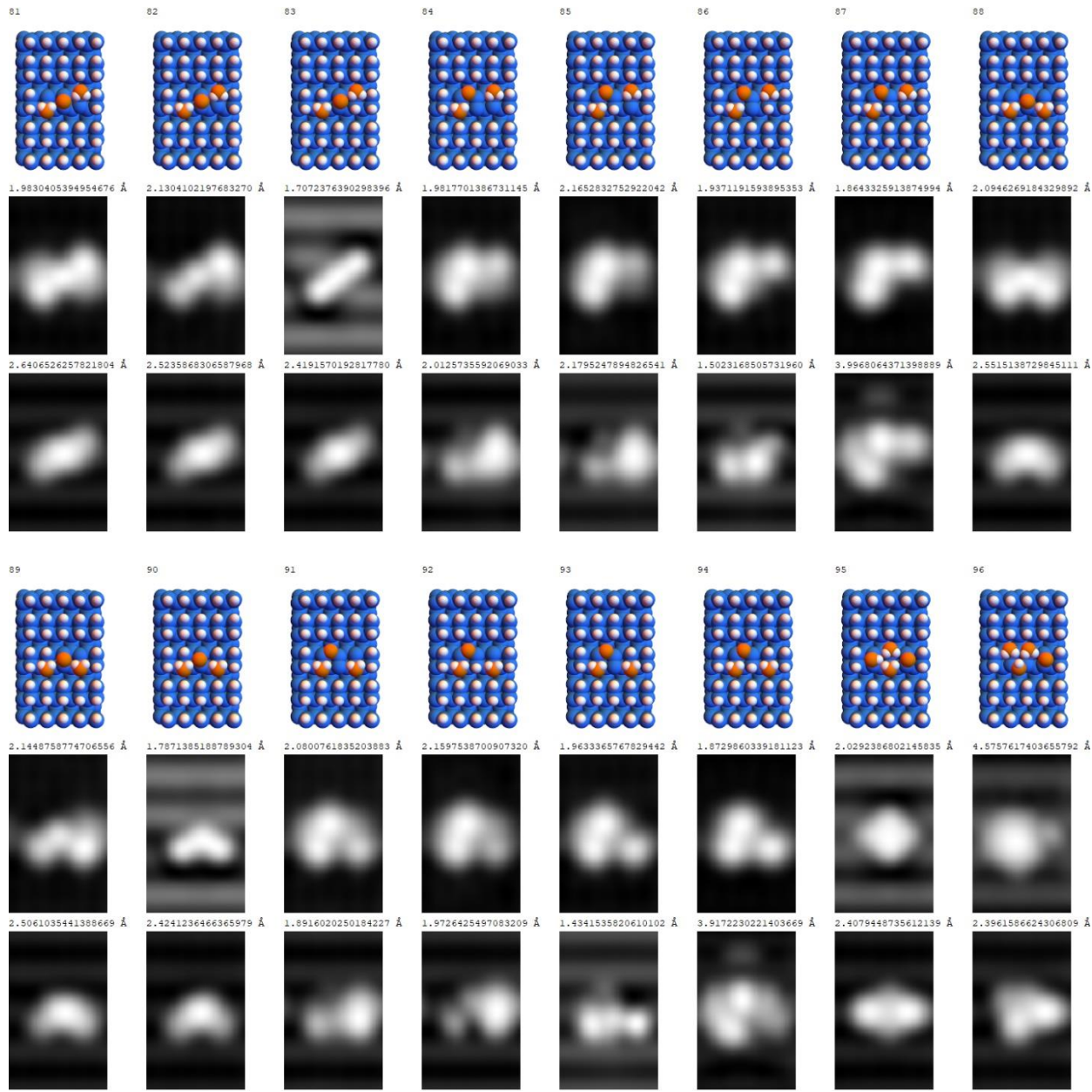

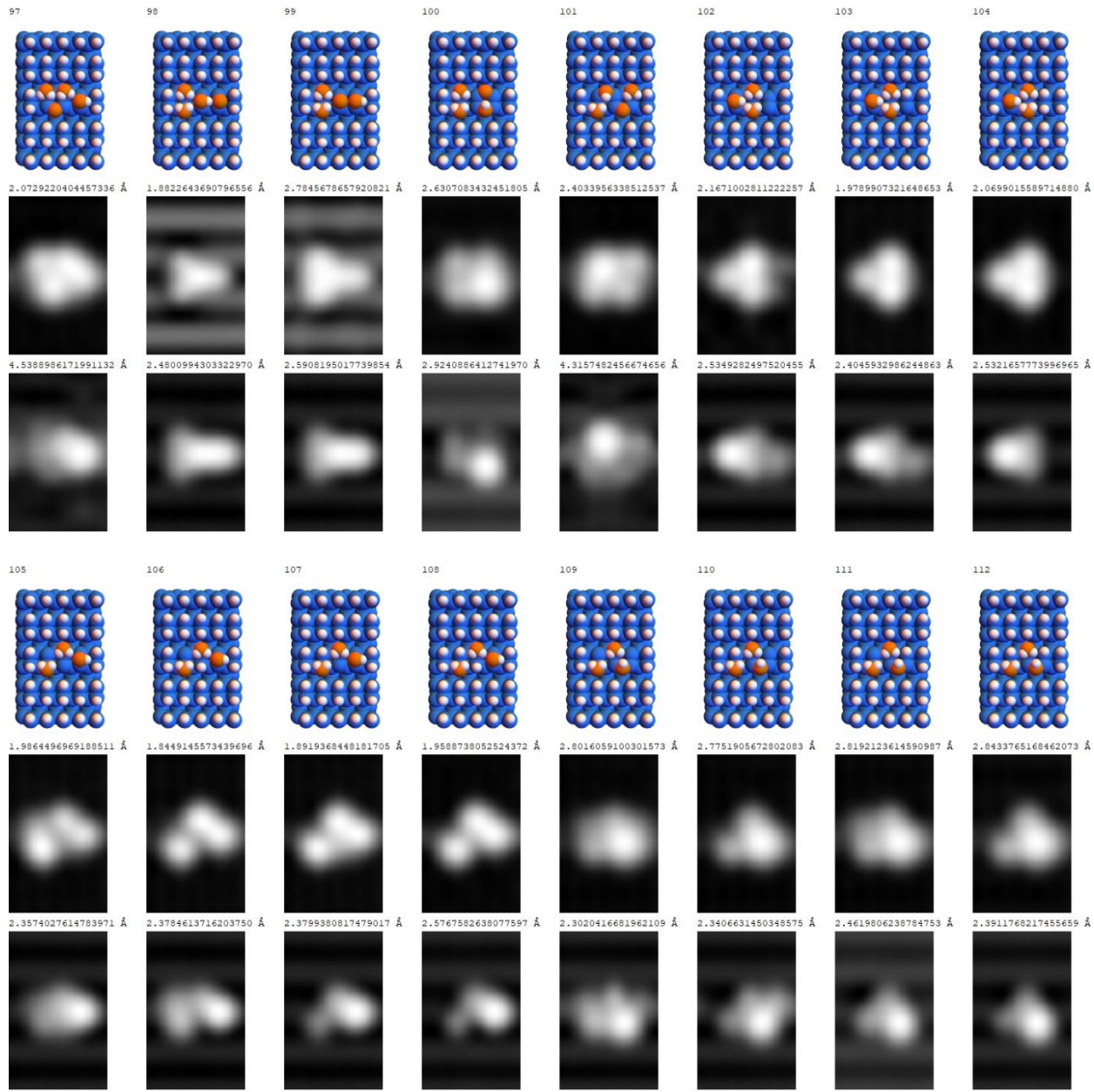

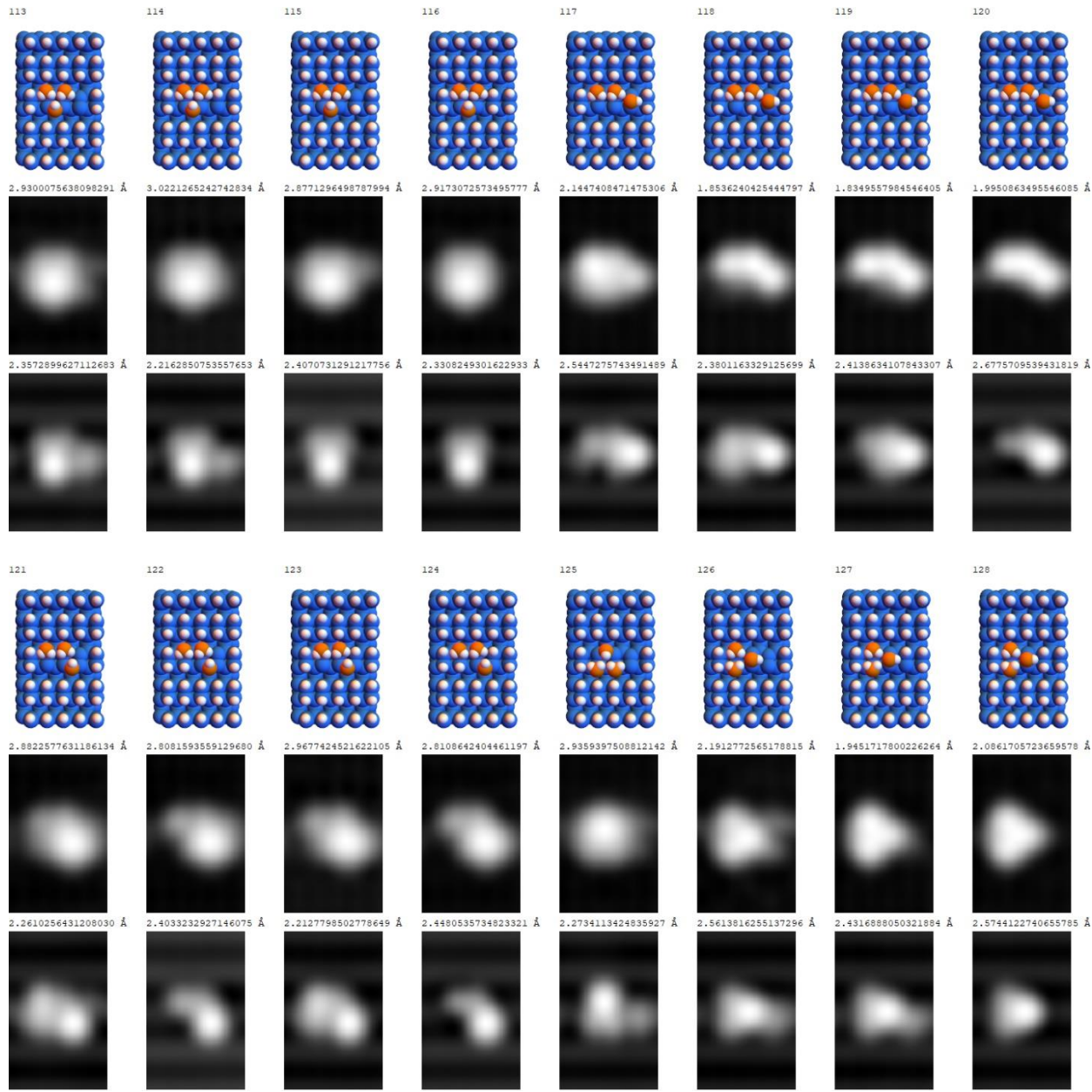

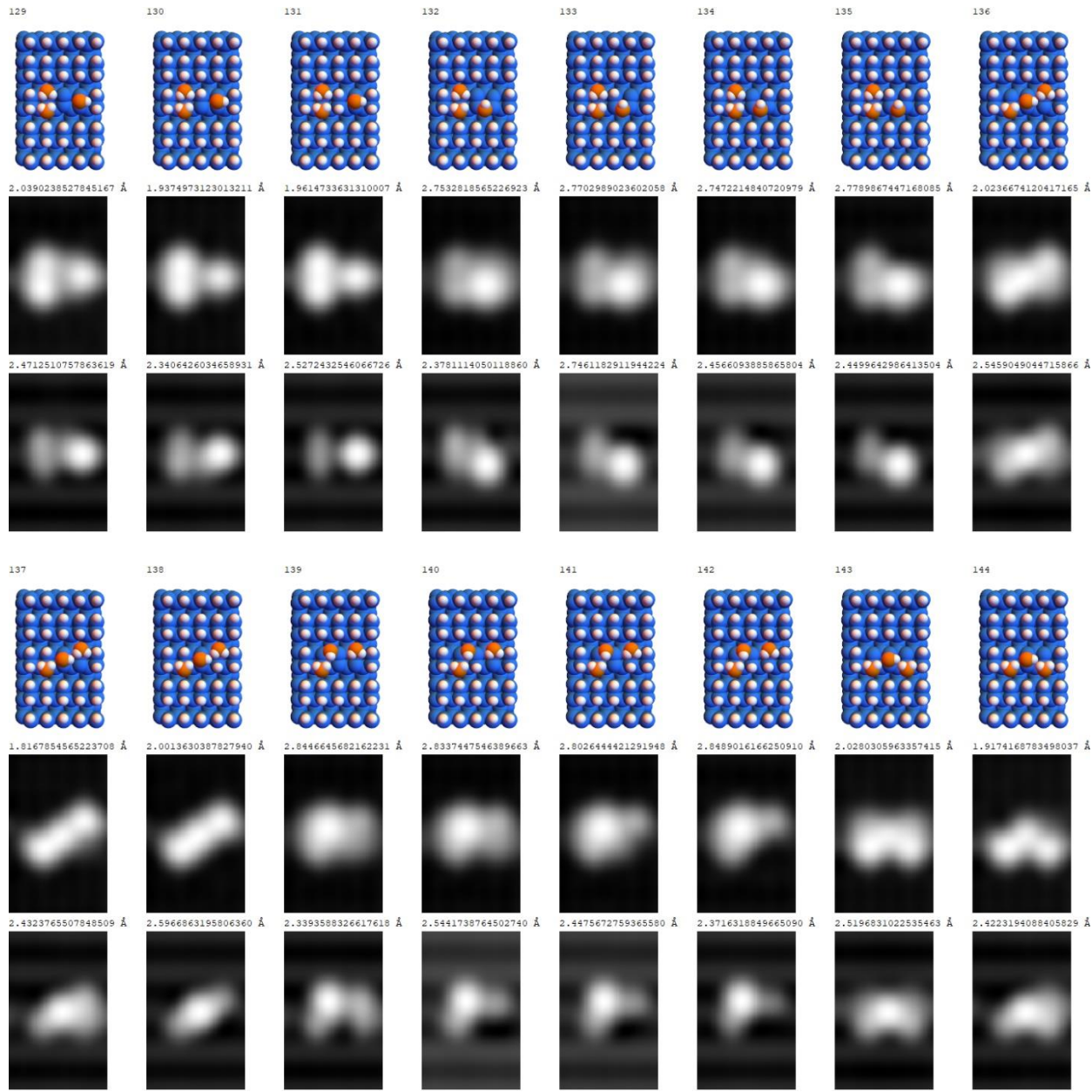

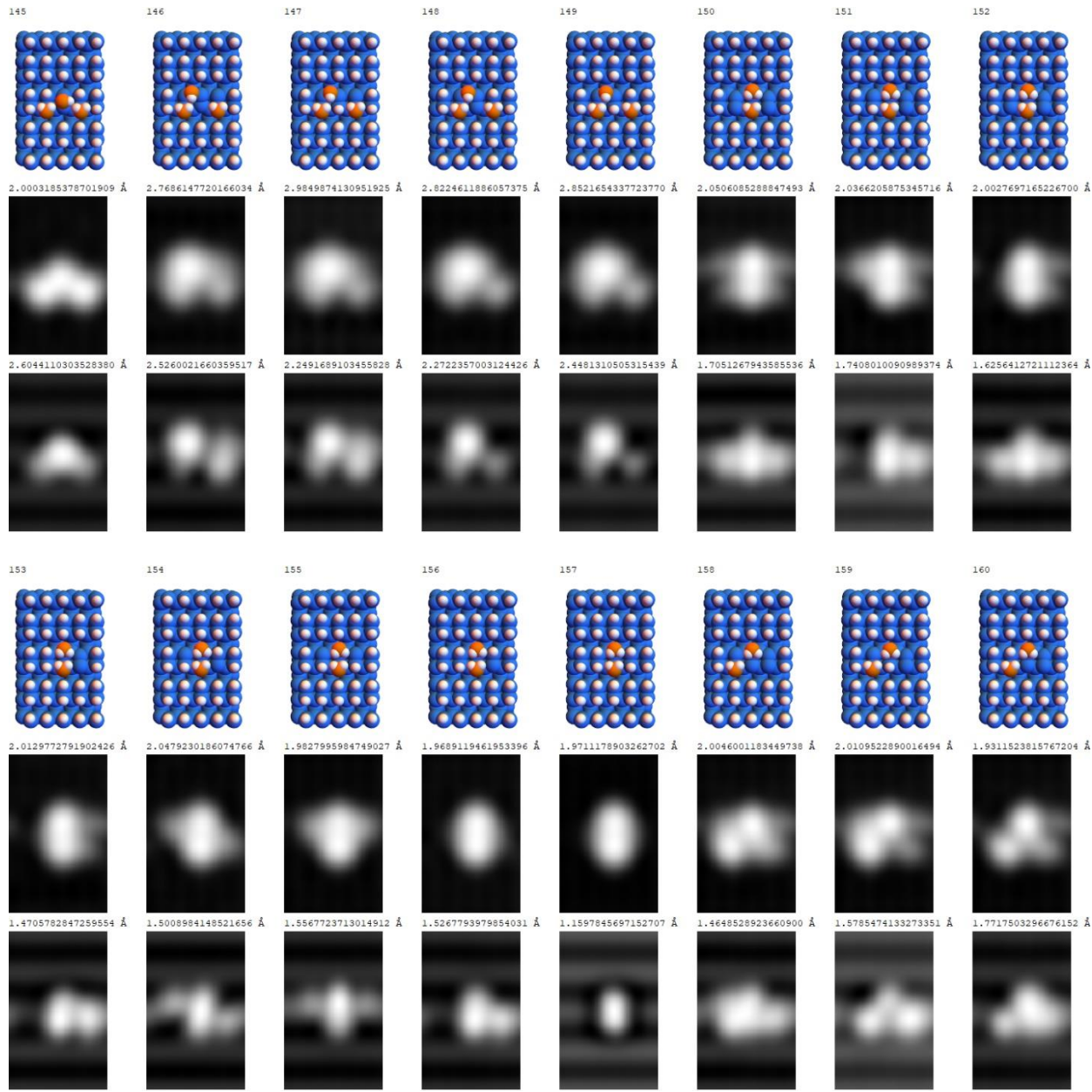

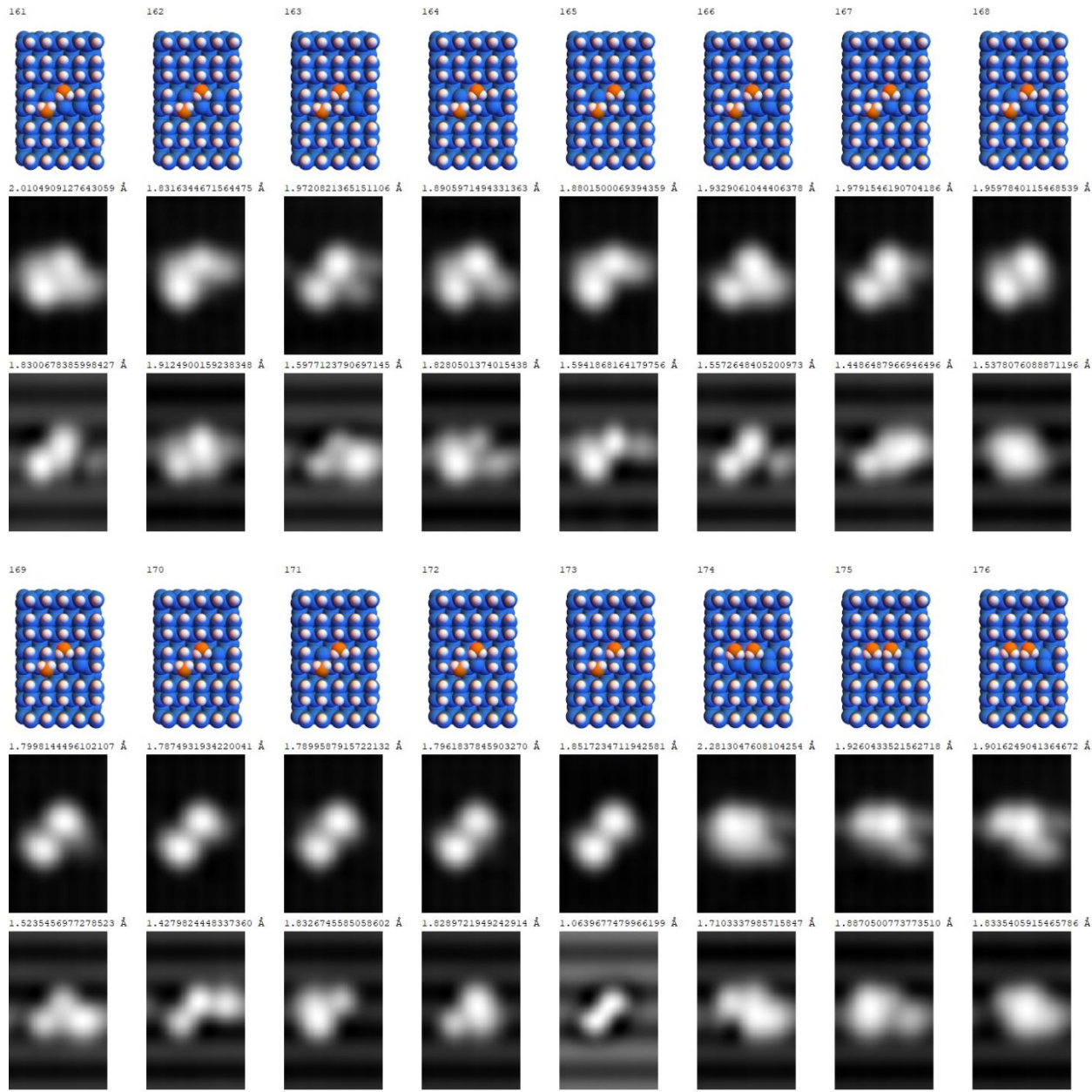

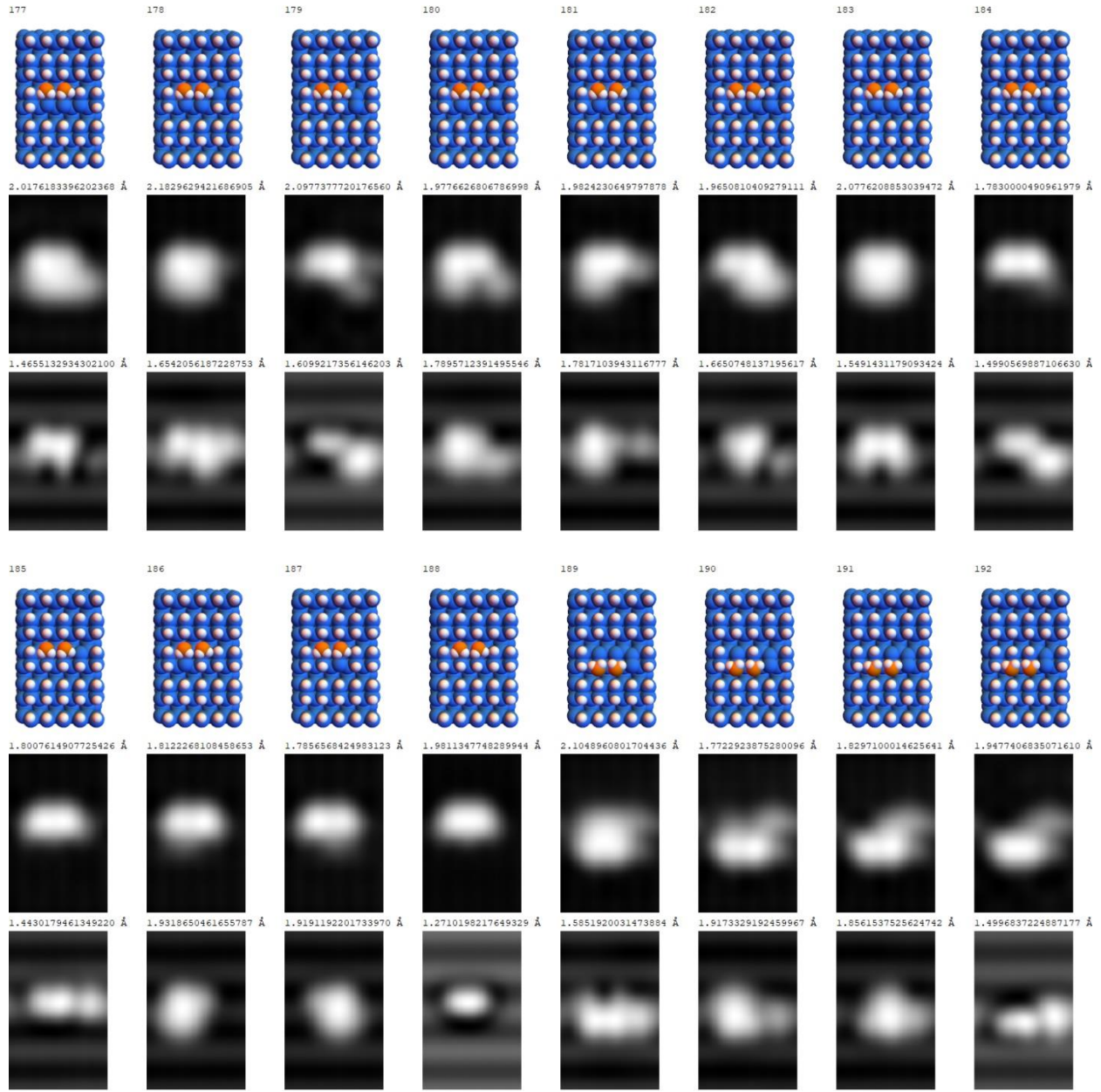

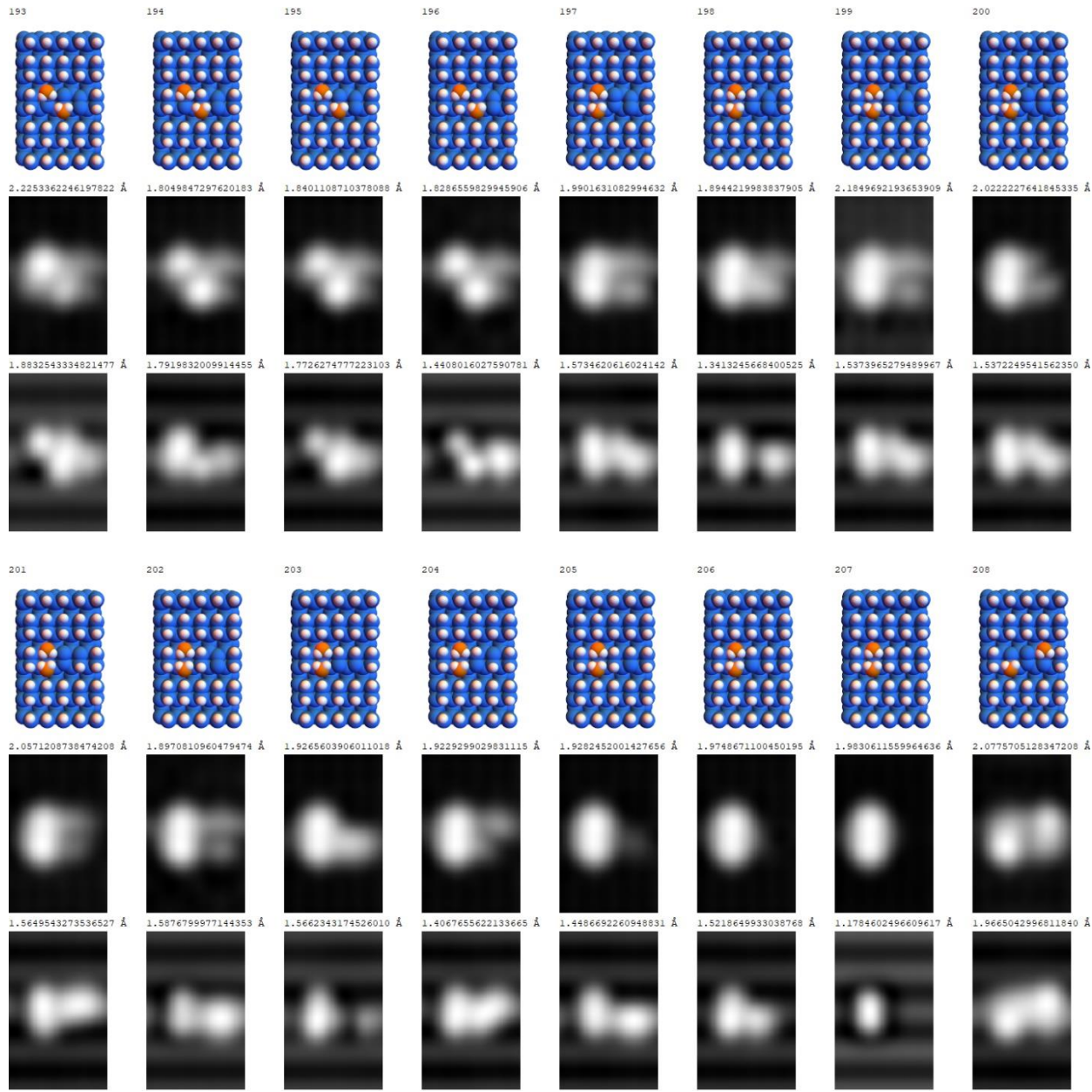

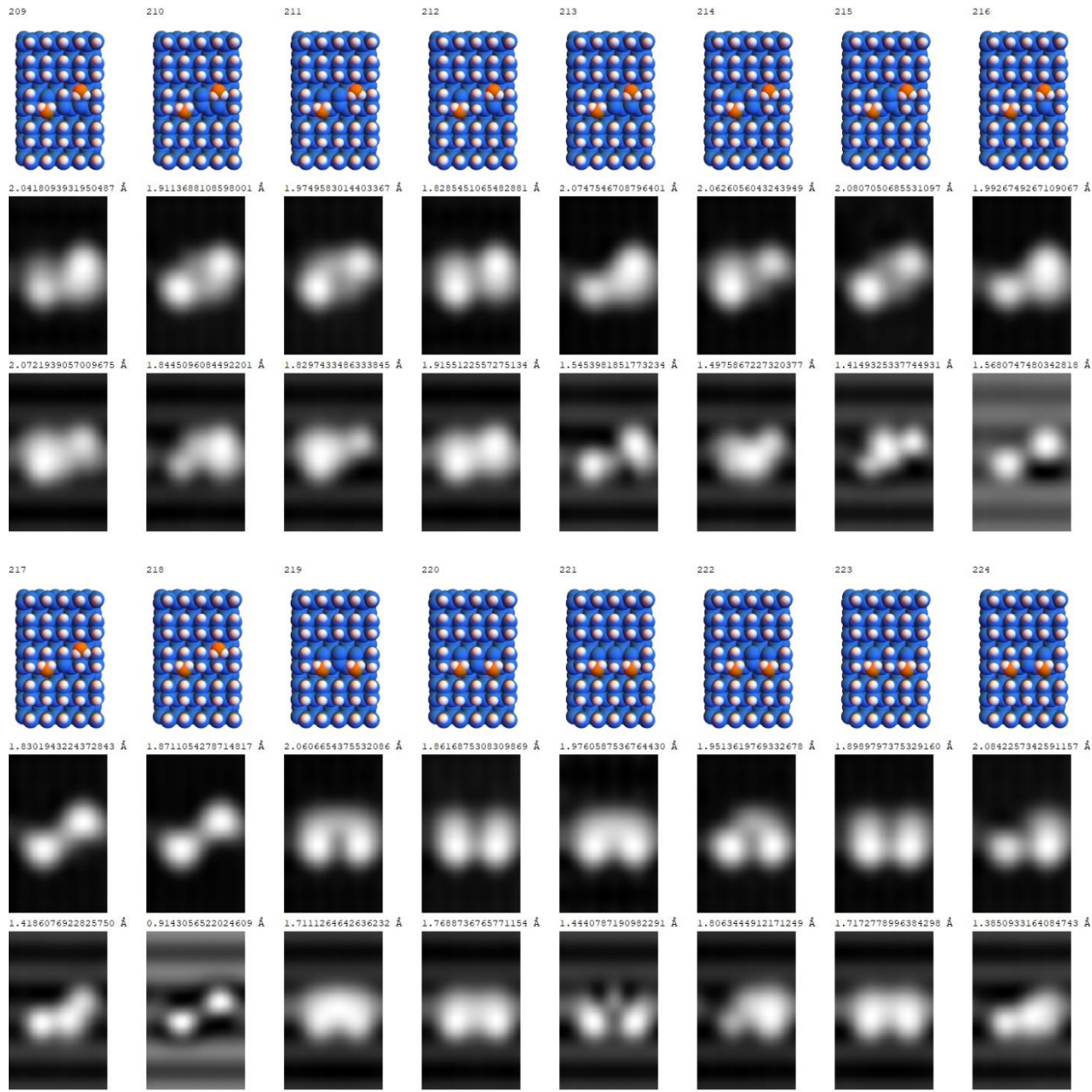

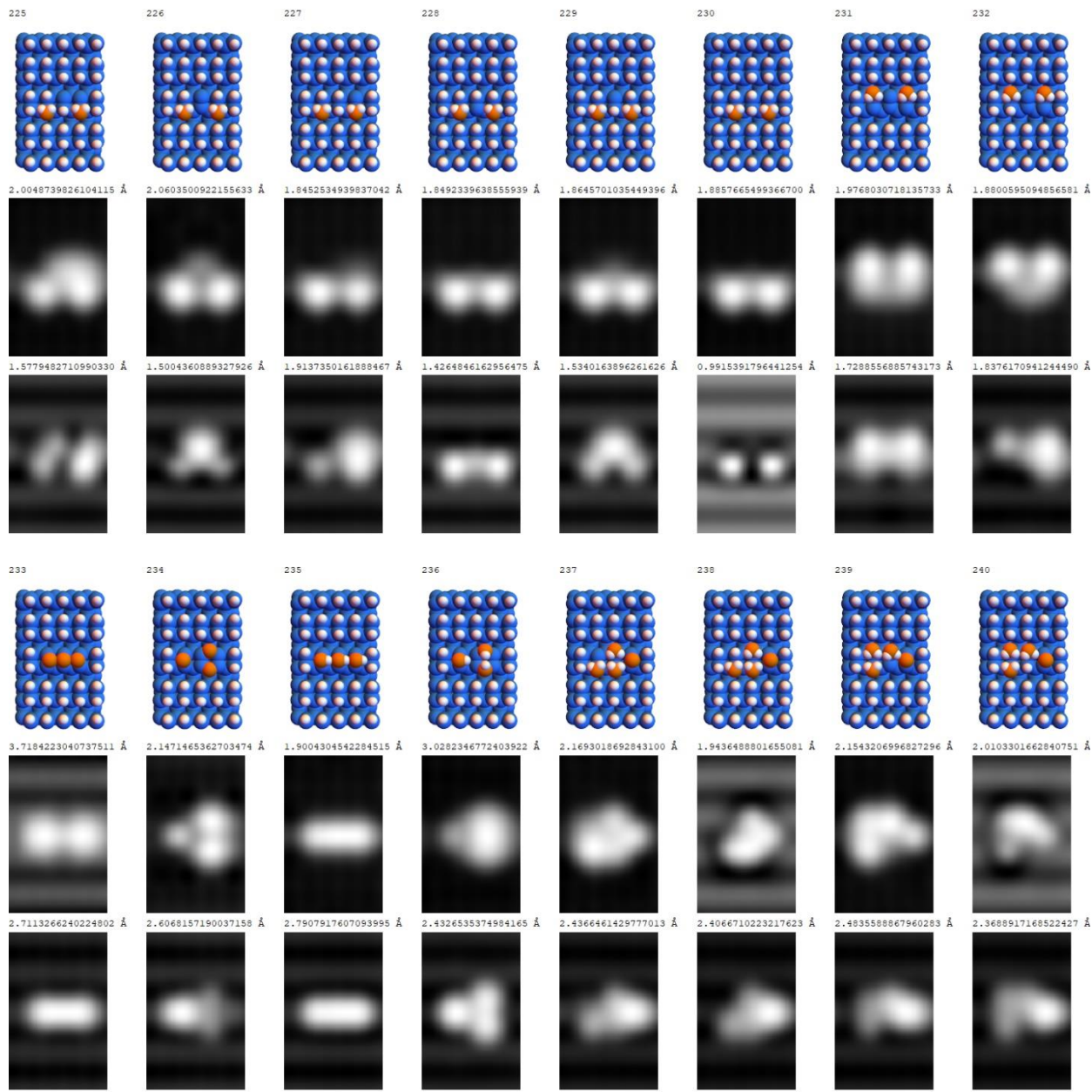

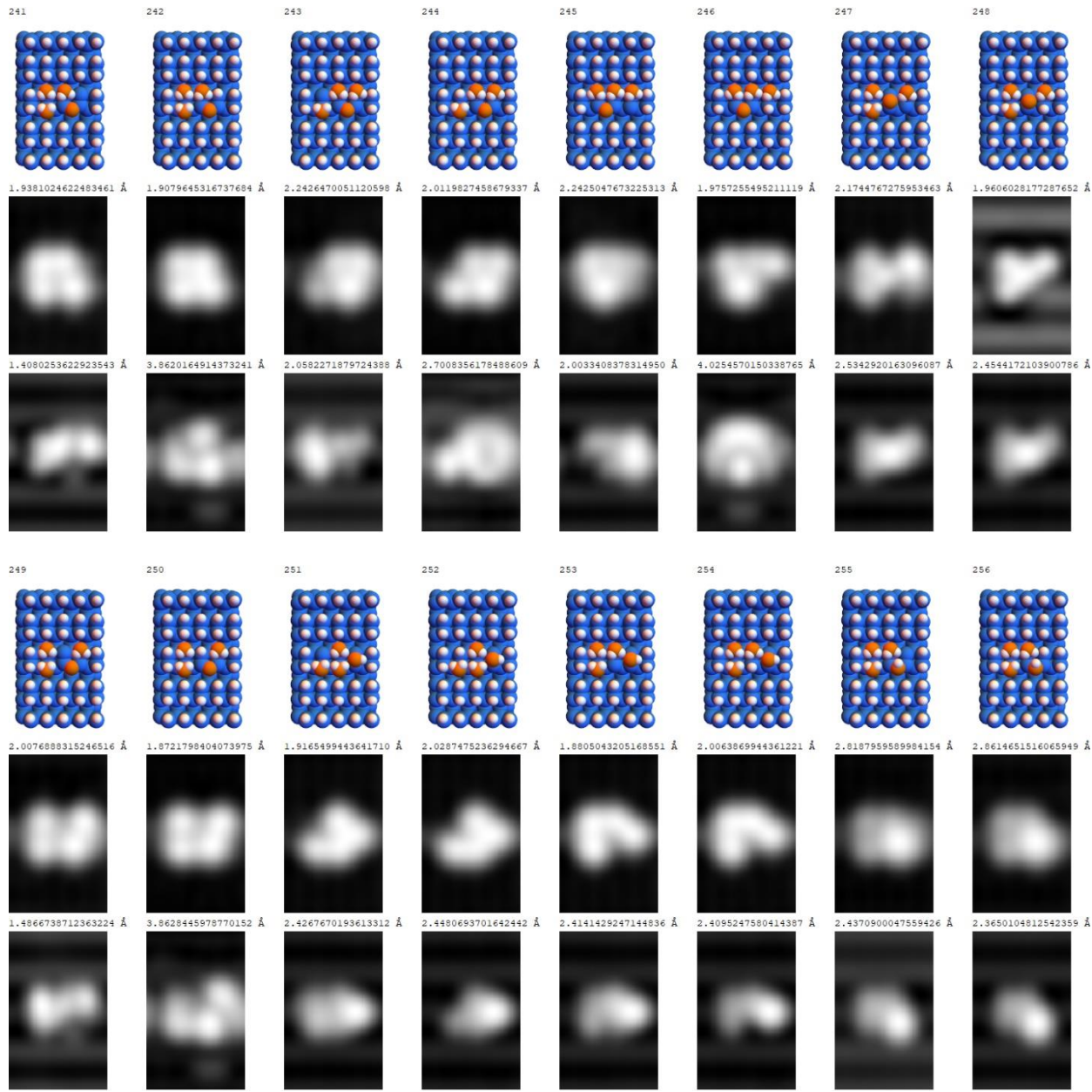

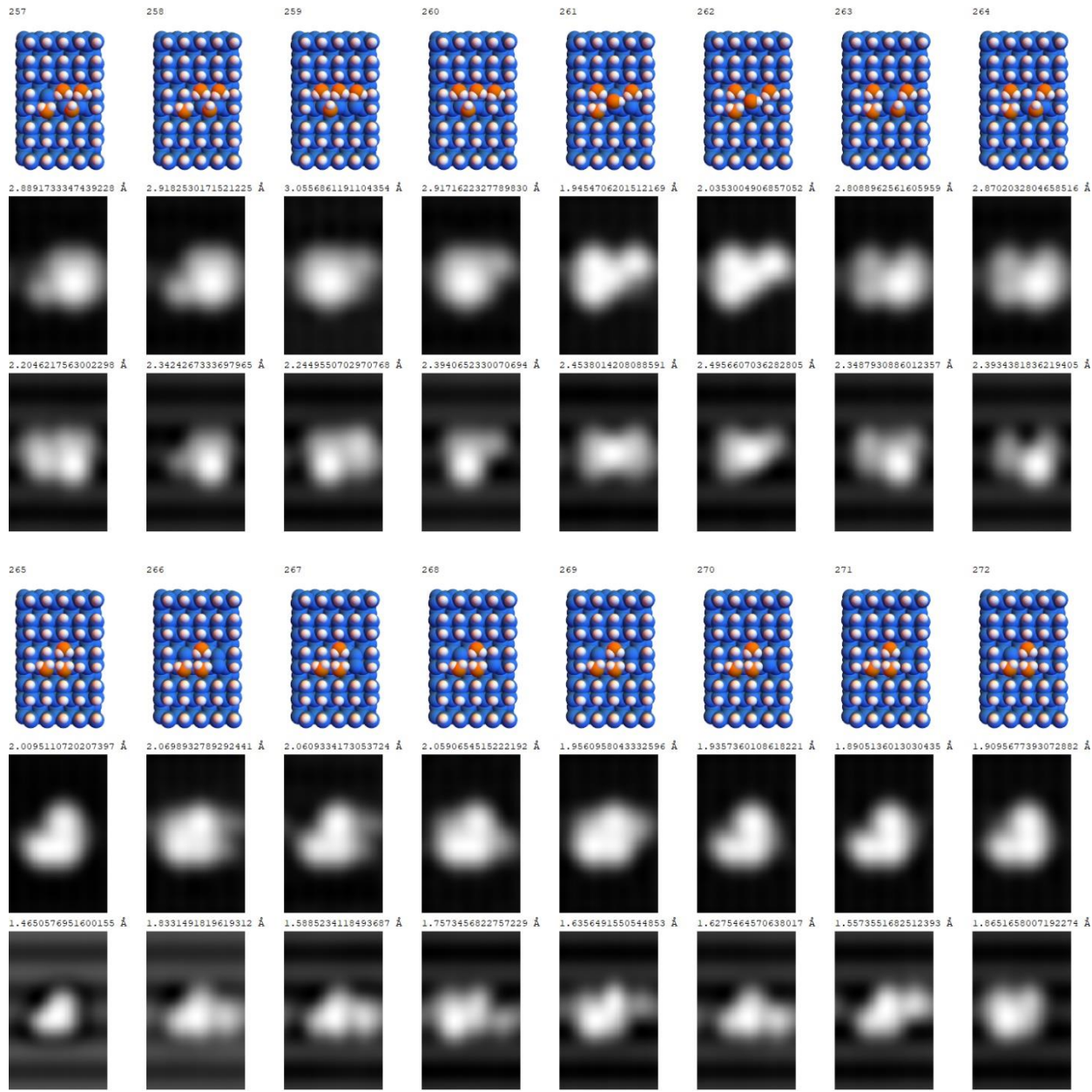

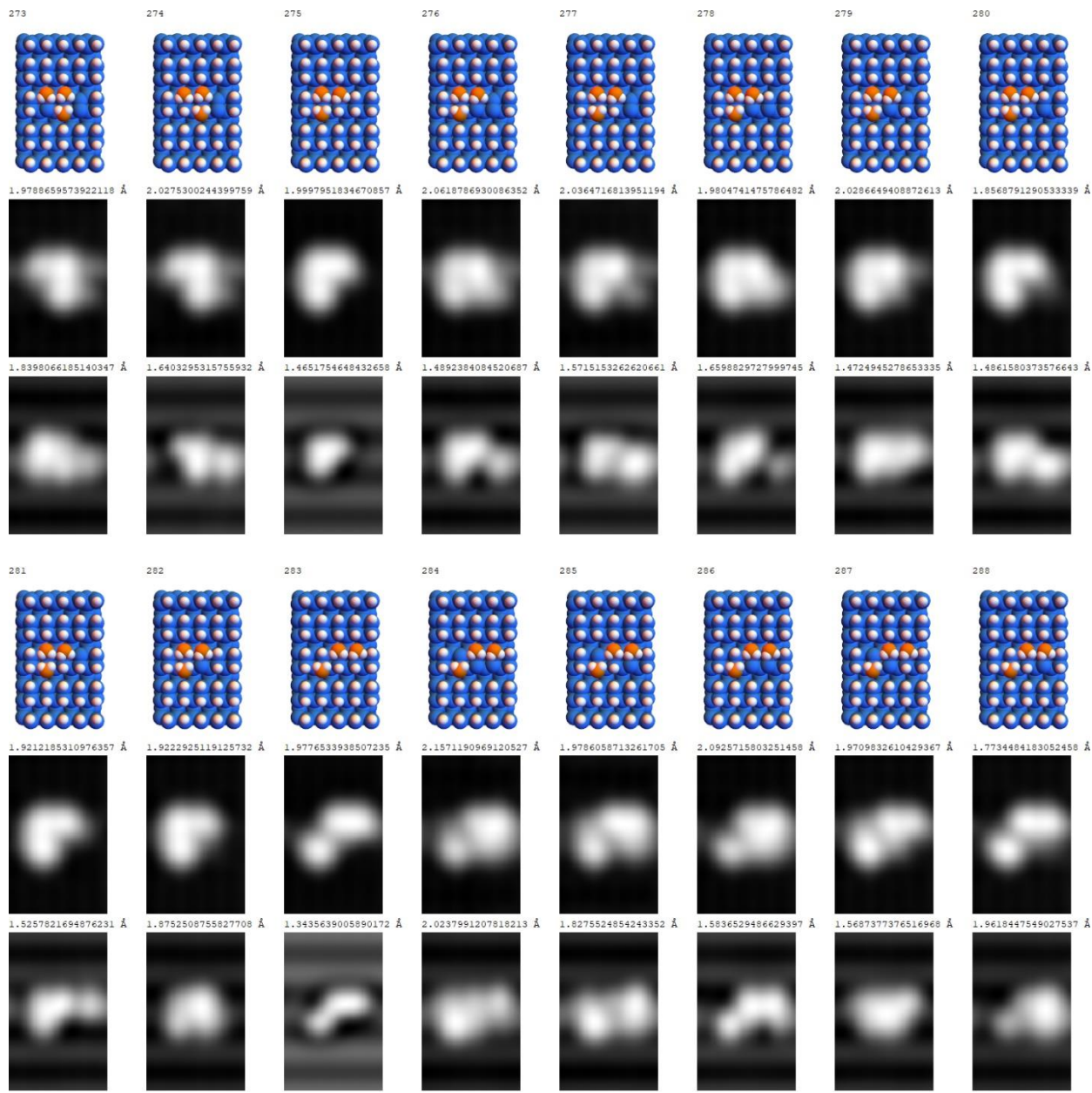

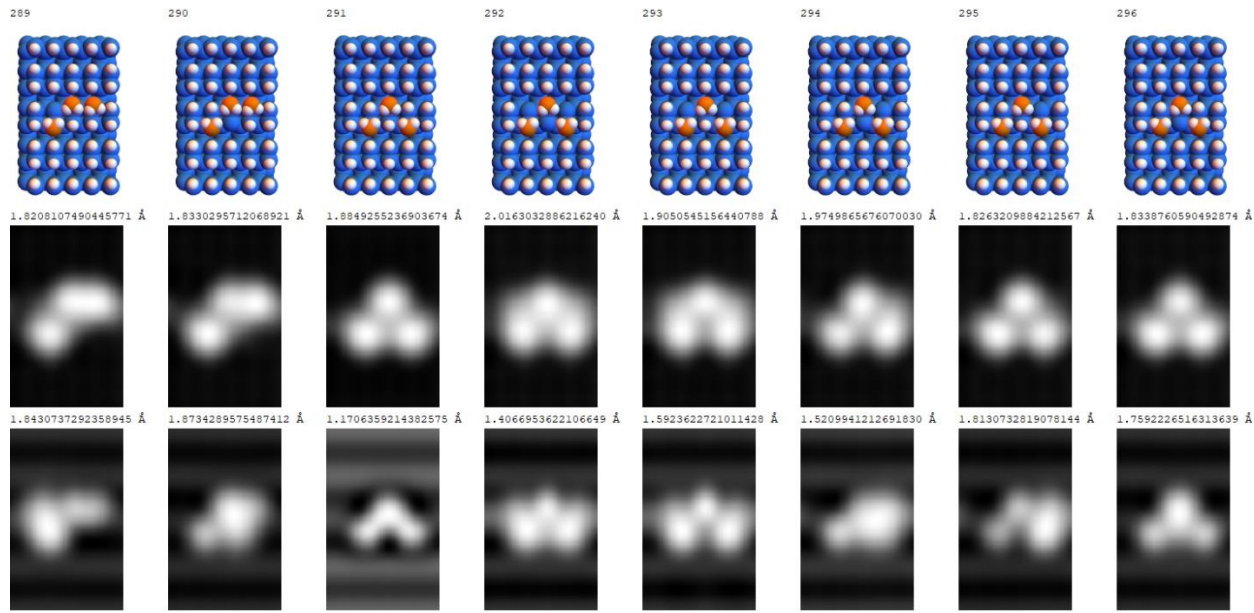
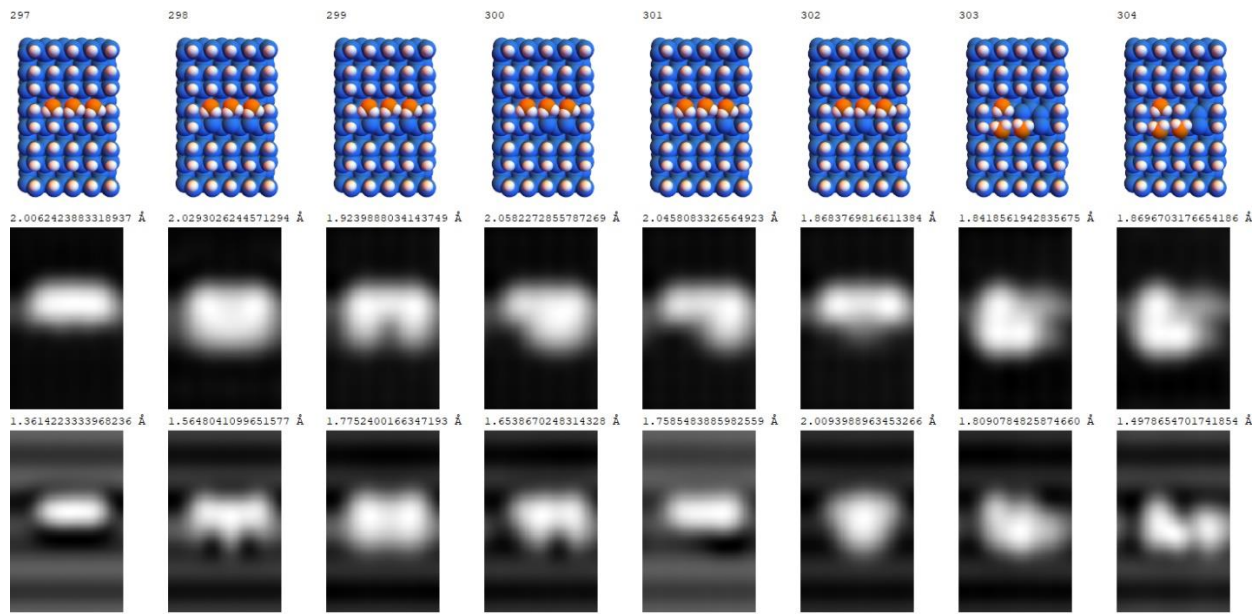

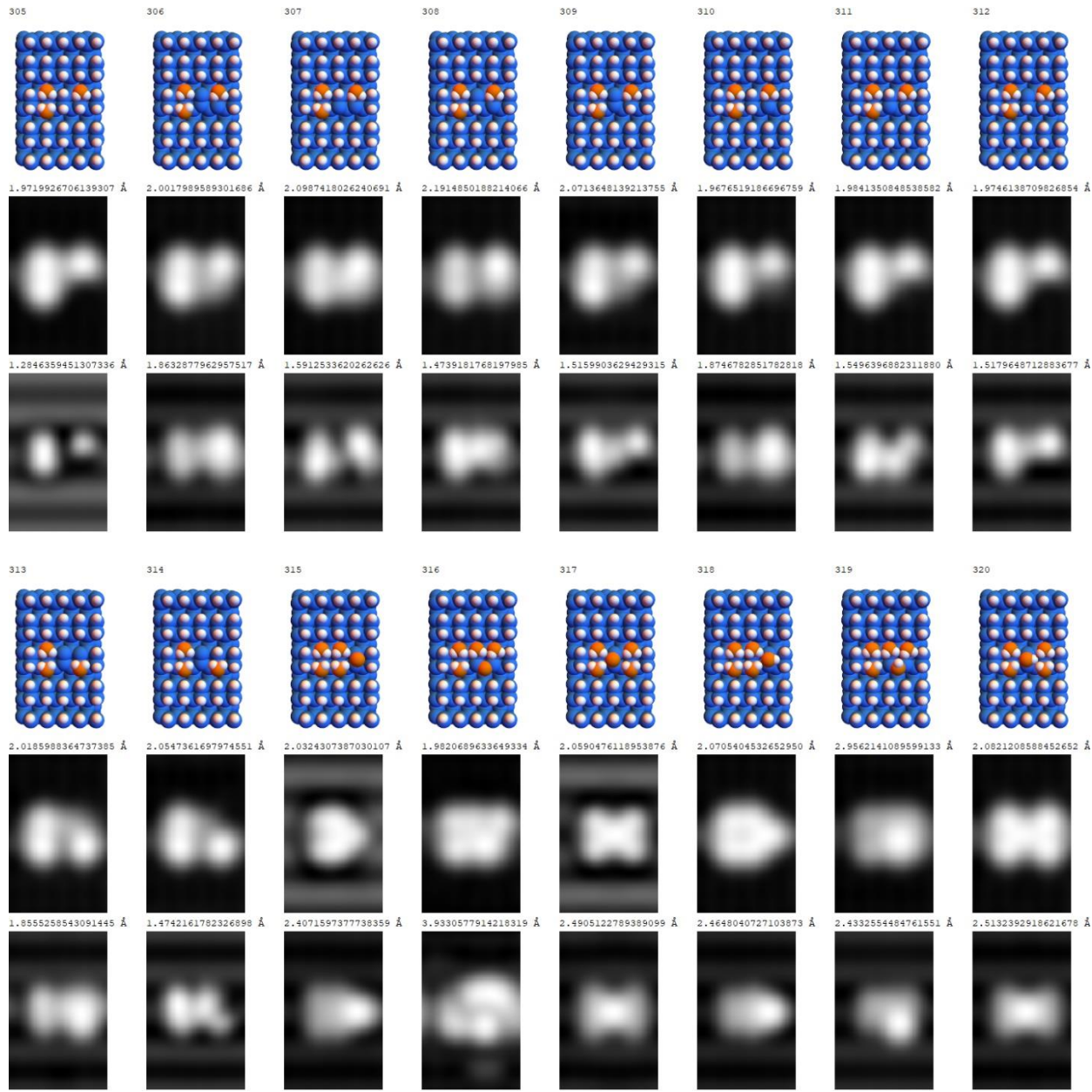

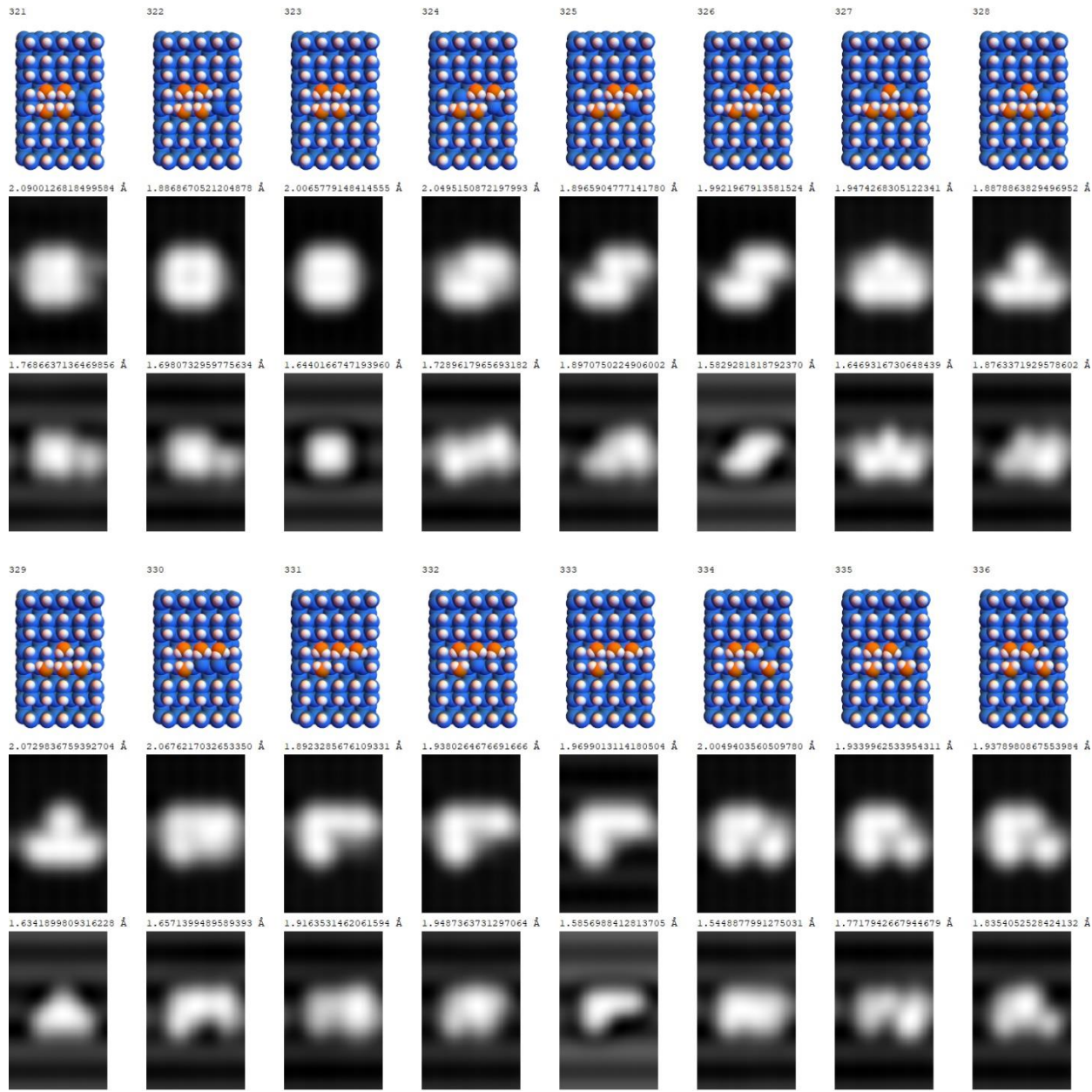

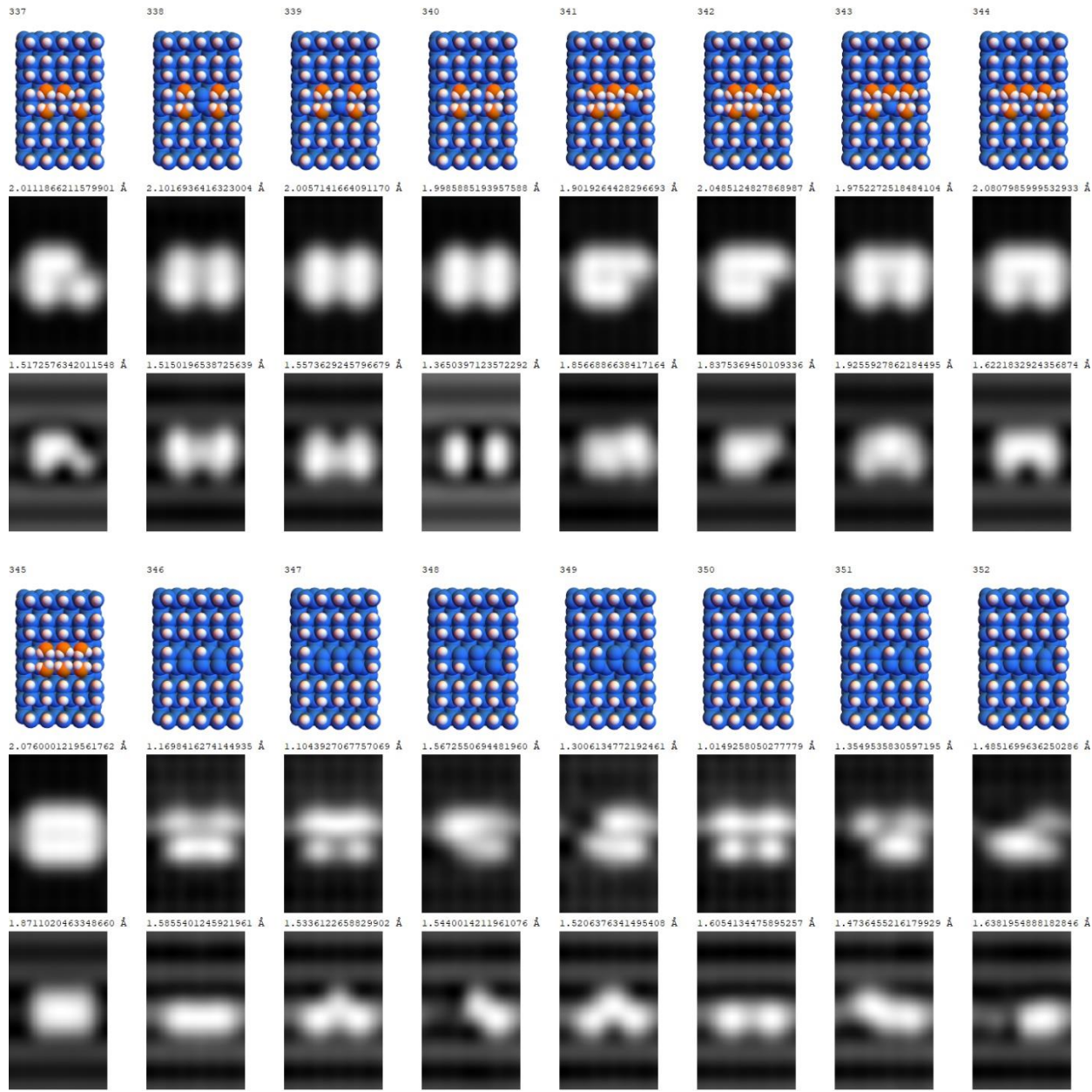

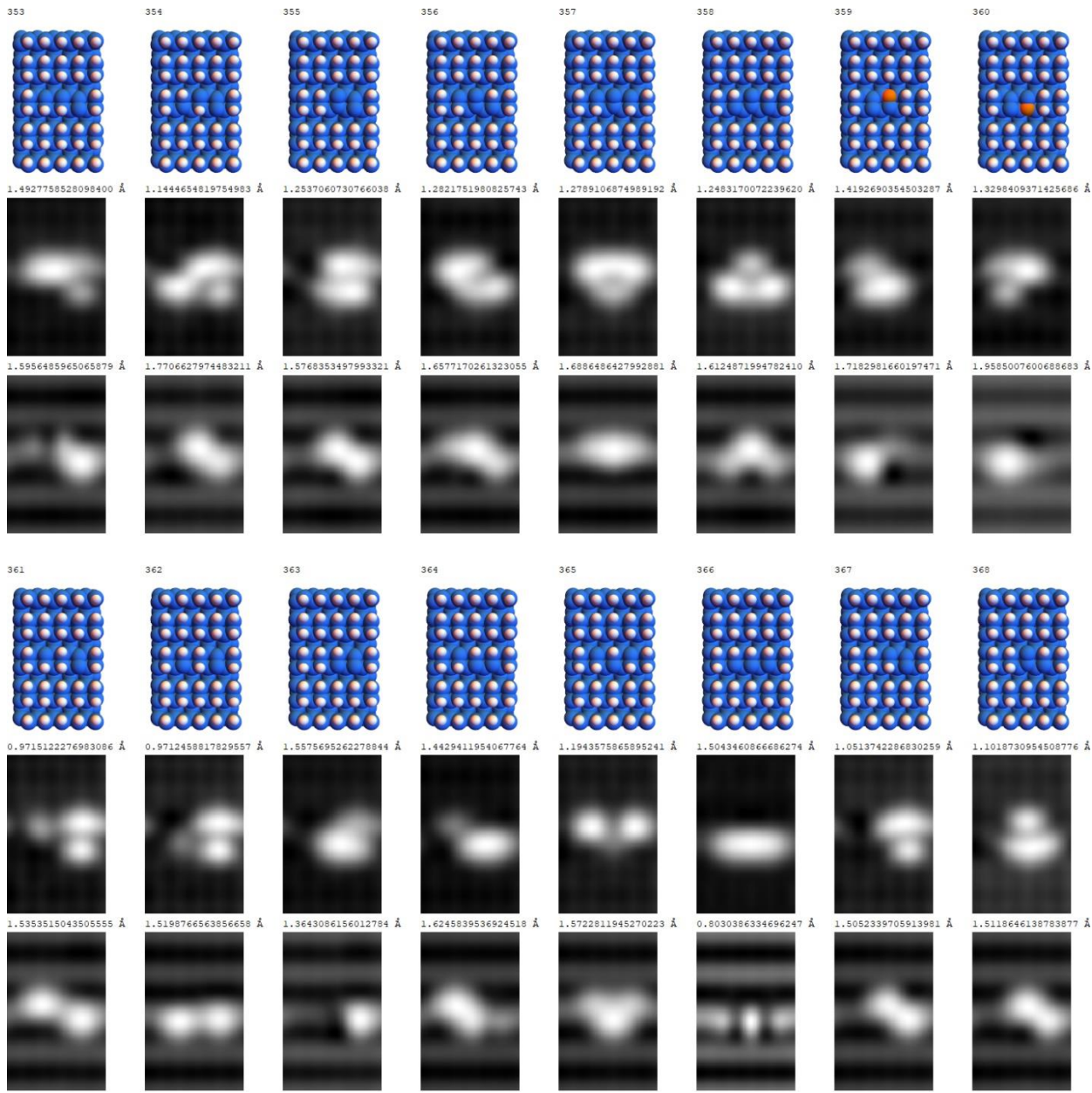

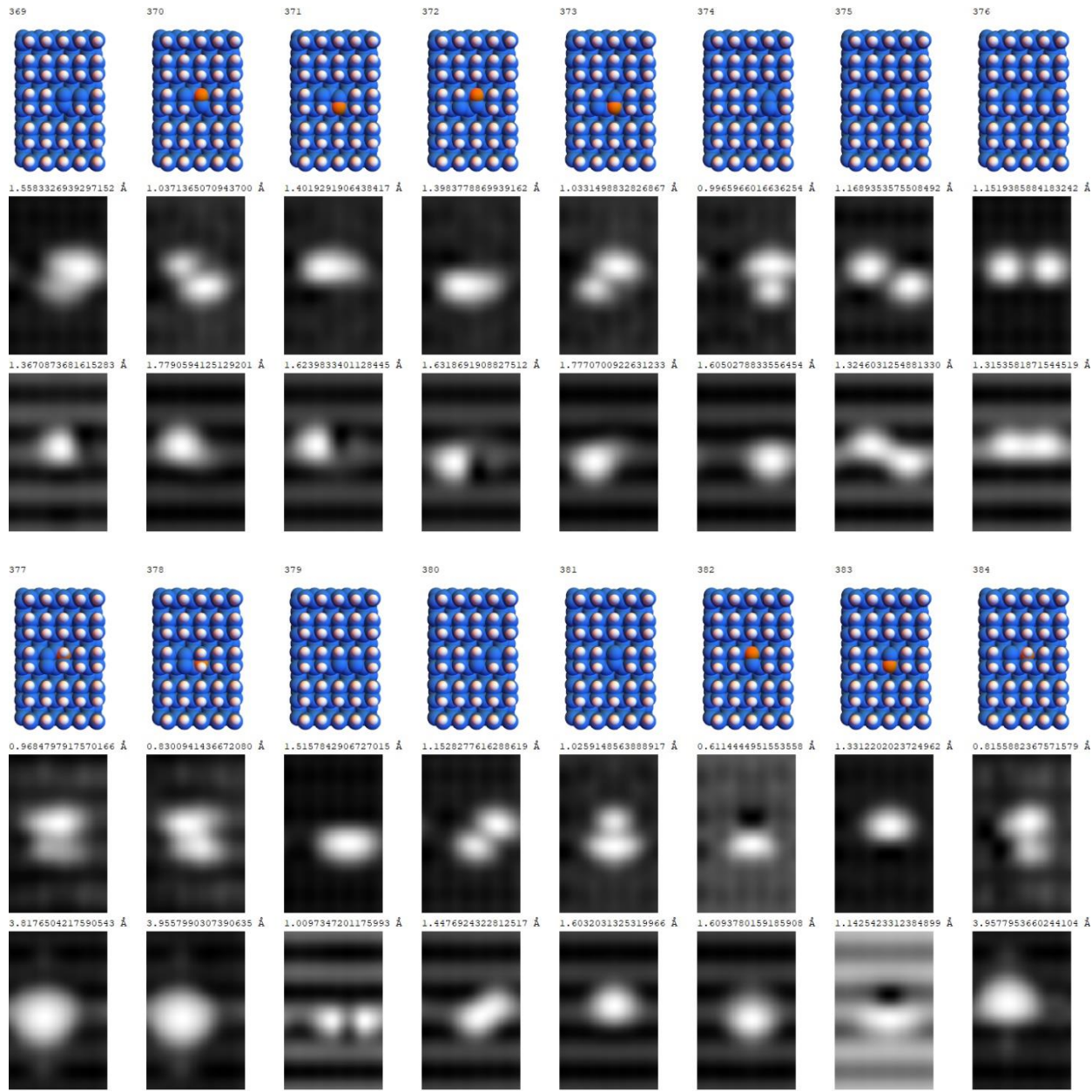

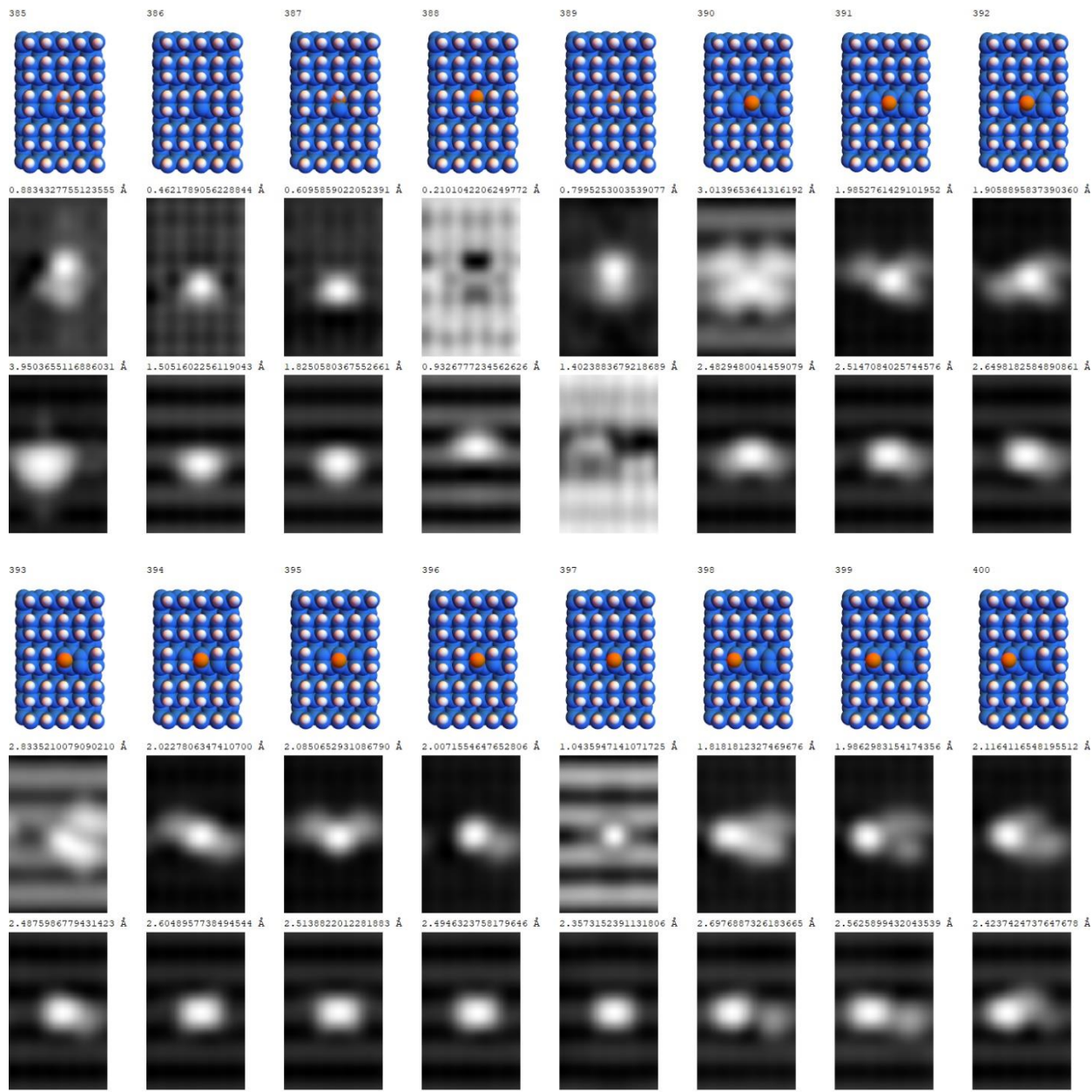

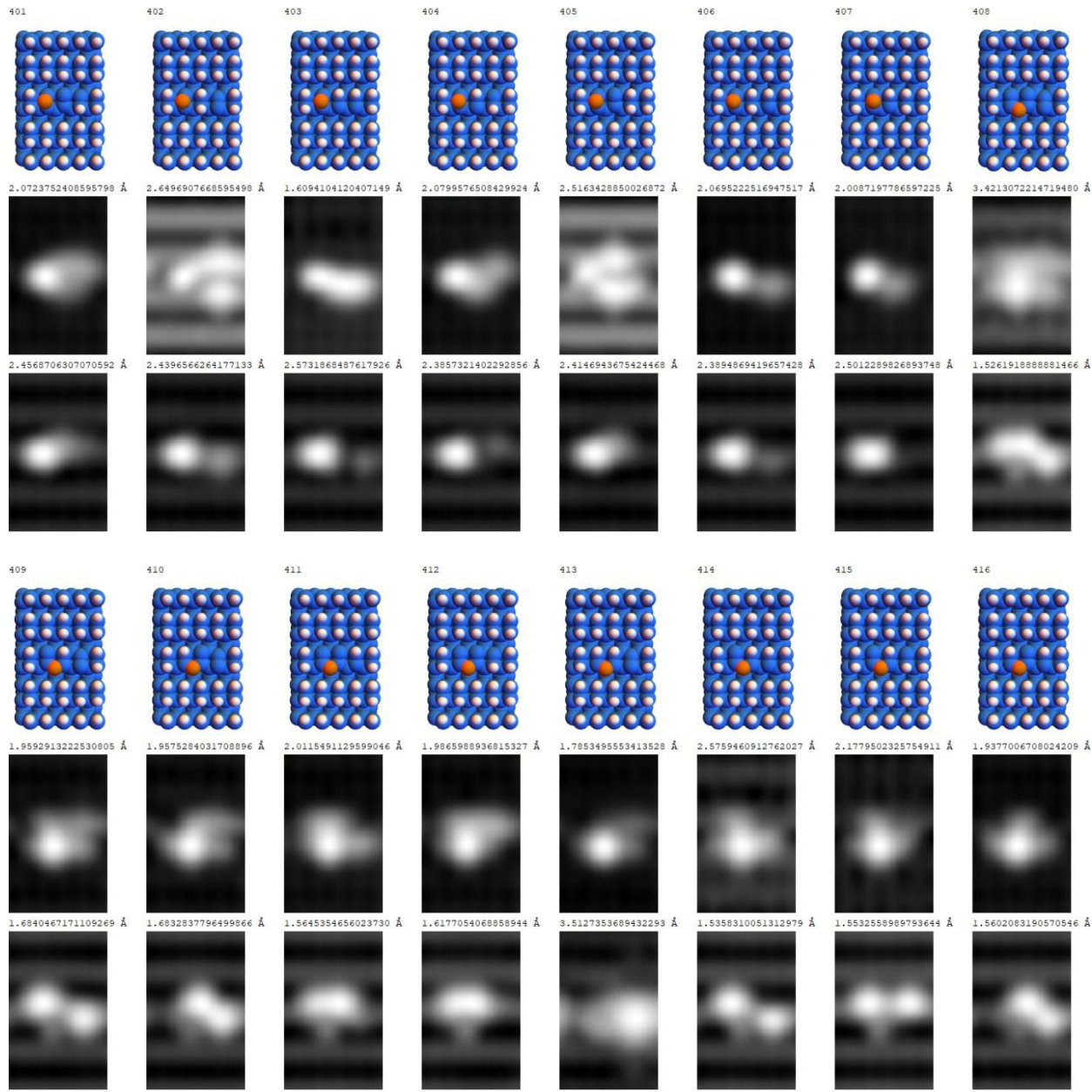

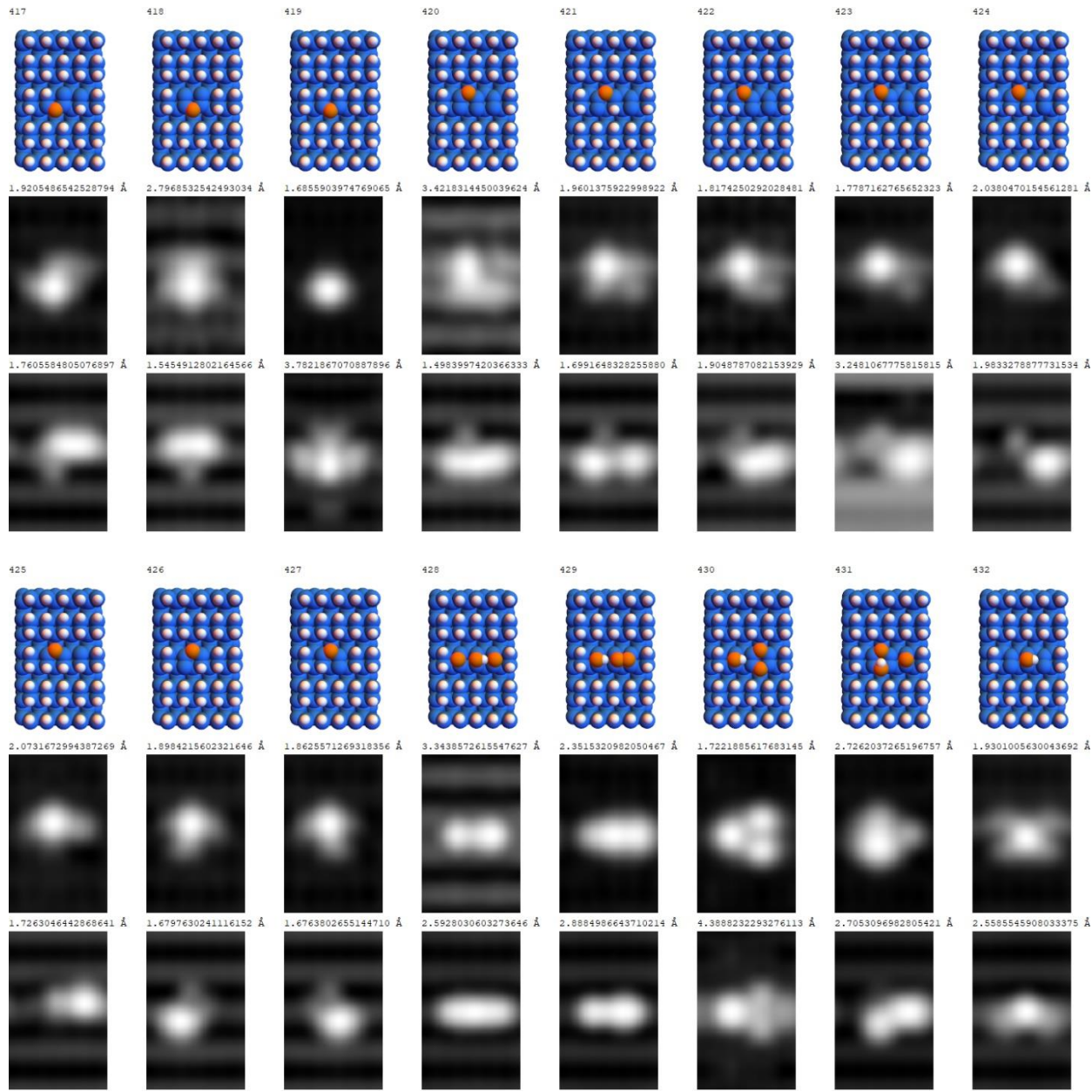

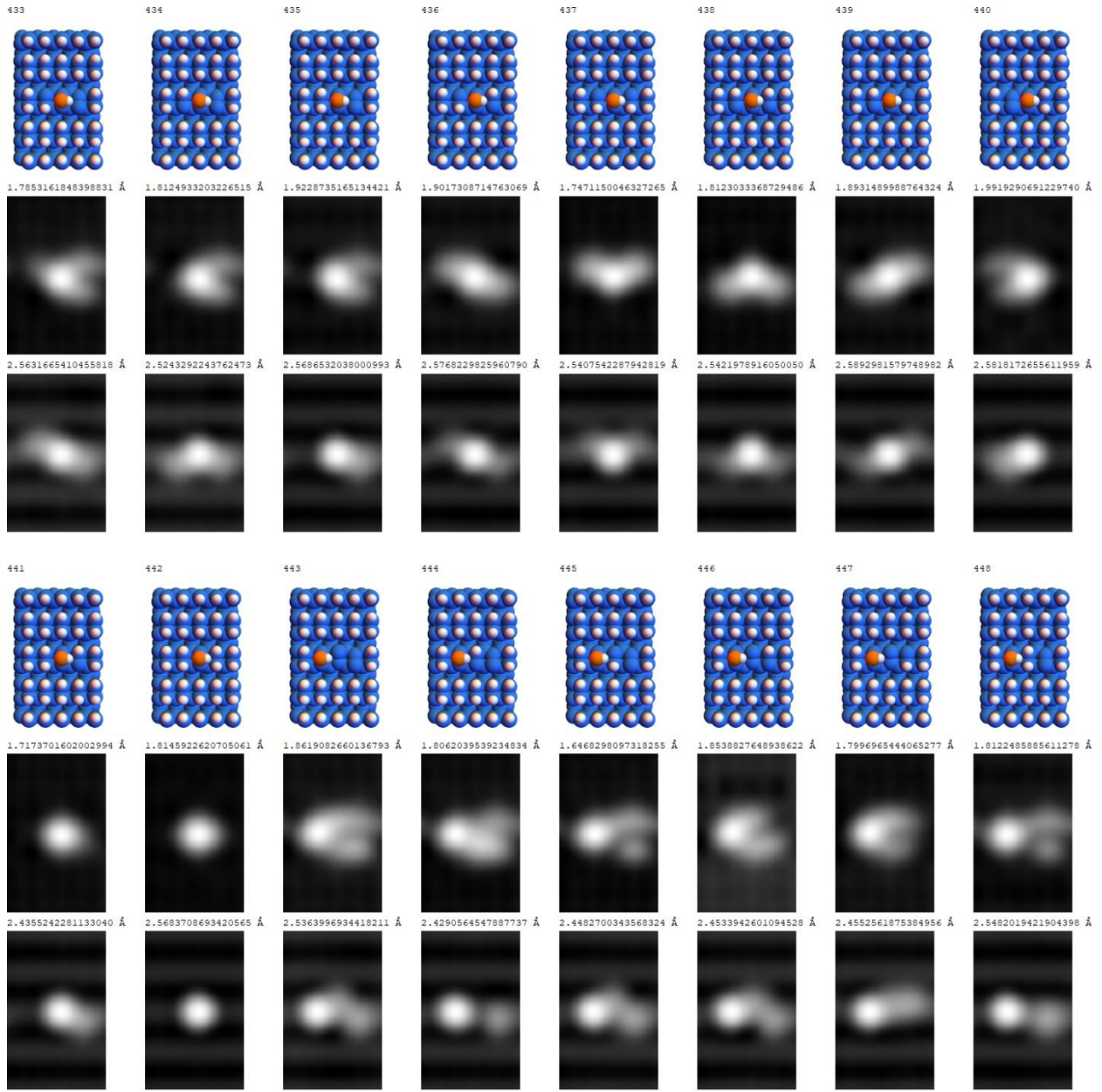

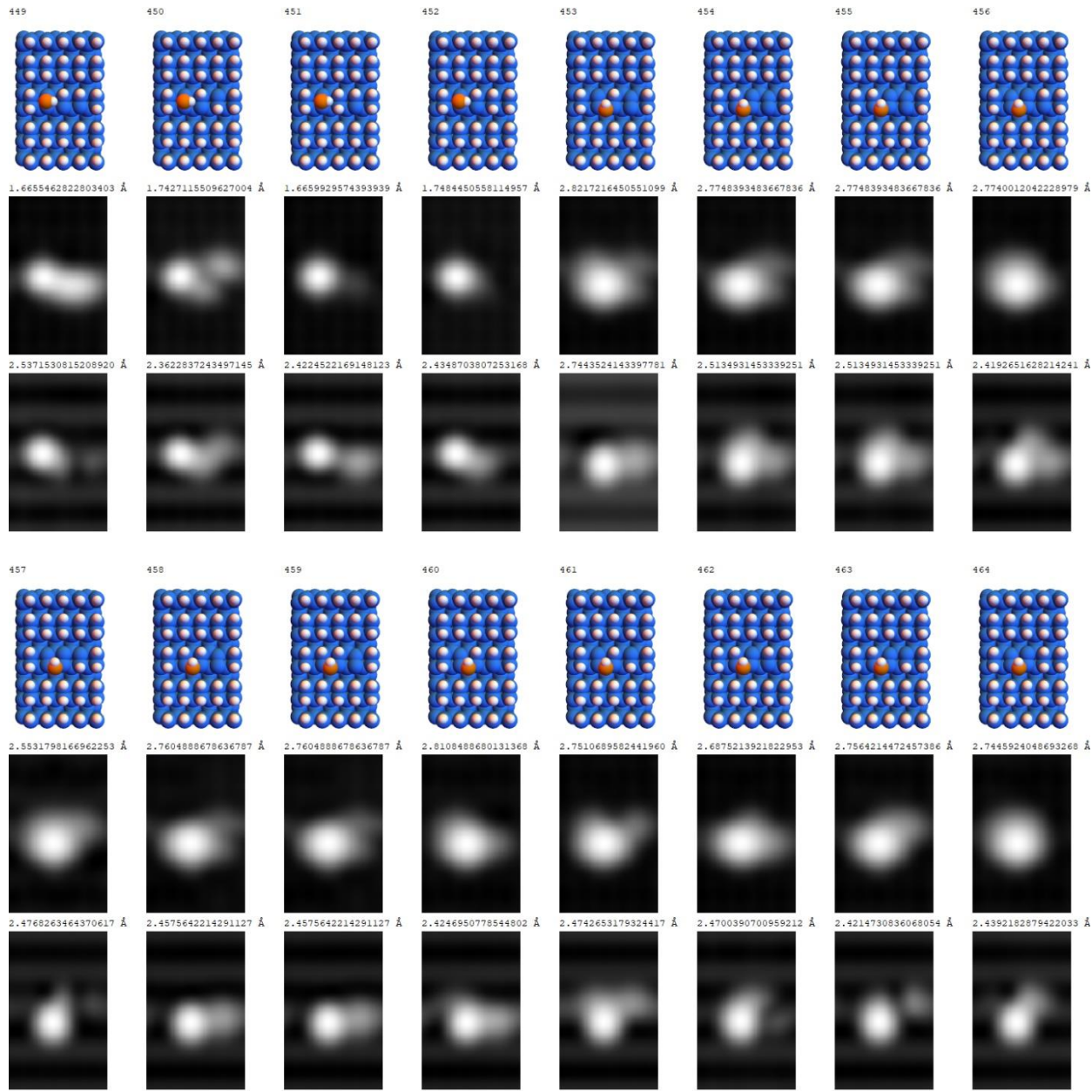

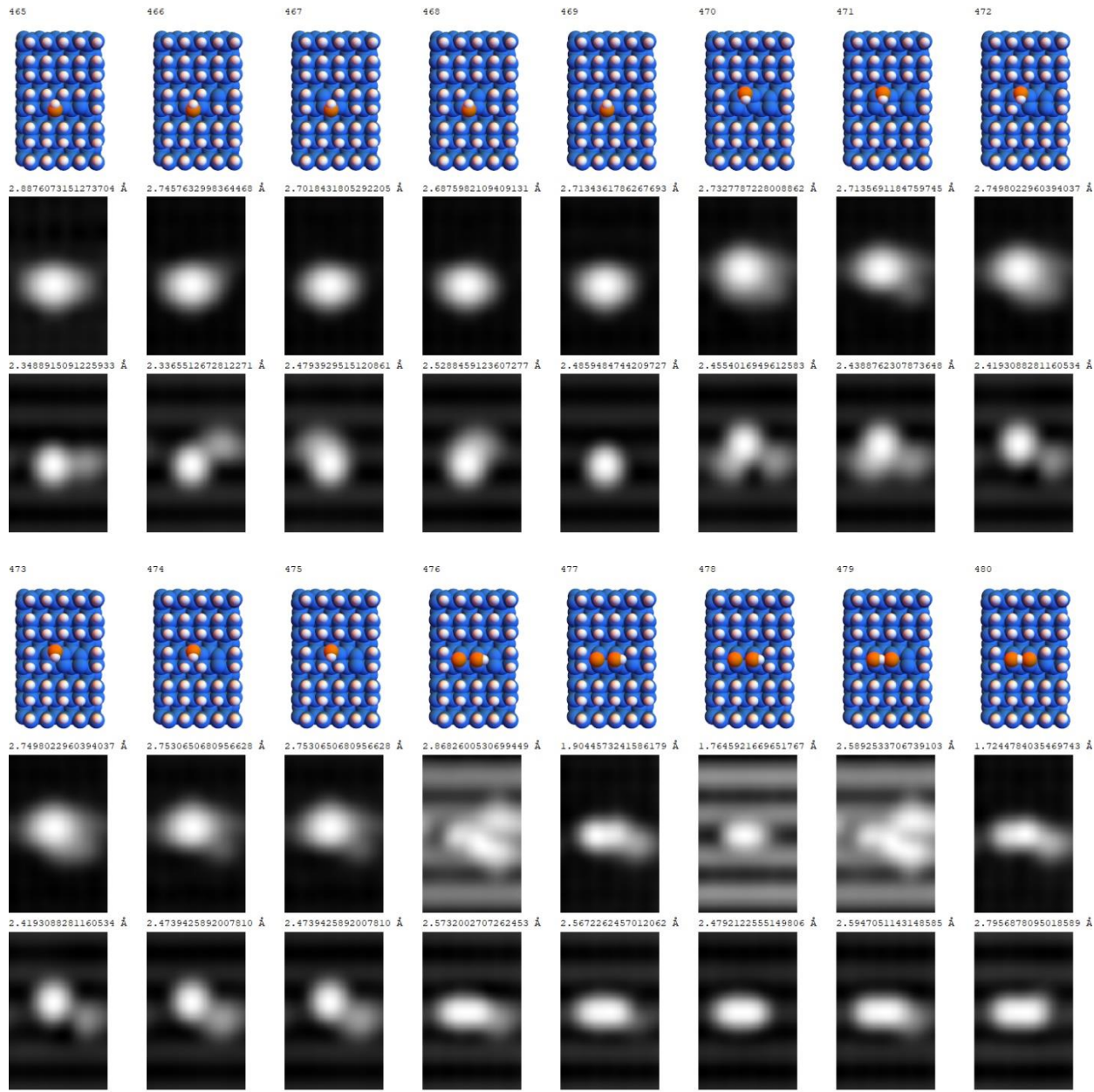

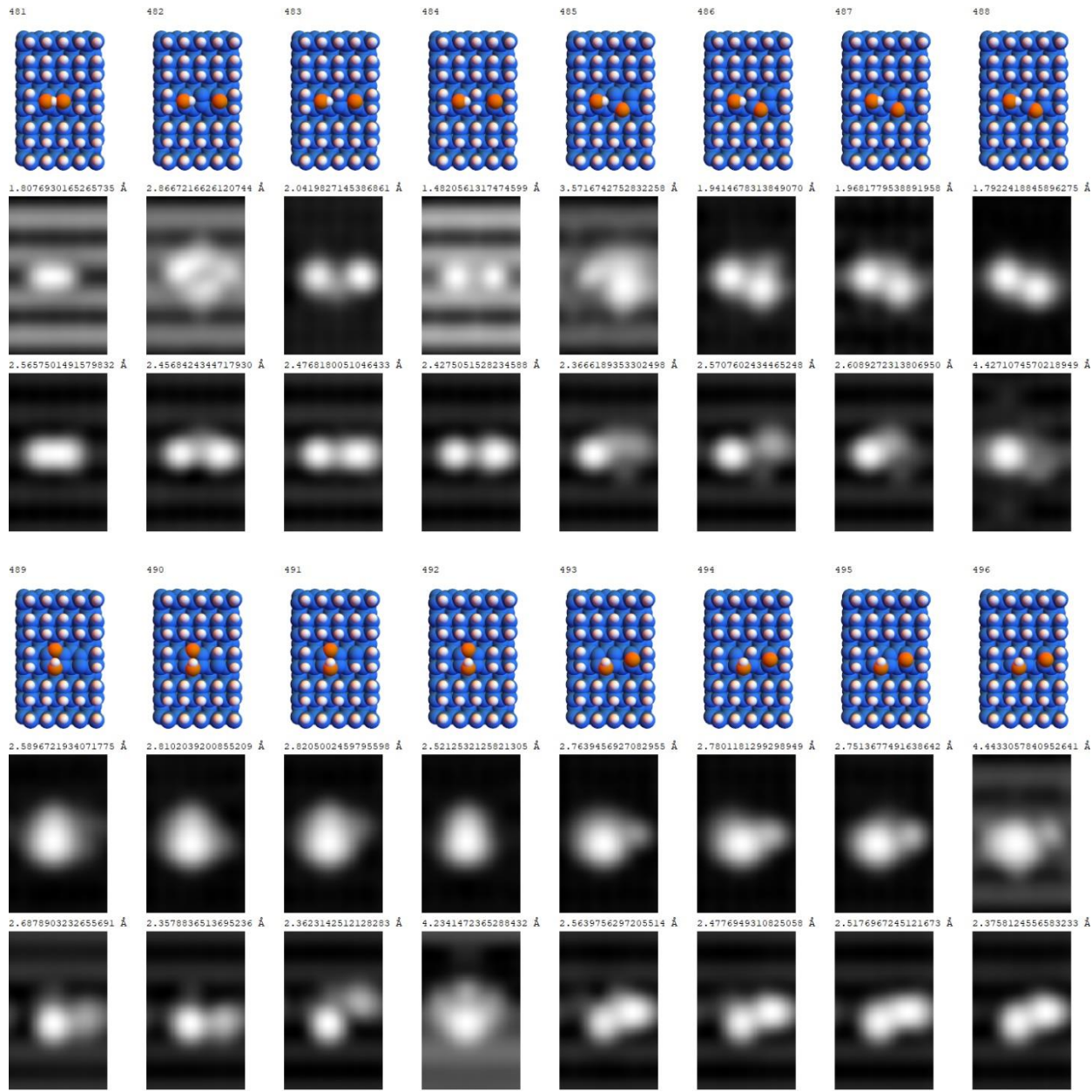

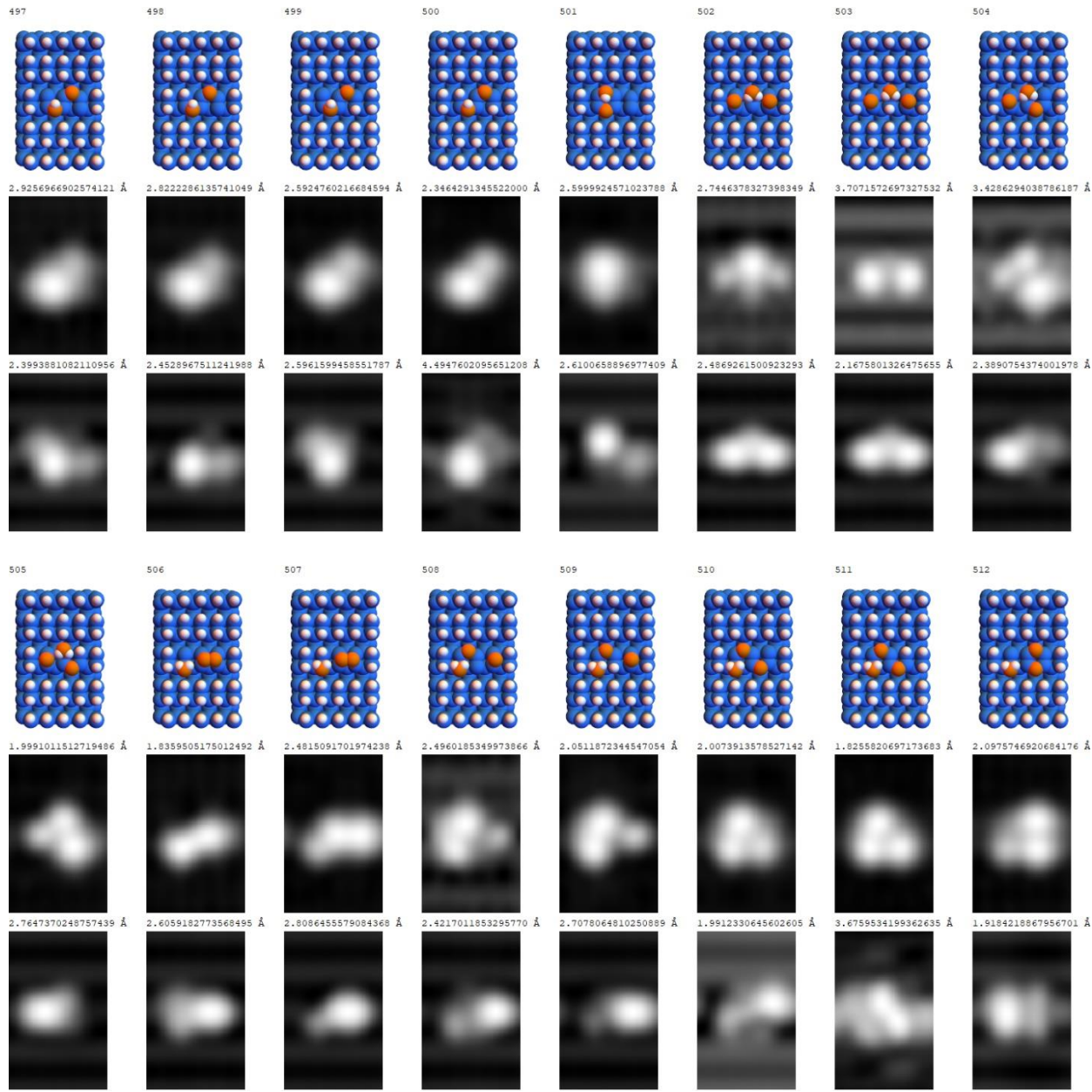

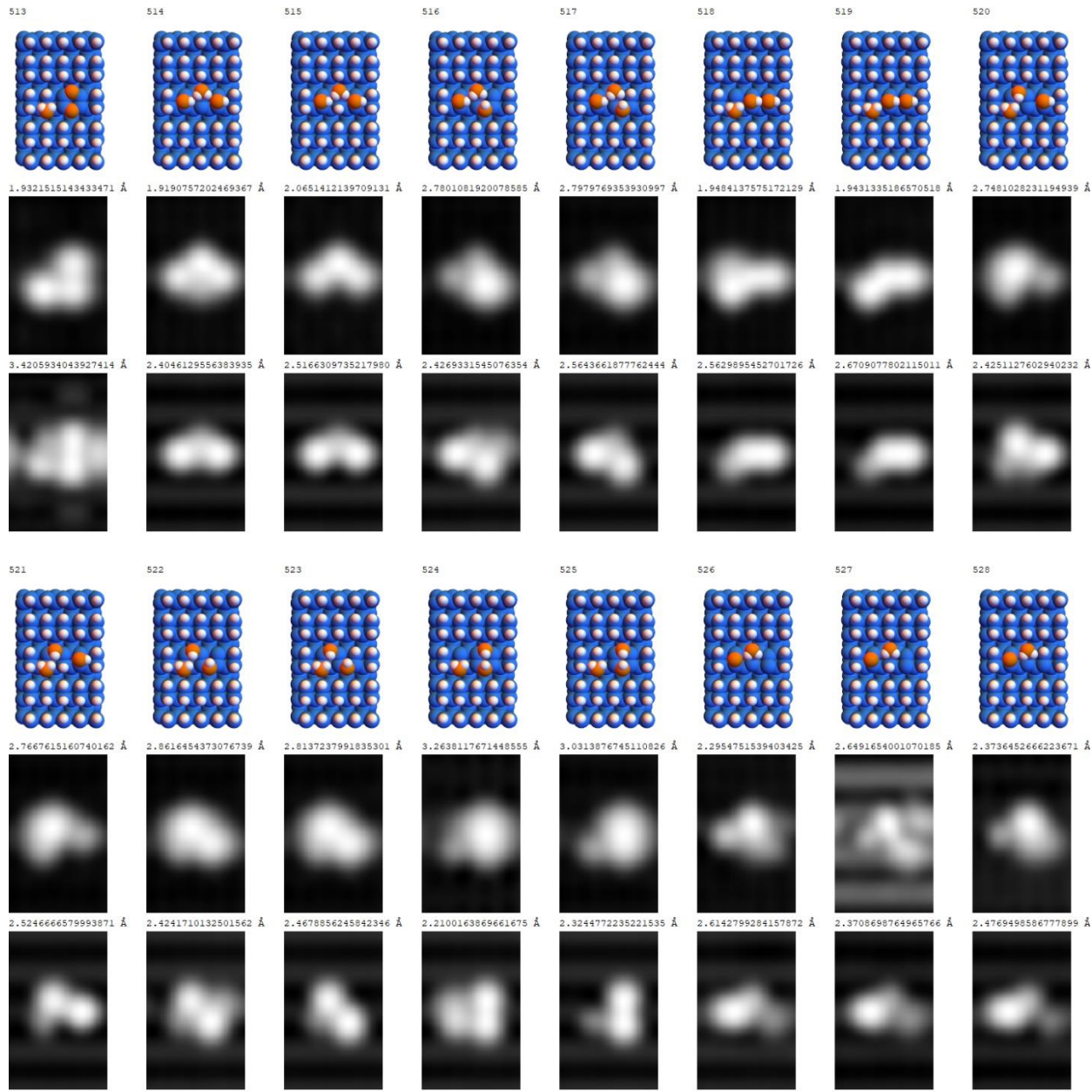

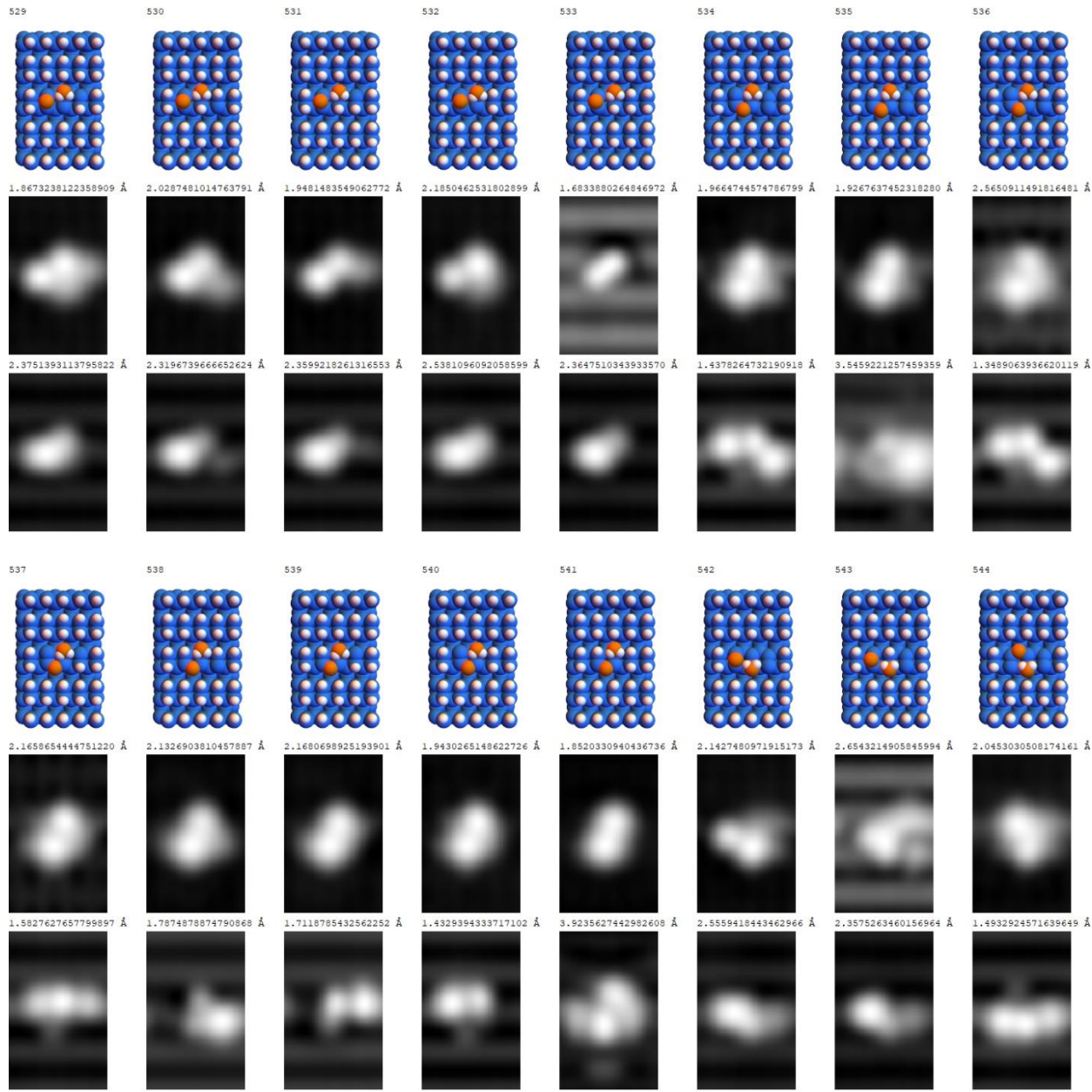

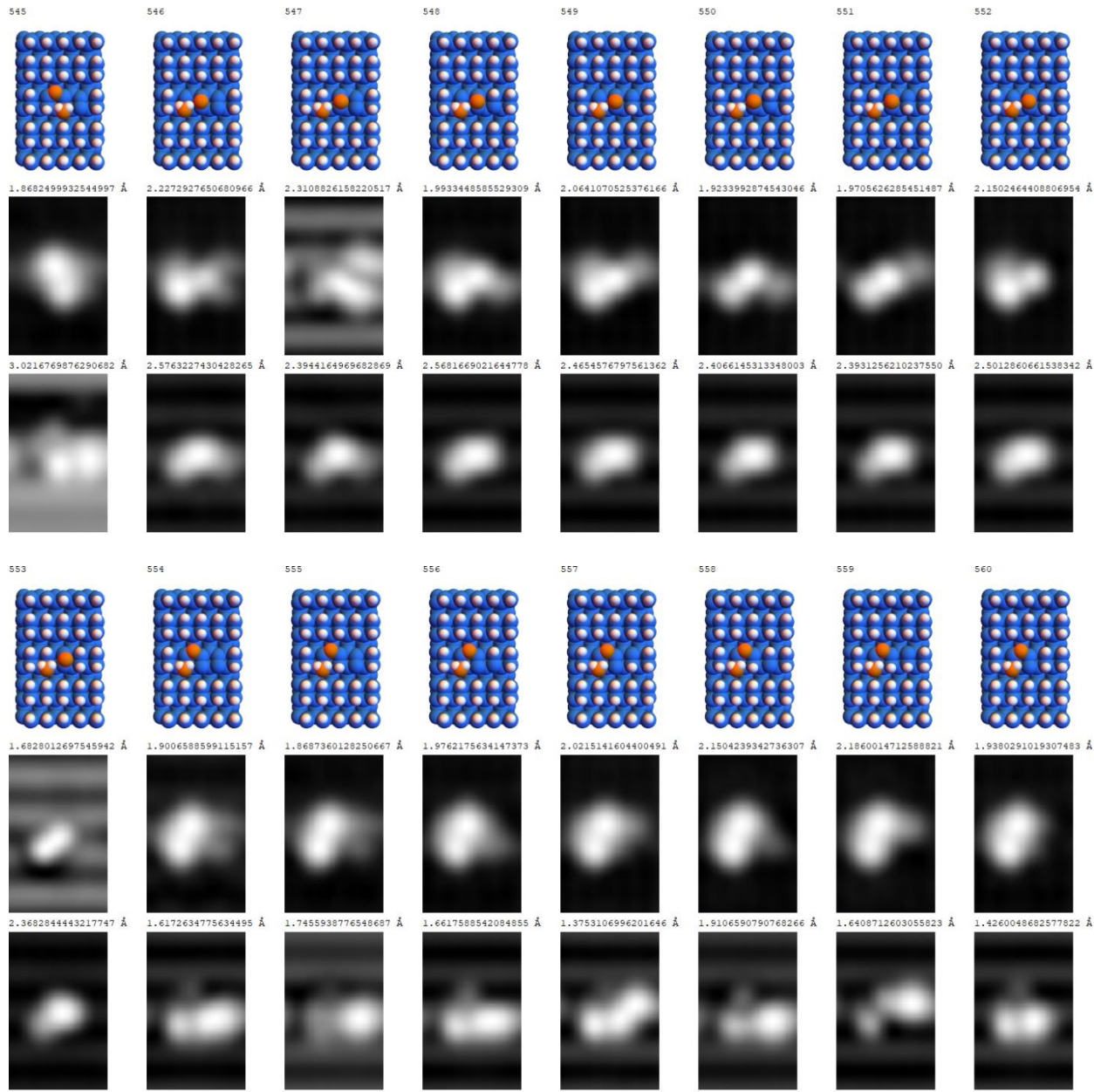

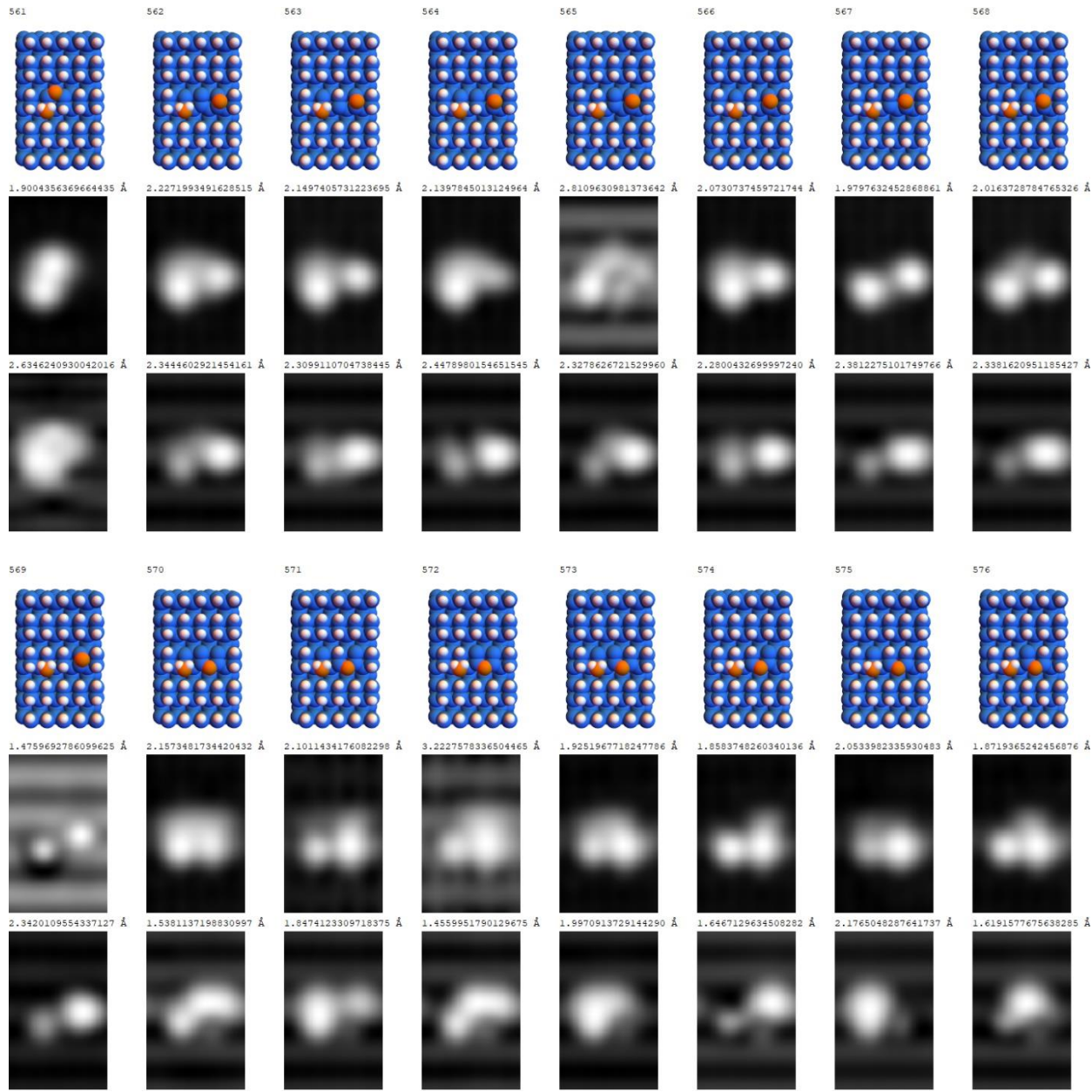

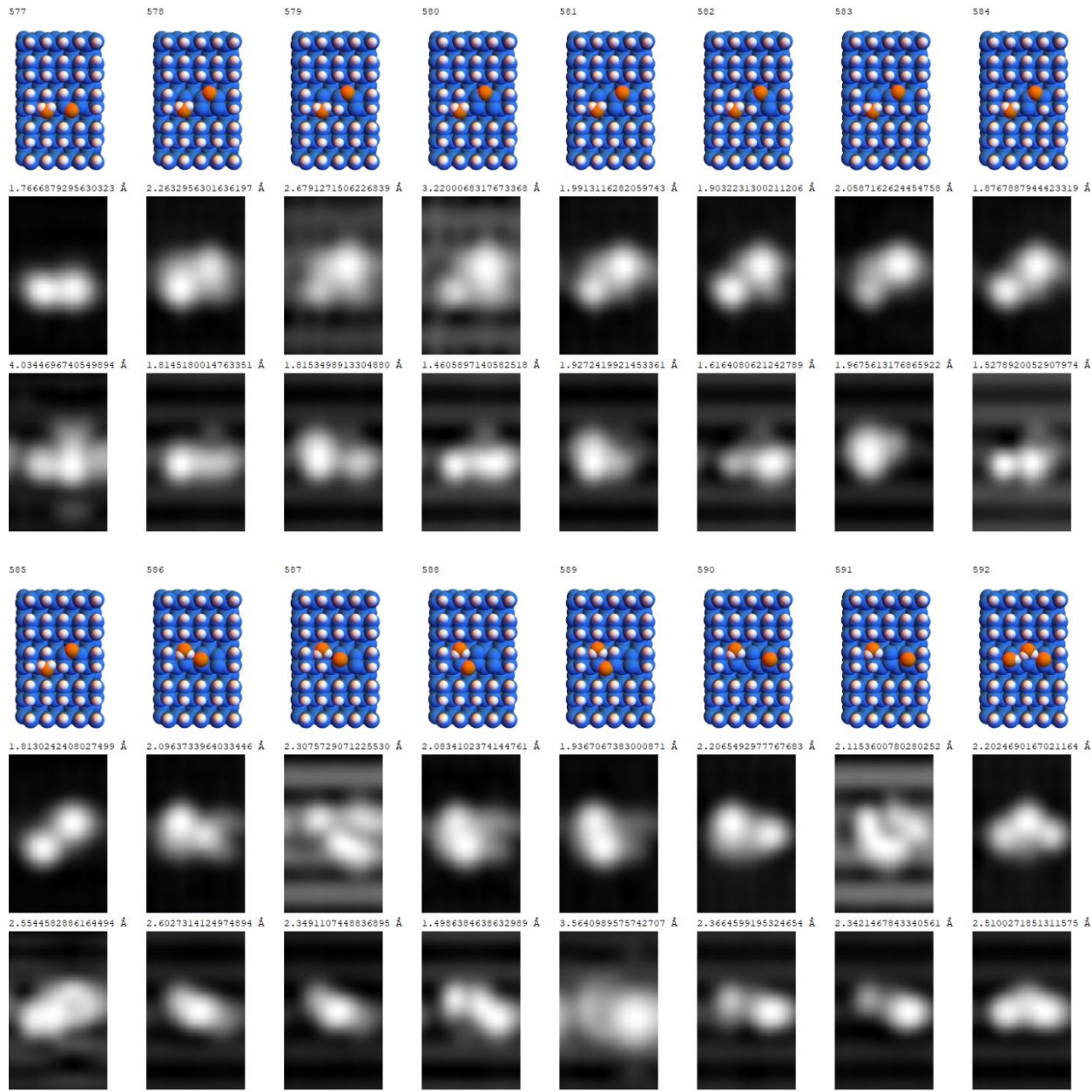

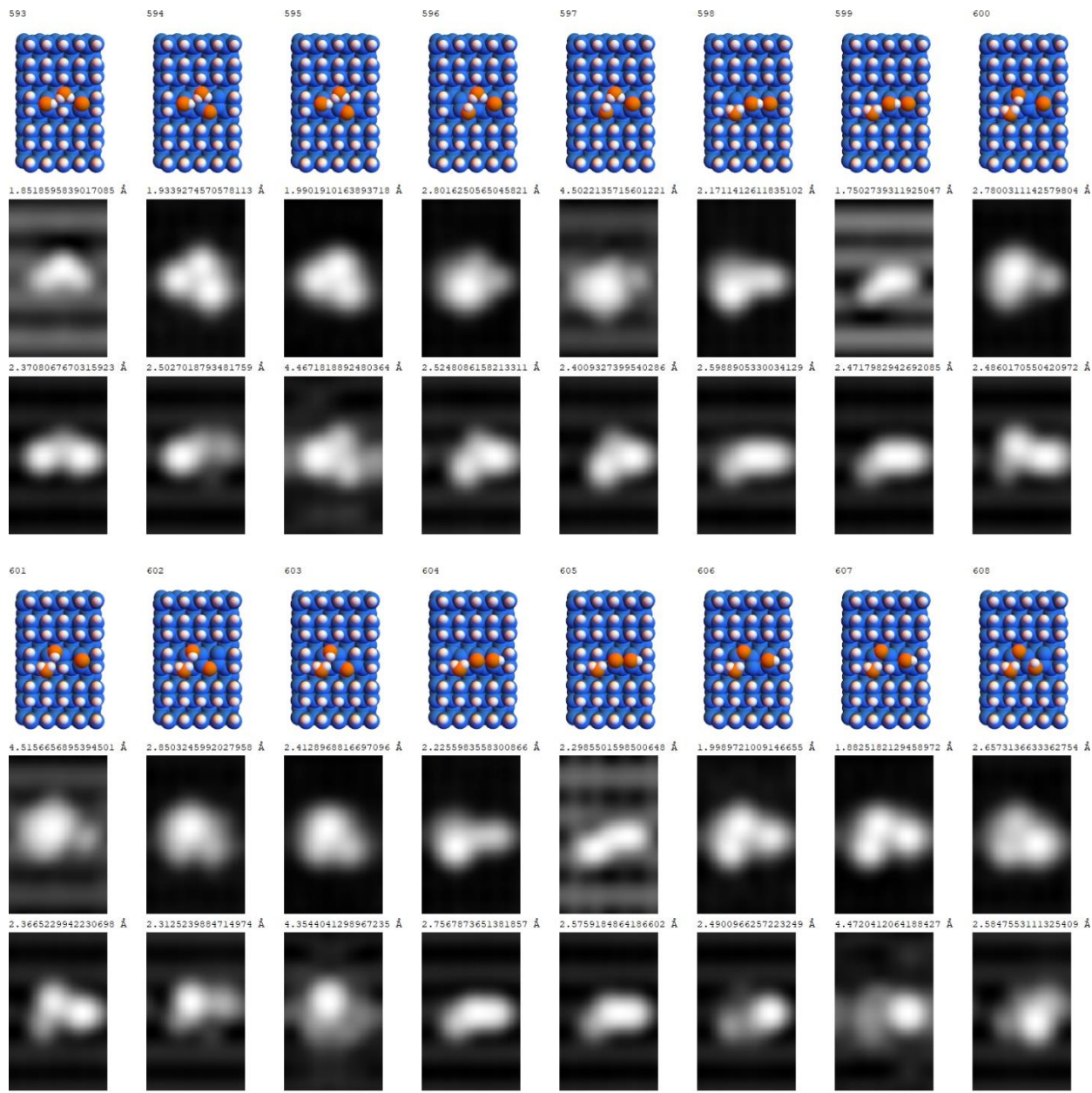

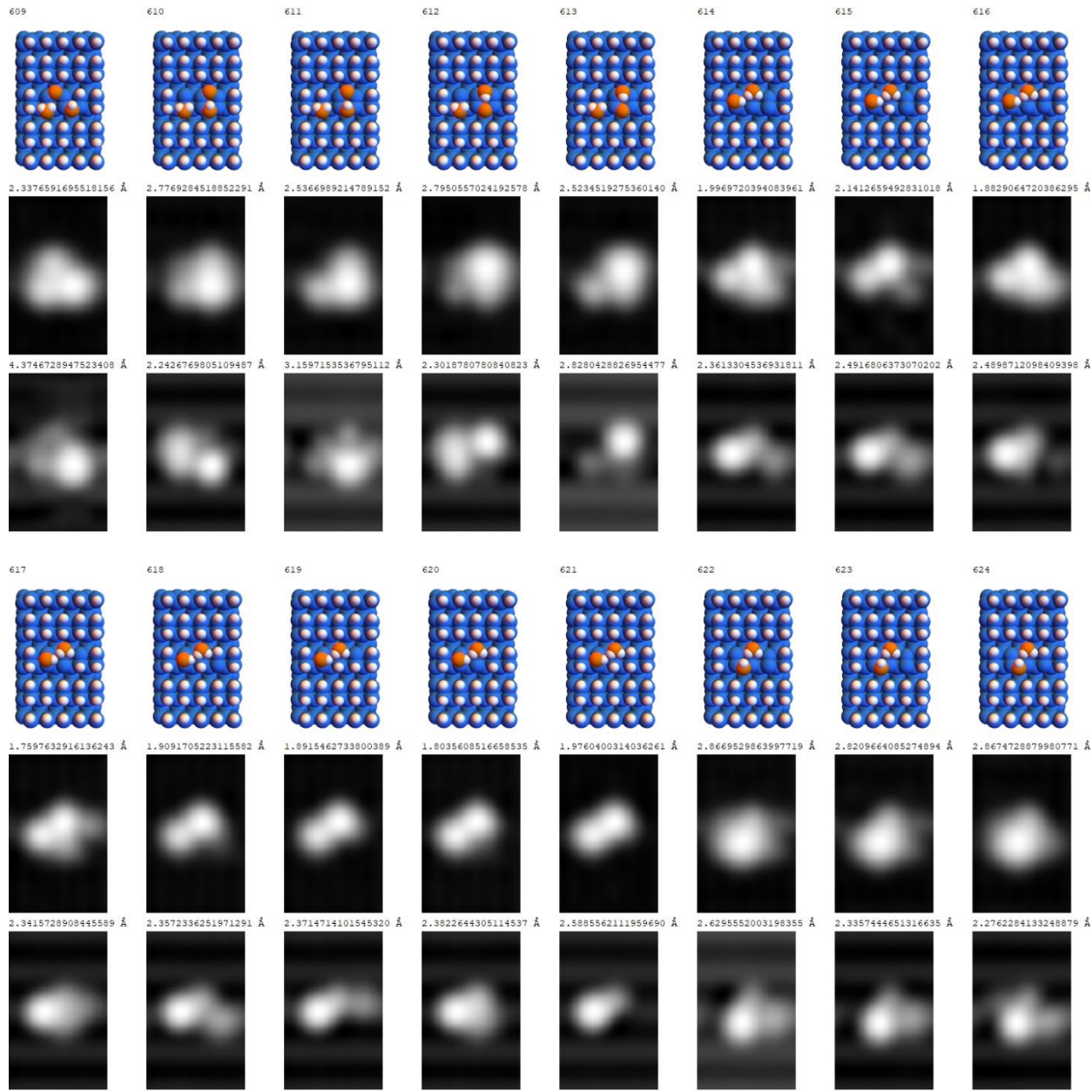

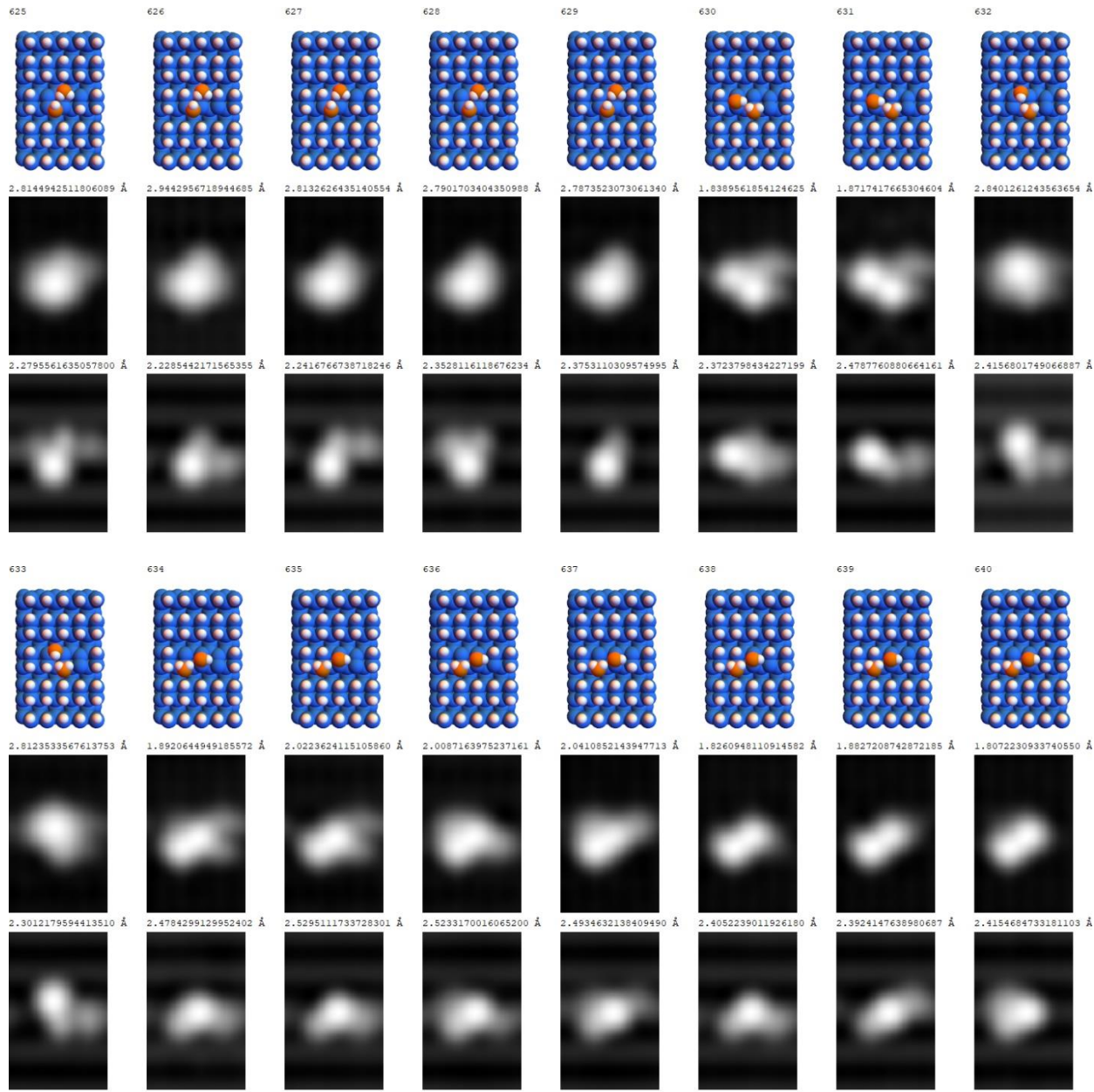

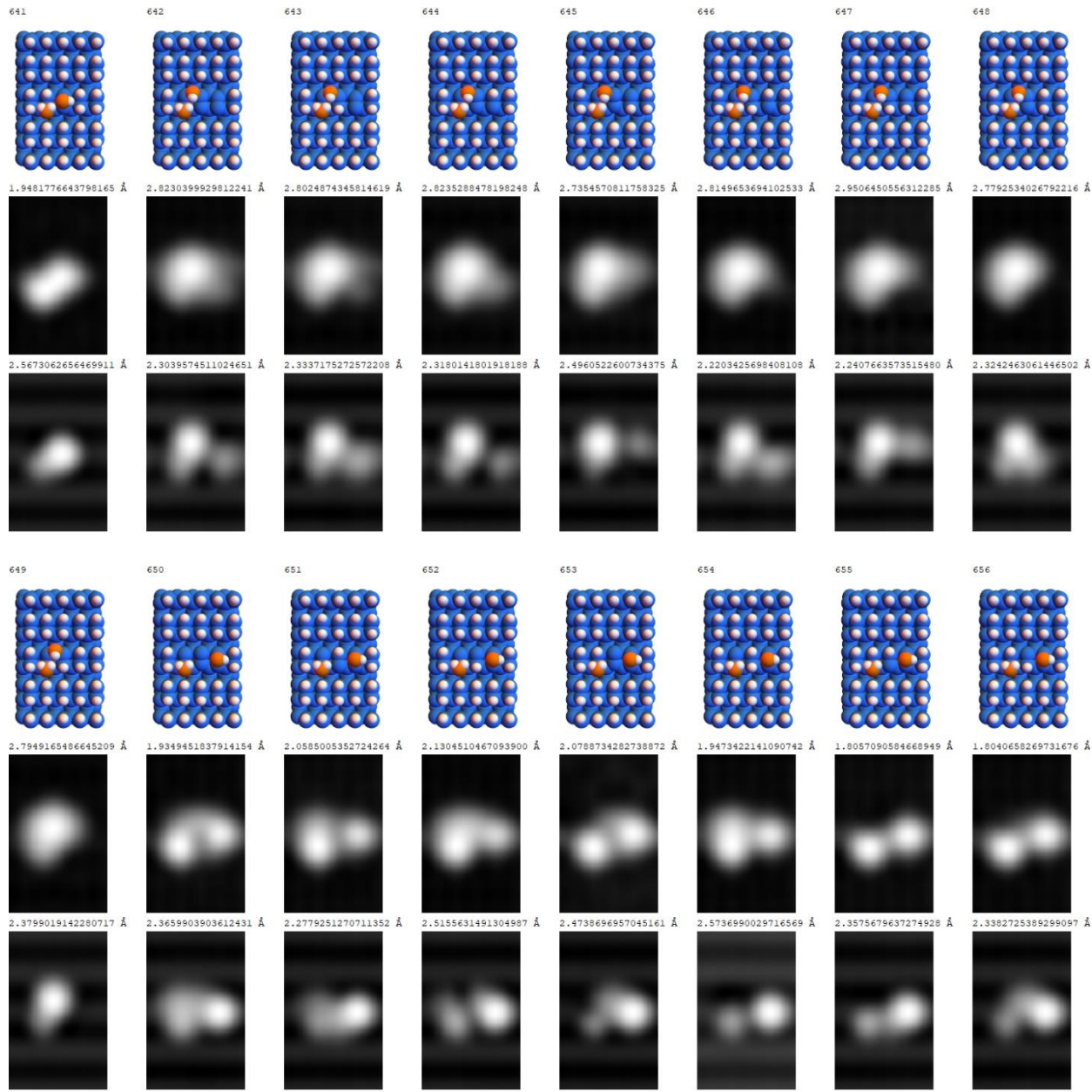

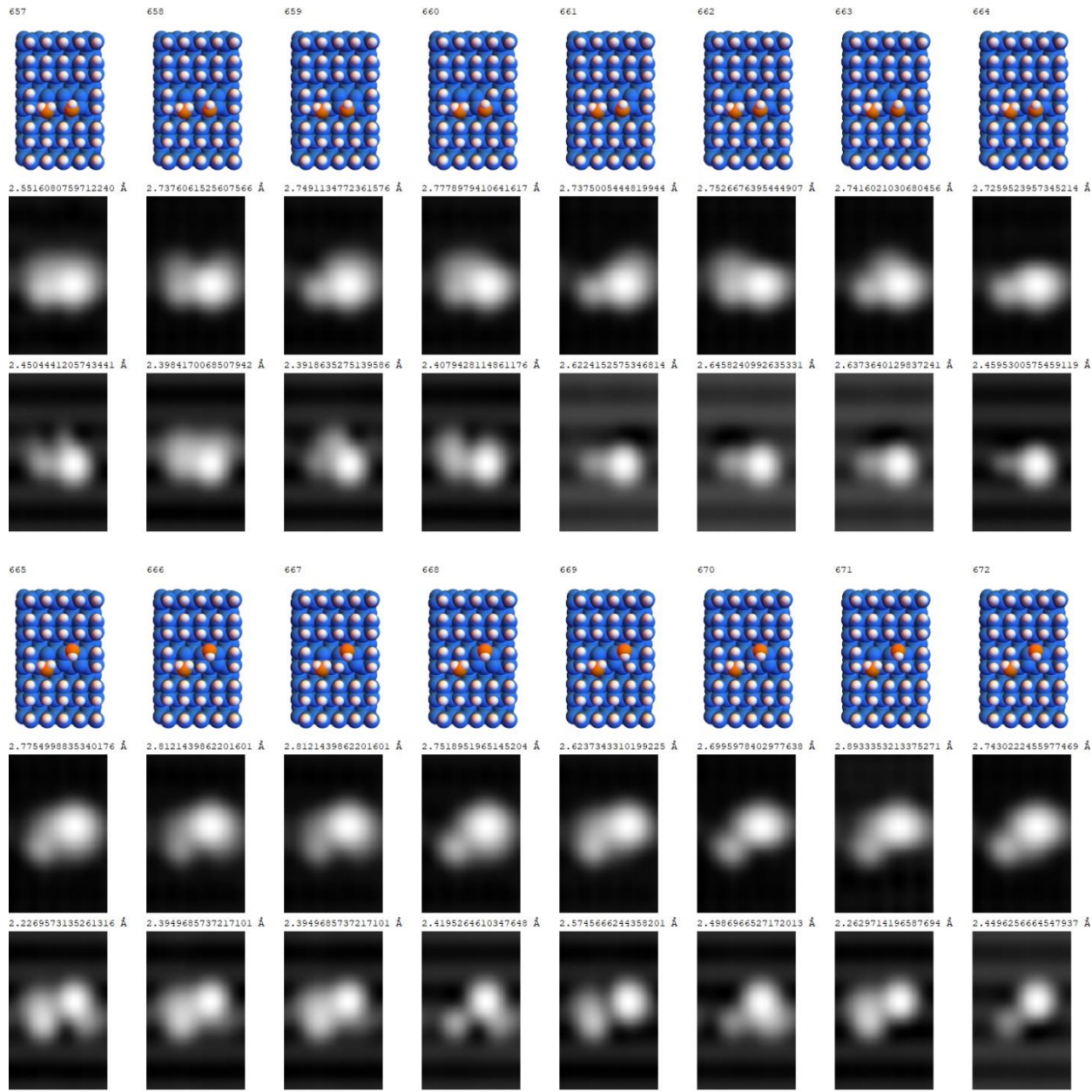

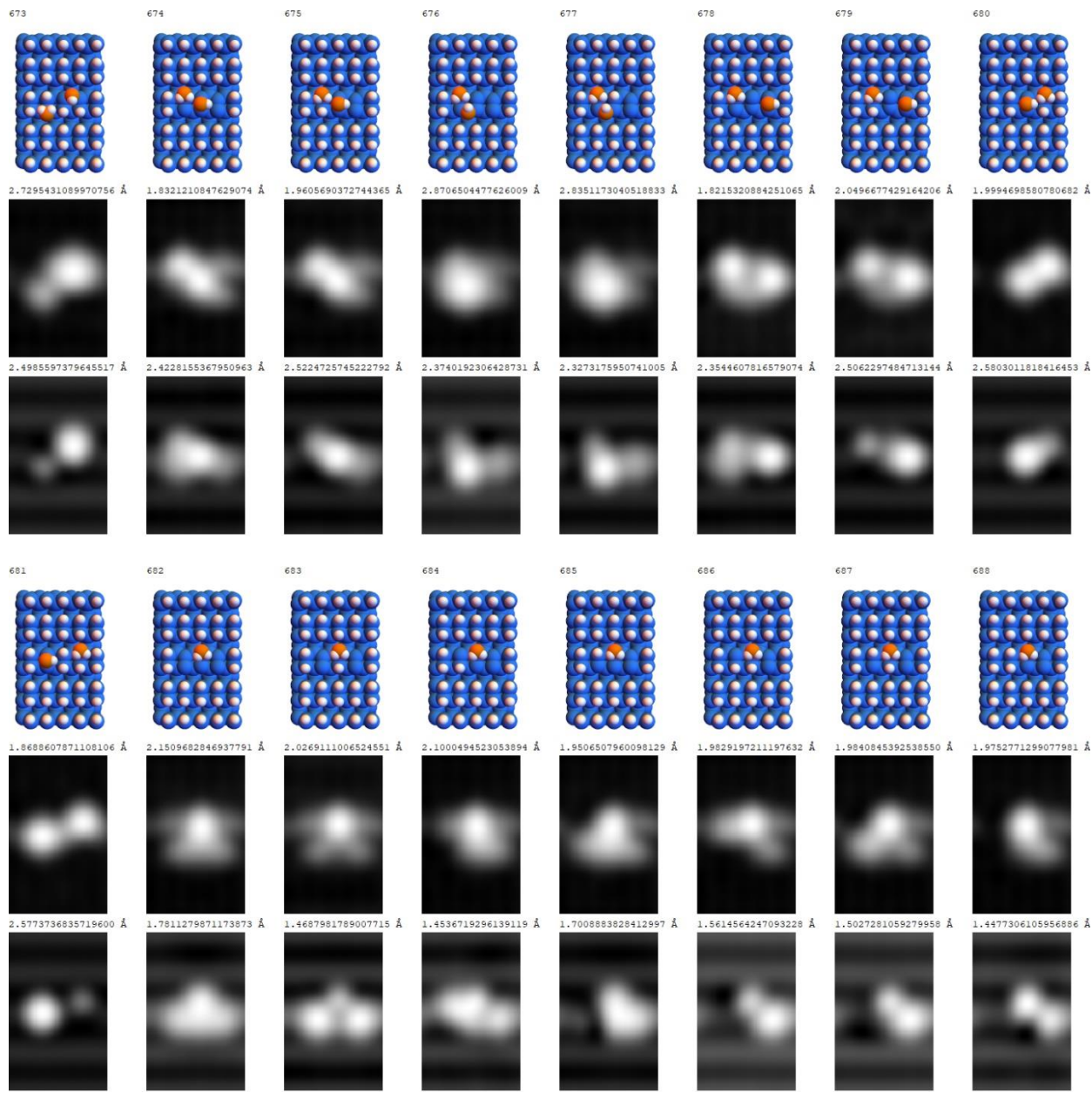

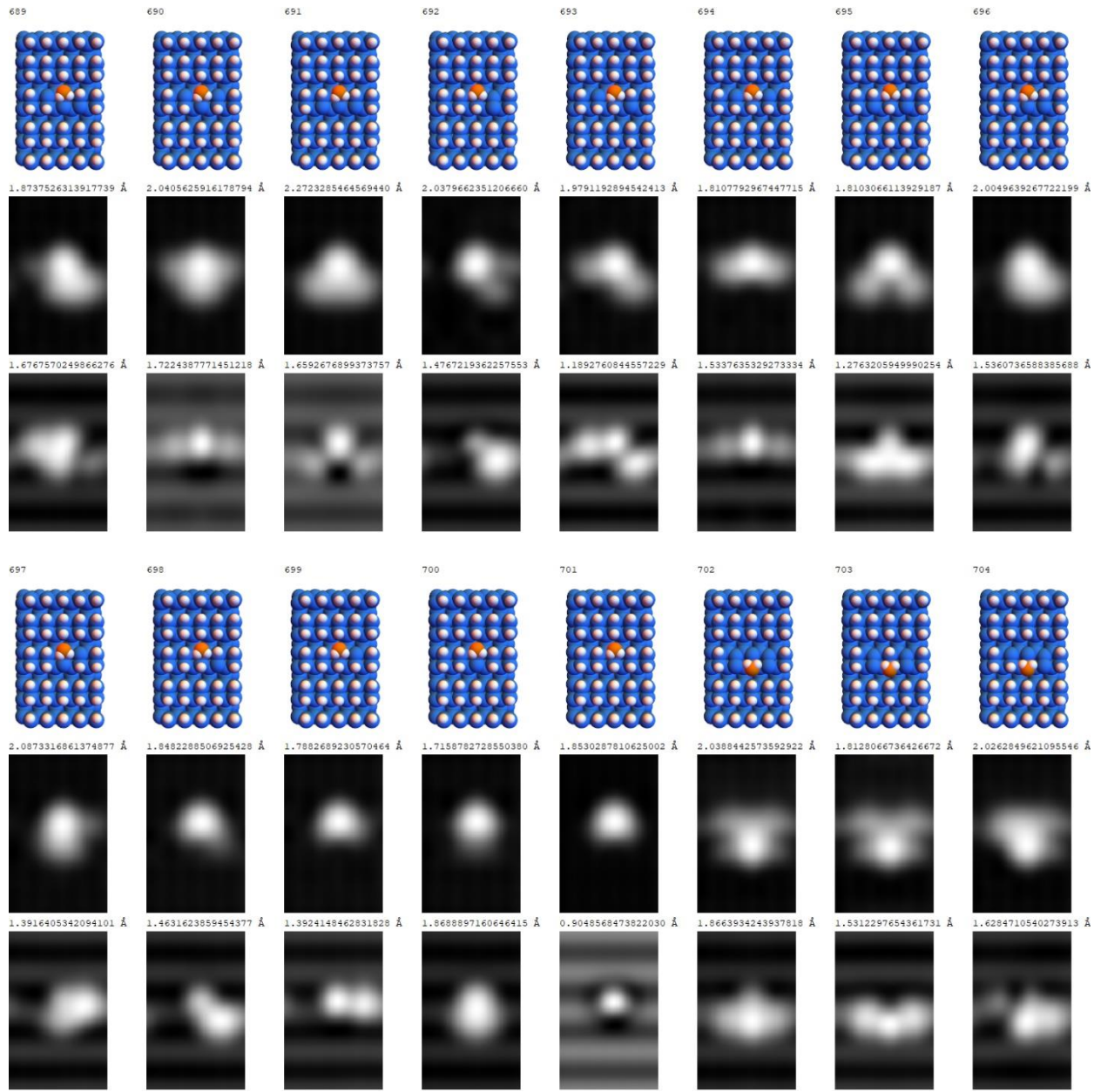

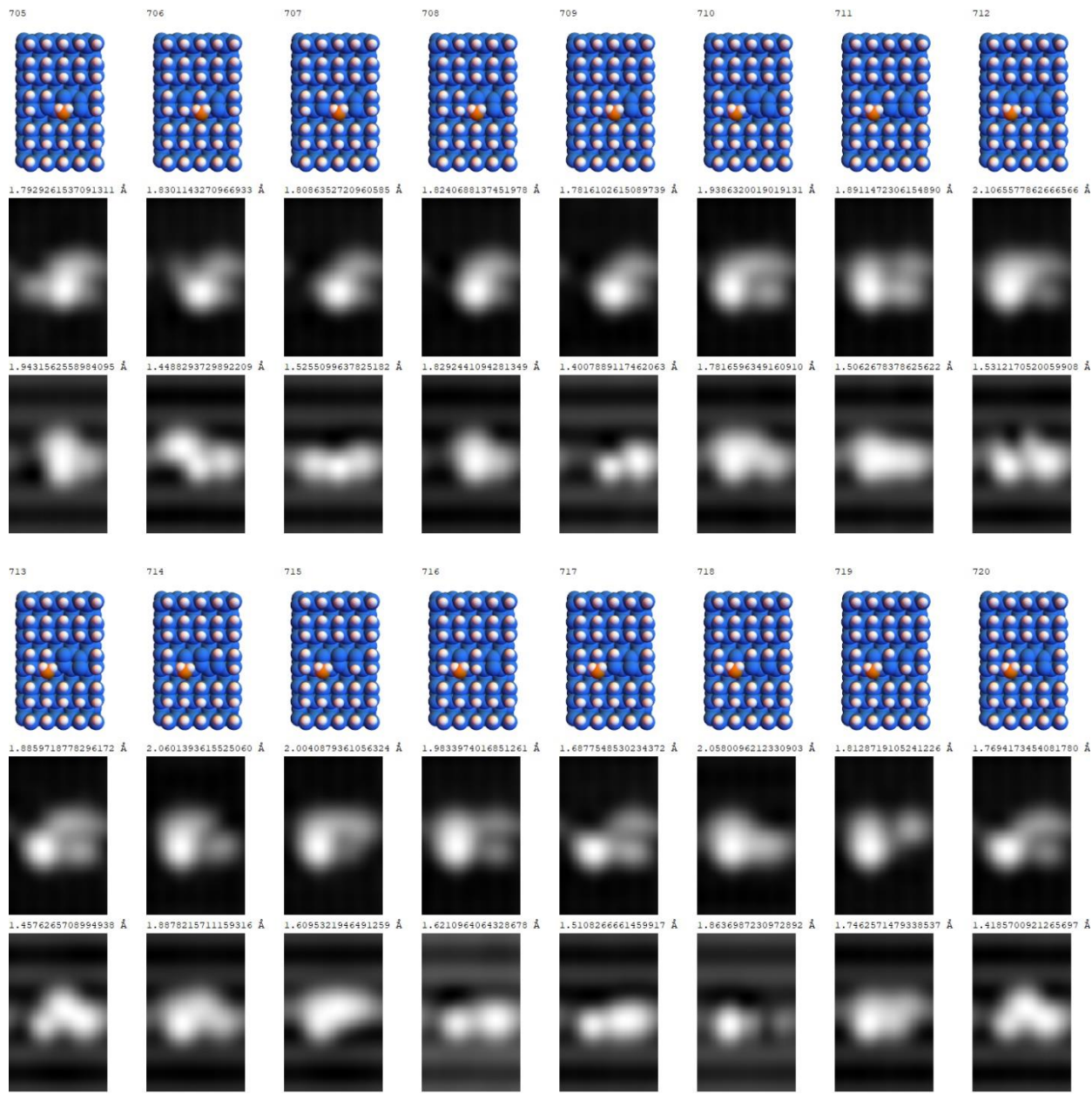

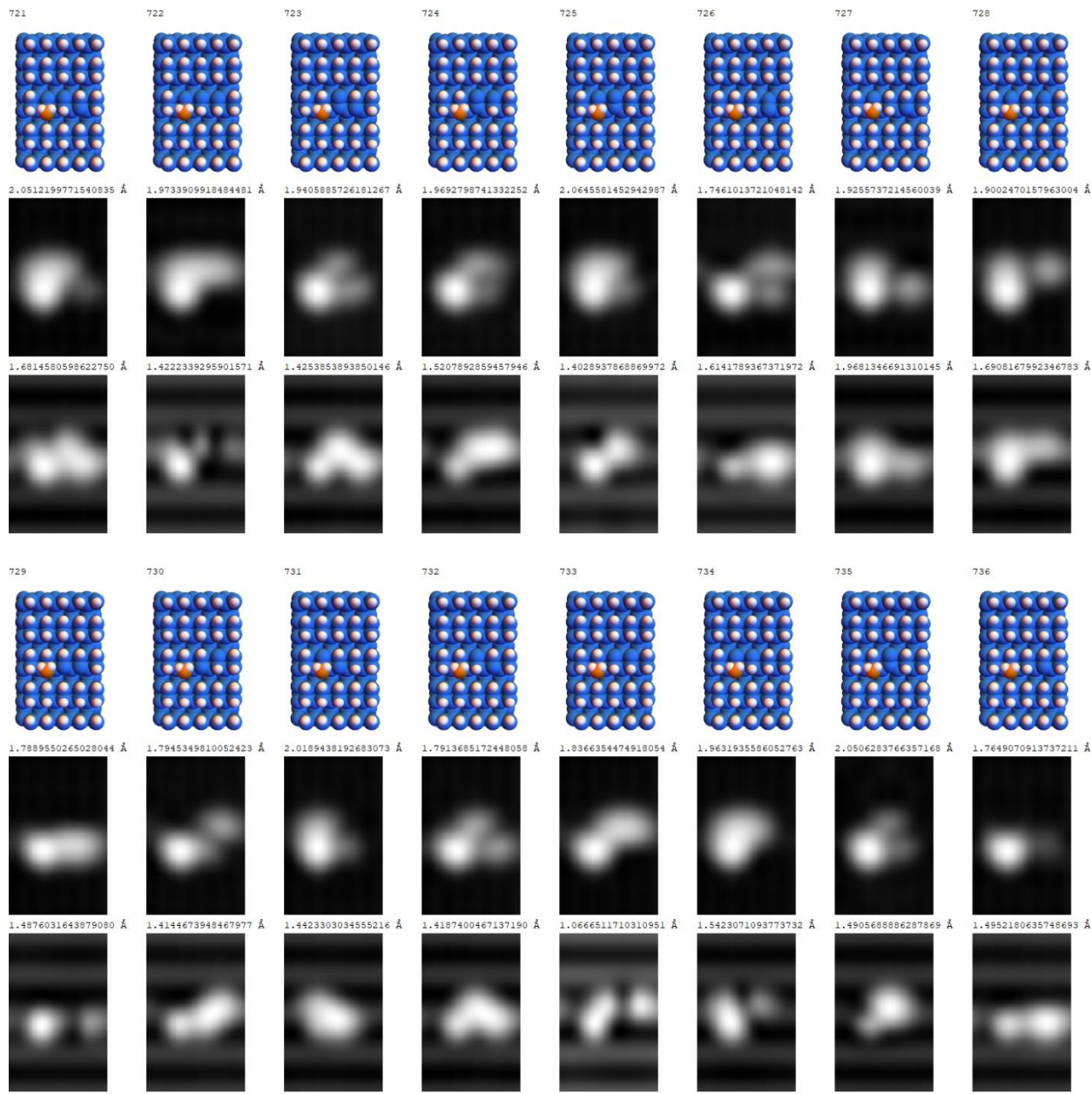

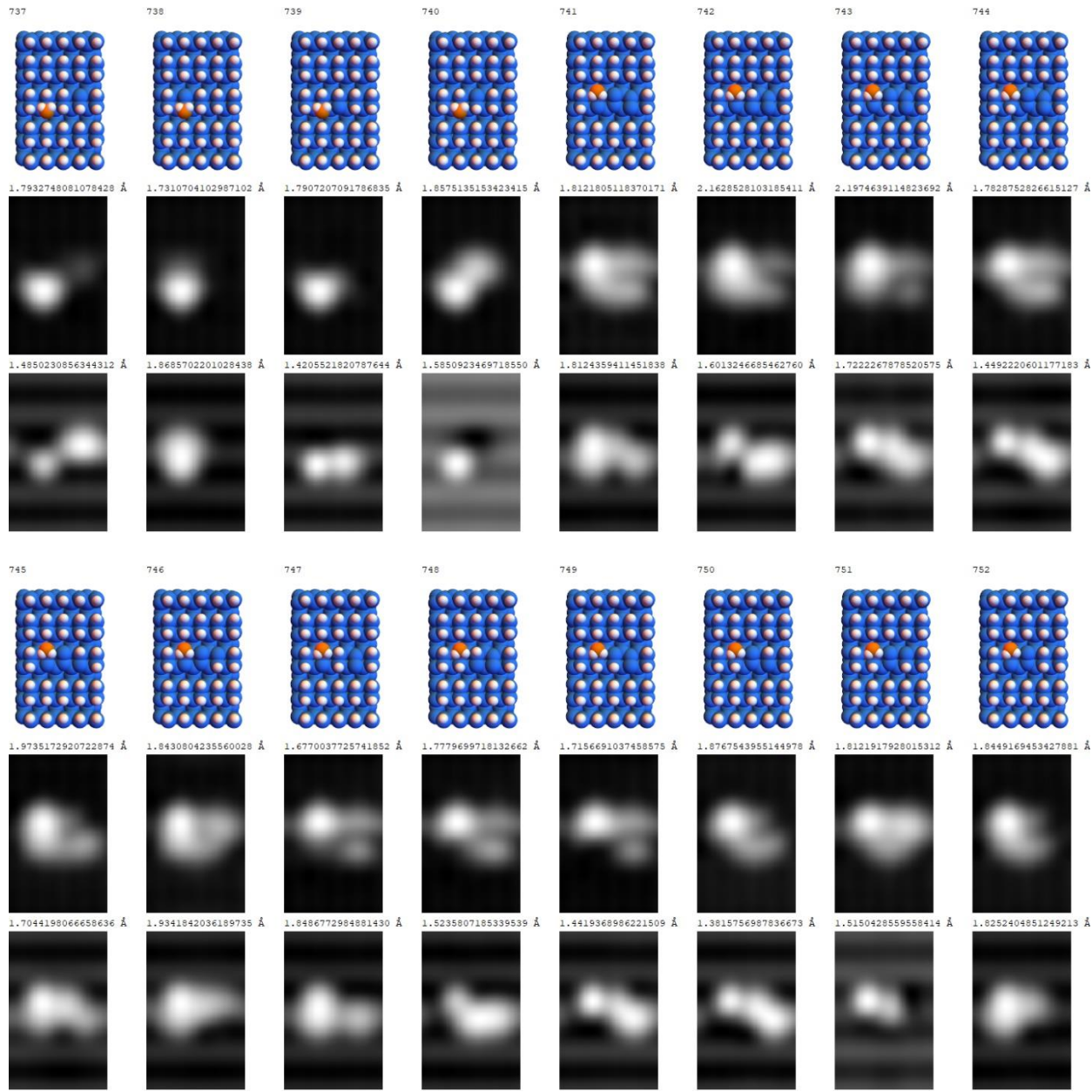

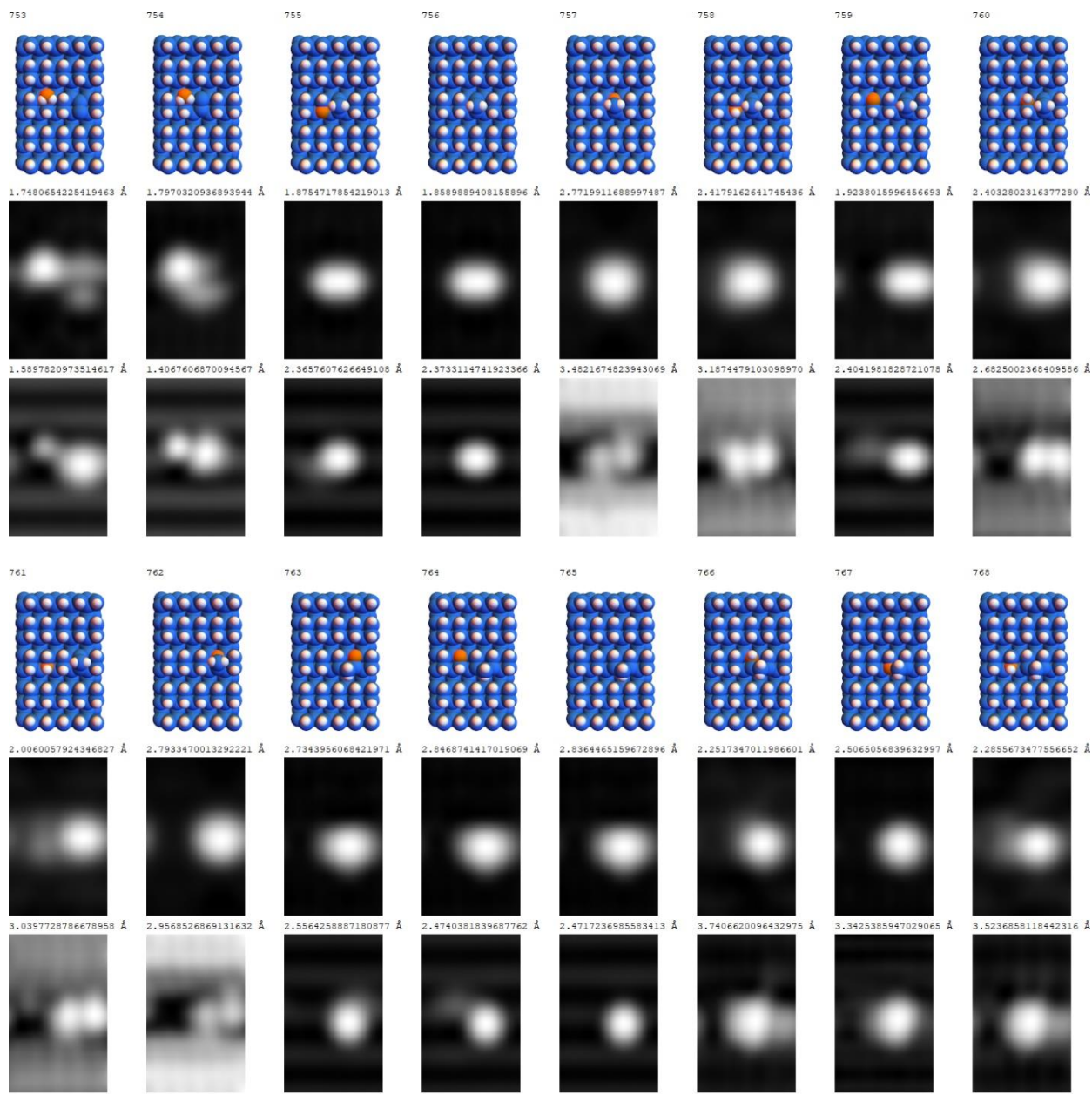

**Figure S13.** Simulated images for all possible configurations having the numbers of species specified in Table S2. The first row shows an overhead view of the atomic setup used to generate the simulated images. The second row shows the corresponding positive bias images, while the third row shows negative bias. Numbers above each simulated image give the height difference between black and white for the respective image.